\documentclass{jfm}
\usepackage{graphicx}
\usepackage{amsmath}
\usepackage{relsize}
\usepackage{amssymb}  
\usepackage{dcolumn}% Align table columns on decimal point
\usepackage{bm}% bold math
\usepackage{bigints}
\usepackage{enumitem}
\usepackage{float}
\usepackage{verbatim}
\usepackage{makecell}
\usepackage{enumitem}
\usepackage[export]{adjustbox}
\usepackage{scalerel}[2016-12-29]
\usepackage[dvipsnames]{xcolor}
\usepackage{multirow,bigdelim}
\usepackage{lscape}
\usepackage{lineno}
\usepackage{tikz}
\usepackage{siunitx}
\usepackage{makerobust}
\MakeRobustCommand\rotatebox
\usepackage[colorlinks=true, linkcolor=blue, urlcolor=blue, citecolor=blue]{hyperref}

\newcount\ndots
\def\drwln#1#2{\raise 2.5pt\vbox{\hrule width #1pt height #2pt}}
\def\spc#1{\hskip #1pt}
\def\solid{\drwln{24}{1}\ }
\def\thick{\drwln{24}{2.0}\ }

\def\dashed{\hbox {\drwln{4}{1}\spc{2}
                   \drwln{4}{1}\spc{2}\drwln{4}{1}}\nobreak\ }
\def\dasheddotted {\hbox {\drwln{4}{1.0}\spc{2}
                   \drwln{1}{1.0}\spc{2}\drwln{4}{1.0}}\nobreak\ }

\def\trian    {$\triangle$\nobreak\ }

\def\filsqr   {${\vcenter{\hrule height 2pt
                          \hbox{\vrule width 2.2pt height 0.2pt \kern 0.1pt
                                \vrule width 2.2pt}
                                \hrule height 2.2pt}}$\nobreak\ }

\def\lsqr{\hbox {\drwln{8}{1}\spc{1}$\square$
                  \drwln{8}{1}}\nobreak\ }

\newcommand\black[1]{{\color{black}#1}}

\newcommand\ar[1]{{\color{black}#1}}
\definecolor{forestgreen}{rgb}{0.13, 0.5, 0.13}

\definecolor{mygray}{gray}{0.6}
\definecolor{mypink3}{cmyk}{0, 0.7808, 0.4429, 0.1412}

\definecolor{hotmagenta}{rgb}{1.0, 0.11, 0.81}
\definecolor{aqua}{rgb}{0.0, 1.0, 1.0}

\makeatletter
\newcommand{\ostar}{\mathbin{\mathpalette\make@circled{\color{aqua}\times}}}
\newcommand{\ostarr}{\mathbin{\mathpalette\make@circled{\color{magenta}\times}}}
\newcommand{\make@circled}[2]{%
  \ooalign{$\m@th#1\smallbigcirc{#1}$\cr\hidewidth$\m@th#1#2$\hidewidth\cr}%
}
\newcommand{\smallbigcirc}[1]{%
  \vcenter{\hbox{\scalebox{0.77778}{$\m@th#1\bigcirc$}}}%
}

\newcommand{\blackdiam}{\rotatebox[origin=c]{45}{$\blacksquare$}}

\makeatother

\begin{document}
\setlength{\unitlength}{1.2cm}

\title{Turbulent drag reduction by spanwise wall forcing. Part 1: Large-eddy simulations}
\shortauthor{Rouhi, Fu, Chandran, Zampiron, Smits and Marusic}

\shorttitle{Turbulent drag reduction by spanwise wall forcing. Part 1: LES}

\author{A. Rouhi$^{1}$ \corresp{\email{amirreza.rouhi@ntu.ac.uk}}, M. K. Fu$^{2}$, D. Chandran$^3$, A. Zampiron$^{4}$, A. J. Smits$^5$ \and I. Marusic$^3$}
\affiliation{$^1$Department of Engineering, School of Science and Technology\\
                Nottingham Trent University, Nottingham NG11 8NS, United Kingdom\\[5pt]
             $^2$Graduate Aerospace Laboratories (GALCIT),
                Caltech, Pasadena, CA 91125, USA\\ [5pt]
             $^3$Department of Mechanical Engineering,
                University of Melbourne, Victoria 3010, Australia\\[5pt]
             $^4$School of Engineering, University of Aberdeen,\\
                King's College, Aberdeen AB24 3FX, United Kingdom\\ [5pt]                 
             $^5$Department of Mechanical and Aerospace Engineering,
             Princeton University,\\ Princeton, NJ 08544, USA}
\maketitle
% \linenumbers

\begin{abstract}
Turbulent drag reduction through streamwise travelling waves of spanwise wall oscillation is investigated over a wide range of Reynolds numbers.  Here, in Part 1, wall-resolved large-eddy simulations in a channel flow are conducted to examine how the frequency and wavenumber of the travelling wave influence the drag reduction at friction Reynolds numbers \ar{$Re_\tau = 951$} and $4000$.  The actuation parameter space is restricted to the inner-scaled actuation (ISA) pathway, where drag reduction is achieved through direct attenuation of the near-wall scales.  The level of turbulence attenuation, hence drag reduction, is found to change with the near-wall Stokes layer protrusion height $\ell_{0.01}$.  A range of frequencies is identified where the Stokes layer attenuates turbulence, lifting up the cycle of turbulence generation and thickening the viscous sublayer; in this range, the drag reduction increases as $\ell_{0.01}$ increases up to $30$ viscous units. Outside this range, the strong Stokes shear strain enhances near-wall turbulence generation leading to a drop in drag reduction with increasing $\ell_{0.01}$.  We further find that, within our parameter and Reynolds number space, the ISA pathway has a power cost that always exceeds any drag reduction savings.  This motivates the study of the outer-scaled actuation (OSA) pathway in Part 2, where drag reduction is achieved through actuating the outer-scaled motions.

\end{abstract}


\section{Introduction}\label{sec:intro}
Flow control aims to reduce drag on vehicles, enhance their efficiency, manoeuvrability, and possibly modify the heat transfer. Techniques for flow control cover a variety of fields, and have been extensively reviewed~\citep{white2008,dean2010,luchini2022}.  Flow control devices are usually divided into two groups: passive devices that are fixed in place and do not change their shape or function in time, such as vortex generators~\citep{lin2002,koike2004,aider2010} and riblets~\citep{garcia2011a,garcia2012,endrikat2021,endrikat2020,modesti2021dispersive,endrikat2022reorganisation,rouhi2022riblet}, and active devices that can be actuated in some way, such as targeted blowing \citep{abbassi2017} or intermittent blowing and suction~\citep{segawa2007,hasegawa2011,yamamoto2013,schatzman2014,kametani2015}. 

Here, we are interested in a particular form of active control for drag reduction in wall-bounded flows based on spanwise oscillation of the surface, leading to the generation of a streamwise travelling wave \citep{jung1992suppression, quadrio2009,viotti2009,quadrio2011b,quadrio2011,gatti2012,gatti2013,gatti2016, ricco2021review}. The wall motion is described by
\begin{equation}
	w_s(x,t)=A \sin{(\kappa_x x - \omega t)},
	\label{eq:wallmotion}
\end{equation}
where $w_s$ is the instantaneous spanwise velocity of the wall surface, $A$ is the amplitude of the spanwise forcing, $\omega$ is the angular frequency of oscillation, and $\kappa_x = 2\pi/\lambda$ is the wavenumber of the travelling wave with wavelength $\lambda$. Negative frequencies result in an upstream travelling wave, and vice versa.  With an appropriate choice of $A, \kappa_x$ and $\omega$, turbulent drag reduction beyond $40\%$ can be achieved~\citep{quadrio2000numerical,quadrio2009,hurst2014,gatti2016}. The actuation mechanism (\ref{eq:wallmotion}) has been mostly investigated in a turbulent channel flow. So far, the only studies that investigate this mechanism in a turbulent boundary layer are the numerical work by \cite{skote2022drag}, and the experimental work by \cite{bird2018experimental} and \cite{Chandran2022part2} in Part 2.

The amount of drag reduction, $DR$, is defined as
\begin{equation}
 DR = \frac{C_{f_0} - C_f}{C_{f_0}}, \label{eq:dr}
\end{equation}
where $C_f \equiv 2\overline{\tau_w}/(\rho U^2_{b,\infty})$ and $C_{f_0} \equiv 2\overline{\tau_{w_0}}/(\rho U^2_{b,\infty})$ are the skin-friction coefficients of the drag-reduced flow (with wall shear-stress $\overline{\tau_w}$) and the non-actuated flow (with wall shear-stress $\overline{\tau_{w_0}}$) and $\rho$ is the fluid density. The overbar in $\overline{\tau_w}$ and $\overline{\tau_{w_0}}$ indicates averaging over the homogeneous directions and time. In a fully-developed channel flow (considered here in Part 1), the averaging dimensions are the streamwise and spanwise directions, as well as time, and in a boundary layer (considered in Part 2), the averaging dimensions are the spanwise direction and time. Further, in a channel flow the drag-reduced flow and the non-actuated flow are exposed to the same bulk velocity $U_b$ (present Part 1, \citealt{quadrio2009,gatti2013}) or pressure gradient~\citep{quadrio2011,ricco2012}, however, in a boundary layer the two flows are exposed to the same free-stream velocity $U_\infty$ (Part 2, \citealt{bird2018experimental}). Accordingly, there are two friction velocities $u_\tau \equiv \sqrt{\overline{\tau_w}/\rho}$ and $u_{\tau_0} \equiv \sqrt{\overline{\tau_{w_0}}/\rho}$, corresponding to the drag-reduced and non-actuated cases, respectively, leading to two choices of normalisation. In the current study, following \cite{gatti2016}, the viscous-scaled quantities that are normalised by $u_{\tau_0}$ are denoted by the `$+$' superscript, and those normalised by $u_\tau$ are denoted by the `$*$' superscript. The friction Reynolds number $Re_\tau$ in a channel flow (Part~1) is defined based on $u_{\tau_0}$ and the channel half-height $h$ $(Re_\tau \equiv u_{\tau_0} h/\nu)$. In a boundary layer (Part~2), is defined based on $u_{\tau_0}$ and the boundary layer thickness $\delta$ $(Re_\tau \equiv u_{\tau_0} \delta/\nu)$. By dimensional analysis \citep{gatti2016,marusic2021} we obtain
\begin{equation}
 DR = DR\left( \kappa^+_x, \omega^+, A^+, Re_\tau \right), \label{eq:dr_space}
\end{equation}
where $\kappa^+_x = \kappa_x \nu/u_{\tau_0},\, \omega^+ = \omega \nu/u^2_{\tau_0}$ and $A^+ = A/u_{\tau_0}$. 
%Given the complexity of the parameter space in (\ref{eq:dr_space}), previous studies focused on the response of $DR$ to certain parameters. 

\cite{quadrio2009} studied this flow control problem using direct numerical simulations (DNS) of a turbulent channel flow. Their study acted as a proof of concept for (\ref{eq:wallmotion}) to demonstrate that the introduction of a \ar{streamwise} travelling wave achieves higher $DR$ than a purely oscillating wall mechanism ($\kappa_x = 0$). They fixed $Re_\tau = 200$ and $A^+ = 12$, and populated a map of $DR(\omega^+,\kappa^+_x)$ for $0 \le \kappa^+_x \le +0.04$ and $-0.3 \le \omega^+ \le +0.3$. \cite{gatti2016} extended this work to $Re_\tau = 1000$ and a broader range of actuation parameters ($0 \le \kappa^+_x \le +0.05, -0.6 \le \omega^+ \le +0.6$ and $3 \le A^+ \le 15$) to construct isosurfaces of $DR(\omega^+, \kappa^+_x, A^+)$ in the 3D actuation parameter space (figure~4 in \citealt{gatti2016}). 
They observed that this type of actuation appears to modify the mean velocity profile through a Reynolds number-invariant additive constant, \ar{$\Delta B$}, in the logarithmic region as:
\begin{equation}
 U^* = \frac{1}{\kappa} \ln (y^*) + B + \Delta B, \label{eq:log}
\end{equation}

\noindent where $U^* \equiv U/u_\tau$ and $y^* \equiv y u_\tau / \nu$ are the viscous-scaled velocity and wall distance, %$\kappa \simeq 0.38 - 0.4$ and $B \simeq 4.3 - 5.5$ 
$\kappa$ and $B$ are the von K\'arm\'an and additive constants for the non-actuated channel. %flow~\citep{kim1987,pirozzoli2016,lee2015}. 
This behaviour in $U^*$ implies that the actuation is primarily acting on turbulent structures in the near wall region and that the outer flow effectively perceives the modified inner layer as one that has a lower stress. \ar{This behaviour is similar to the flows over riblets and rough surfaces \citep{chan2015,squire2016b,endrikat2020}} and \citet{gatti2016} used this assumption to propose the modified friction law (hereafter called GQ's model) given by
\begin{equation}
 \Delta B  = \sqrt{\frac{2}{C_{f_0}}} \left[ \left( 1 - DR \right)^{-1/2} - 1 \right] - \frac{1}{2\kappa} \ln{\left( 1 - DR \right)}. \label{eq:gq_model}
\end{equation}

\noindent In this framework, the Reynolds number dependence of the flow is captured by \ar{$C_{f_0}$}, provided that there is a well-defined logarithmic region in the mean velocity profile. The \ar{behaviour} of the log-region is modified by the actuation solely through the offset parameter \ar{$\Delta B$; this parameter} is independent of Reynolds number and can \ar{be parameterised} by the dimensionless actuation parameters so that  $ \Delta B = \Delta B (\kappa^*_x,\, \omega^*,\, A^*)$. The model therefore predicts $DR$ at arbitrarily high Reynolds numbers for a given set of actuation parameters. The model also predicts that $DR$ decreases monotonically with increasing $Re_\tau$, regardless of the actuation parameters.  To date, the predictions from this model have been found to be largely consistent with the existing low-Reynolds number simulations of travelling wave drag reduction~\citep{baron1995,yudhistira2011,ricco2012,touber2012,hurst2014}. 

The findings reported so far are based on DNS of turbulent channel flow. Experiments have also reported the efficacy of spanwise wall forcing for turbulent drag reduction. The configurations are mainly turbulent boundary layer~\citep{choi1998,choi2001mechanism,ricco2004,bird2018experimental} or pipe flow~\citep{choi1998drag,auteri2010experimental}. The experiments mostly consider uniform spanwise wall oscillation (i.e.\ $\kappa_x = 0$ in \ref{eq:wallmotion}). The exceptions are \cite{auteri2010experimental} and \cite{bird2018experimental} that attempt to mimic the travelling wave motion. \cite{auteri2010experimental} subdivide the pipe wall into thin slabs that rotate independently, and \cite{bird2018experimental} pneumatically deform a compliant structure. The experimental findings are consistent with the DNS findings. They report $DR$ between $21\%$~\citep{bird2018experimental} to $45\%$~\citep{choi2001mechanism}. They also observe the shift in the log region (\ref{eq:log}) that underlies GQ's model~\citep{choi1998,choi2001mechanism,ricco2004}. The DNS and experimental studies reviewed so far consider $Re_\tau \lesssim 1500$.

\citet{marusic2021} recently investigated the parameter space (\ref{eq:dr_space}) at much higher Reynolds numbers by conducting experiments up to $Re_\tau = 12800$ and wall-resolved large-eddy simulations (LES) up to $Re_\tau = 2000$.
% In a recent study of travelling wave drag reduction by \citet{marusic2021}, \dc{experiments} were conducted up to $Re_\tau = 12800$ and wall-resolved large-eddy simulation (LES) was carried out up to $Re_\tau = 2000$. 
By covering such a large Reynolds number range, they were able to explore the increasing contribution of \ar{turbulent scales in the log-region and beyond} to the total  drag~\citep{marusic2010,smits2011,mathis2013,chandran2020}. In contrast to previous studies, the drag reduction was found to occur via two distinct physical pathways. The first pathway, which \cite{marusic2021} referred to as the ``small-eddy'' actuation strategy, as was applied in previous studies. It will be more aptly termed {\it inner-scaled actuation} (ISA) in the present work because drag reduction is achieved by actuating at frequencies associated with the near-wall cycle and the near-wall peak in turbulent kinetic energy.
For example, $\omega^+ \approx -0.06$ equates to a time period of oscillation of $T^+_{osc} = 2\pi/ \left|\omega^+ \right| = 100$. 
The $DR$ obtained under this pathway was found to follow GQ's model.
% (e.g.\ $\left| \omega^+ \right| > 0.018$, with peak energy kinetic energy around $T^+_{osc} = 2\pi/\omega^+ = 100$ ($\omega^+ = 0.063$)), which dominate turbulence in the near-wall buffer region and exhibit behaviour consistent with GQ's model \citep{gatti2016}. 
The second pathway, which \cite{marusic2021} referred to as  the ``large-eddy'' actuation strategy, was new. It involved actuating at frequencies comparable to those of the inertia-carrying eddies \ar{in the logarithmic region and beyond} ($T^+_{osc} \gg 100$).
% (e.g.\ $\left| \omega^+ \right| < 0.018$).  
It will be more aptly termed {\it outer-scaled actuation} (OSA) in the present work. Unlike the ISA pathway, the OSA pathway achieves drag reduction that increases with Reynolds number, and requires significantly less input power due to the lower actuation frequencies that are required to target the inertia-carrying eddies. 
%For instance, a net power saving ($NPS$) of order $+10\%$ was demonstrated at $Re_\tau=9700$. Here we use $NPS$ as defined by \cite{gatti2013}; see also \S  \ref{sec:power} and \eqref{eq:nps}.  
%In other words, the total power cost of the actuated case (for driving the flow plus operating \ref{eq:wallmotion}) is less than the power cost of the non-actuated case by $10\%$.
\citet{marusic2021} considered actuation frequencies with $T_{osc}^+<350$ to be primarily along the ISA pathway, and those with $T_{osc}^+>350$ to be primarily along the OSA pathway.  

In conjunction with Part 2 \citep{Chandran2022part2}, we investigate the drag reduction \eqref{eq:dr_space} over a range of parameters that have not been investigated previously, covering both the ISA and OSA pathways, and explain the physics behind the variation of $DR$ with $Re_\tau$, $\kappa^+_x$ and $\omega^+$. In this  Part 1, we focus on the ISA pathway and use wall-resolved LES to extend the parametric study of \cite{gatti2016} at $Re_\tau \approx 1000$, generating a new map of $DR$ at $Re_\tau = 4000$ over $0.002 \le \kappa^+_x \le 0.02$ and $-0.2 \le \omega^+ \le +0.2$ for $A^+ = 12$.  Accurately populating the $DR$ map required a careful study of the LES setup in terms of the subgrid-scale model, grid and computational domain size, to ensure the accuracy of the simulations and computational tractability. The resulting map at $Re_\tau = 4000$ is used to evaluate the predictive accuracy of GQ's model, and by using turbulence statistics, triple decompositions, spectrograms and flow visualisations, we identify and explain the regimes of the flow at different regions of the $DR$ map. We find that the flow regimes change with the extent of the Stokes layer generated by the surface motion. As the Stokes layer grows in size, up to the optimal range of $20-30$ viscous units, the near-wall turbulence is damped, and there is a corresponding  increase in $DR$. In contrast, growth beyond $30$ viscous units amplifies the near-wall turbulence, leading to a decrease in $DR$. Finally, we examine the power cost at $Re_\tau = 4000$ over the range of parameters considered here.

% It is found that GQ's model performs well to the extent that the logarithmic shift $\Delta B$ is computed accurately. In turbulent flows over roughness or riblets, $\Delta B$ is computed at about or above $100$ viscous units~\citep{macdonald2017,endrikat2021,endrikat2020}, as the logarithmic velocity profile appears at that distance. Here, however, we identify a critical region in the $DR(\omega^+,\kappa^+_x)$ map where the logarithmic velocity profile appears beyond $200$ viscous units (\S\ \ref{sec:error_gq_model}). We find that this is due to the Stokes layer that protrudes into the turbulent field by up to $100$ viscous units. In fact, we find that the resulting drag reduction across the actuated parameter space (in the ISA pathway) correlates with the height of the protrusive Stokes layer. The interaction of the Stokes layer with the near-wall turbulence and its effects on the drag are then investigated through Reynolds stress profiles, spectrograms and flow visualisations. 

\black{\section{Numerical flow setup}}\label{sec:setup}

\subsection{Governing equations and solution method}\label{sec:eqns}

\begin{figure}
  \centering
  \includegraphics[width=\textwidth,trim={{0.4\textwidth} {0.0\textwidth} {0.05\textwidth} {0.0\textwidth}},clip]{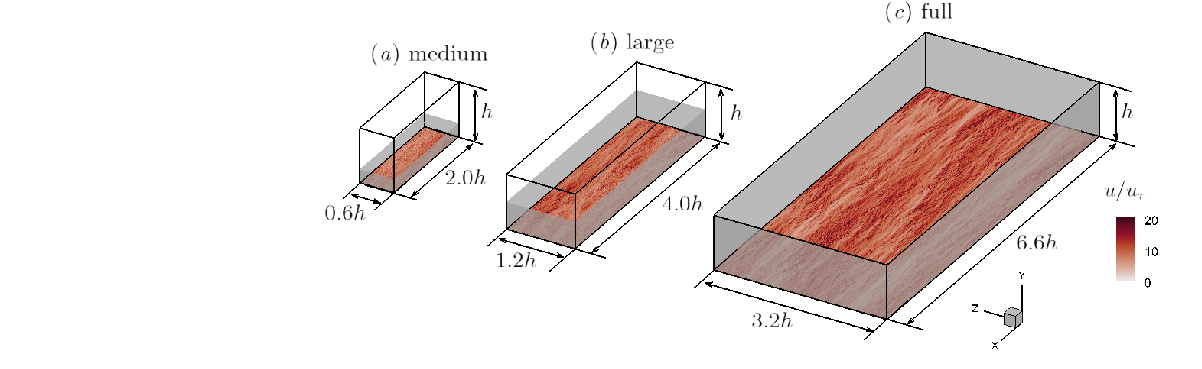}  
\caption{Various domain sizes for LES in a channel configuration - (\textit{a}) medium $2.0h \times 0.6h$, (\textit{b}) large $4.0h \times 1.2h$ and (\textit{c}) full $6.6h \times 3.2 h$. For each domain size, the instantaneous streamwise velocity ($u$) field is visualised at about $15$ viscous units above the bottom wall. The grey shaded zones indicate the wall-heights up to which the flow is resolved for each domain size ($y_\mathrm{res} \simeq 0.4 L_z$, \citealt{chung2015}).}  
  \label{fig:geometry}
\end{figure}

We solve the filtered equations for a channel flow (figure~\ref{fig:geometry}) of an incompressible fluid with constant density $\rho$ and kinematic viscosity $\nu$
\begin{equation}
 \frac{\partial \widehat{u}_i}{\partial x_i} = 0, \quad
 \frac{\partial \widehat{u}_i}{\partial t} + \frac{\partial \widehat{u}_i \widehat{u}_j}{\partial x_j} = -\frac{1}{\rho} \frac{\partial \widehat{p}}{\partial x_i} + \nu \frac{\partial^2 \widehat{u}_i}{\partial x^2_j} - \frac{\partial \tau_{ij}}{\partial x_j} + G\delta_{i1} \tag{2.1\textit{a,b}} \label{eq:mom}
\end{equation} 
The hat $\widehat{(...)}$ indicates the filtered quantity; $x_1, x_2$ and $x_3$ (also referred to as $x, y$ and $z$) are the streamwise, wall-normal, and spanwise directions, corresponding to the velocity components $\widehat{u}_1, \widehat{u}_2$ and $\widehat{u}_3$ (or $\widehat{u}, \widehat{v}$ and $\widehat{w}$), respectively. The pressure gradient in (2.1\textit{b}) is decomposed into the domain and time-averaged driving part $-\rho G$, and the periodic (fluctuating) part $\partial \widehat{p}/\partial x_i$. By averaging (2.1\textit{b}) in time and over the entire fluid domain, we obtain $G = \overline{ \tau_w} /(\rho h) = u^2_\tau/h$ where $h$ is the (open) channel height. $G$ is adjusted based on a target flowrate (i.e.\ target bulk Reynolds number $Re_b \equiv U_b h/\nu$), that is matched between the actuated and non-actuated cases. The unresolved subgrid-scale (SGS) stresses $\tau_{ij} = \widehat{u_i u_j} - \widehat{u}_i \widehat{u}_j$ are modelled using the dynamic Smagorinsky model~\citep{germano1991} incorporating Lilly's improvement~\citep{lilly1992}. For the model coefficient, we perform $xz$-plane averaging of the inner products of the identity stresses (equation 11 in \citealt{lilly1992}).

Equations (\ref{eq:mom}) are solved using an LES extension of the DNS code by \citet{chung2014}. We perform wall-resolved LES in a channel flow (figure~\ref{fig:geometry}) by applying periodic boundary conditions in the streamwise and spanwise directions. At the bottom wall we apply $\widehat{u}=\widehat{v} = 0$ and $\widehat{w}(x,z,t) = A\sin(\kappa_x x - \omega t)$, and at the top boundary we apply free-slip and impermeable conditions ($\partial \widehat{u}/\partial y = \partial \widehat{w}/\partial y = \widehat{v} = 0$). The present channel flow with free-slip top boundary conditions and domain height $h$ (also known as open channel flow) requires less computational cost to converge compared to the conventional channel flow with no-slip top boundary conditions and domain height $2h$ (also known as full channel flow). Except for a small outer region, the mean velocity profiles and turbulence statstics are very similar between the two channel configurations (figures 2 and 3 in \citealt{yao2022direct}). Compared to the boundary layer, we speculate small differences with (open) channel flow when we focus on the ISA pathway. This is supported by extensive comparison of channel flow with the boundary layer~\citep{monty2009comparison,mathis2009comparison,chin2014reynolds}. The two configurations have identical mean velocity profiles up to the end of the logarithmic region (see figure~1\textit{a} in \citealt{monty2009comparison}). Up to the fourth-order statistics are in agreement between the two configurations to a height of half the boundary layer thickness (half the channel height), e.g.\ see figure~3 in \cite{mathis2009comparison}. However, differences appear in the outer region due to the differences in the large-scale motions. Nevertheless, in the ISA pathway these large-scale motions do not contribute to $DR$.

% Therefore, the present (open) channel is a cost-effective approach for studying ISA, as is also used for studying turbulent flows over roughness~\citep{macdonald2018direct,aghaei2019turbulence,rouhi2019} and riblets~\citep{endrikat2020}.

% are solved with periodic boundary conditions  using  
% 
% The  uses the bottom-wall boundary conditions . The boundary conditions at the 

% $G$ is adjusted based on a target friction Reynolds number $Re_\tau \equiv u_{\tau_0} h/\nu$ of the non-actuated case. 

\subsection{Simulation cases}\label{sec:cases}

Table~\ref{tab:production} lists all the simulations completed for $Re_\tau = 951$ and $Re_\tau = 4000$, where $Re_\tau \equiv u_{\tau_0} h/\nu$ represents the friction Reynolds number of the non-actuated case.  At each $Re_\tau$, a parametric sweep of $7 \times 8$ combinations of streamwise wavenumber ($\kappa^+_x$) and oscillation frequency ($\omega^+$) is conducted over  $0.00238 \le \kappa^+_x \le 0.02$ and  $-0.2 \le \omega^+ \le +0.2$. The spanwise velocity amplitude is fixed at $A^+ = 12$. The seven non-zero values of $\omega^+$ give the oscillation time periods $T_{osc}^+=126$, 63, 42, and 31 (all within the ISA pathway), and $\omega^+=0$ corresponds to a time-invariant standing wave in the streamwise direction. In Table~\ref{tab:production}, N/A denotes the specifications of the non-actuated simulation which serves as the reference case for calculating $DR$. For each actuated case, the $DR$ is computed by matching the bulk Reynolds number $Re_b = U_b h /\nu$ between the actuated and non-actuated cases and substituting the respective values of $C_f$ and $C_{f_0}$ into \eqref{eq:dr}. We consider matched bulk Reynolds numbers $Re_b = 19700$ and $94450$, which correspond to $Re_\tau = 951$ and $4000$ for the non-actuated channel flow. \cite{quadrio2011} and \cite{ricco2012} compute $C_f$ and $C_{f_0}$ at matched $Re_\tau$ (instead of matched $Re_b$) by driving the actuated and non-actuated cases with a constant pressure-gradient. Several differences exist between matching $Re_b$ (constant flowrate) and matching $Re_\tau$ (constant pressure-gradient), see \cite{quadrio2011}, \cite{quadrio2011b} and \cite{ricco2012}. With matched $Re_b$, $C_f$ and $C_{f_0}$ are obtained at different $Re_\tau$. However, for our considered parameter space, the maximum $DR$ is about $30\%$, which leads to a maximum deviation of about $16\%$ in $Re_\tau$ between $C_f$ and $C_{f_0}$. Another source of difference between matched $Re_b$ and matched $Re_\tau$ is in the actuation amplitude $A$ (\ref{eq:wallmotion}). With constant $A^+ = 12$, $A^* = 12$ for the actuated cases with matched $Re_\tau$. However, with matched $Re_b$, $A^* > 12$ when $DR > 0$, and vice versa. Nevertheless, $DR$ weakly depends on $A^* \gtrsim 12$~\citep{quadrio2009,gatti2016,Chandran2022part2}. Overall, we speculate marginal differences in $DR$ between matched $Re_b$ and matched $Re_\tau$ for our parameter space.
%\ar{These values of $Re_b$ correspond to $Re_\tau = 951$ (for $Re_b \simeq 19,700$) and $Re_\tau = 4000$ (for $Re_b \simeq 94,450$) for a non-actuated smooth wall.} 
% For the actuated cases, the value of $Re_\tau$ is a nominal one based on the non-actuated $u_{\tau_o}$. The actual friction Reynolds number for each actuated case will vary with $DR$.

 \begin{table}
\centering
 \begin{tabular}[t]{cccccccc}
  domain & $Re_\tau, Re_b$ & $y^+_\mathrm{res}$ & $\kappa^+_x$  & $\omega^+$ & $L_x/h,,L_z/h$ & $N_x , N_y, N_z$ & $\Delta^+_x , \Delta^+_z$   \\  \\
  full & $951, 19700$ & $951$  & $0.00238$ & $0, \pm 0.05, \pm 0.1, +0.15, \pm 0.2$ & $8.33 , 3.14 $ & $360,48,96$ & $22 , 31$  \\
  full & $951, 19700$ & $951$ & $0.004$ & $0, \pm 0.05, \pm 0.1, +0.15, \pm 0.2$ & $6.61 , 3.14$ & $288,48,96$ & $22 , 31$  \\
  full & $951, 19700$ & $951$ & $0.007$ & $0, \pm 0.05, \pm 0.1, +0.15, \pm 0.2$ & $6.61 ,3.14 $ & $288,48,96$ & $22 , 31$  \\
  full & $951, 19700$ & $951$ & $0.010$ & $0, \pm 0.05, \pm 0.1, +0.15, \pm 0.2$ & $6.61 ,3.14 $ & $288,48,96$ & $22 , 31$  \\
  full & $951, 19700$ & $951$ & $0.012$ & $0, \pm 0.05, \pm 0.1, +0.15, \pm 0.2$ & $6.61 ,3.14 $ & $288,48,96$ & $22 , 31$  \\
  full & $951, 19700$ & $951$ & $0.017$ & $0, \pm 0.05, \pm 0.1, +0.15, \pm 0.2$ & $6.61 ,3.14 $ & $288,48,96$ & $22 , 31$  \\  
  full & $951, 19700$ & $951$ & $0.021$ & $0, \pm 0.05, \pm 0.1, +0.15, \pm 0.2$ & $6.29 ,3.14 $ & $256,48,96$ & $23 , 31$   \\ \\
  full & $951, 19700$ & $951$ & N/A & N/A & $6.28,3.14$ & $192,64,128$ & $47 , 23$   \\ \hline 
  medium & $4000, 94450$ & $1000$ & $0.00238$ & $0, \pm 0.05, \pm 0.1, +0.15, \pm 0.2$ & $1.99,0.63$ & $384 , 192, 80$ & $21 , 31$  \\
  medium & $4000, 94450$ & $1000$ & $0.004$ & $0, \pm 0.05, \pm 0.1, +0.15, \pm 0.2$ & $1.96,0.63$ & $384 , 192, 80$ & $20 , 31$  \\
  medium & $4000, 94450$ & $1000$ & $0.007$ & $0, \pm 0.05, \pm 0.1, +0.15, \pm 0.2$ & $2.04,0.63$ & $384 , 192, 80$ & $21 , 31$  \\
  medium & $4000, 94450$ & $1000$ & $0.010$ & $0, \pm 0.05, \pm 0.1, +0.15, \pm 0.2$ & $2.04,0.63$ & $384 , 192, 80$ & $21 , 31$  \\
  medium & $4000, 94450$ & $1000$ & $0.012$ & $0, \pm 0.05, \pm 0.1, +0.15, \pm 0.2$ & $1.96,0.63$ & $384 , 192, 80$ & $20 , 31$  \\
  medium & $4000, 94450$ & $1000$ & $0.017$ & $0, \pm 0.05, \pm 0.1, +0.15, \pm 0.2$ & $2.03,0.63$ & $384 , 192, 80$ & $21 , 31$  \\  
  medium & $4000, 94450$ & $1000$ & $0.021$ & $0, \pm 0.05, \pm 0.1, +0.15, \pm 0.2$ & $2.04,0.63$ & $384 , 192, 80$ & $21 , 31$  \\ \\
  medium & $4000, 94450$ & $1000$ & $0.007$ & $0, \pm 0.05, + 0.1, + 0.2$ & $2.04,0.63$ & $576 , 288, 120$ & $14 , 21$  \\  \\
  medium & $4000, 94450$ & $1000$ & N/A & N/A & $2.04,0.63$ & $576 , 288, 120$ & $14 , 21$  \\
\end{tabular}
\caption{Summary of the parameters of computational runs. Cases above the separating line are conducted at \ar{$Re_\tau = 951$ ($Re_b = 19700$)}, and below the line at \ar{$Re_\tau = 4000$ ($Re_b = 94450$)}. The last row at each $Re_\tau$ indicated with N/A for $\kappa^+_x$ and $\omega^+$, corresponds to the non-actuated reference case. For all the actuated cases, $A^+ = 12$. Each row of the actuated cases consists of a set of cases with equal domain size, $Re_\tau$, $\kappa^+_x$ and grid size, but $\omega^+$ is different for each case \ar{(as listed in the fifth column). Those values of $\omega^+$ with $\pm$ sign indicate two separate simulations, one with a positive sign (downstream travelling wave) and one with a negative sign (upstream travelling wave).} The first column indicates the domain size (see figure~\ref{fig:geometry}); at $Re_\tau = 951$ we use the full domain and at $Re_\tau = 4000$ we use the medium domain. The \ar{third} column $y^+_\mathrm{res}$ is the maximum resolved height by the simulation domain ($\simeq 0.4 L^+_z$, \citealt{chung2015}). The eighth row at $Re_\tau = 4000$ repeats some of the cases with $\kappa^+_x = 0.007$ (the third row at $Re_\tau = 4000$), but with a finer grid resolution.}  
\label{tab:production}
\end{table}

The grid resolutions were chosen based on extensive validation studies as presented in Appendices~\ref{sec:grid_drag} and \ref{sec:aliasing}. In these appendices, we compare our LES results with DNS data of \cite{gatti2016} at $Re_\tau \approx 1000$, experimental data of \cite{marusic2021} at $Re_\tau = 6000$, and our self-generated DNS data at $Re_\tau = 590$.  For $DR$ and the mean velocity profile, we used the same viscous-scaled grid resolution at $Re_\tau = 951$ and $4000$, corresponding to the streamwise and spanwise grid sizes of $\Delta^+_x \times \Delta^+_z \simeq 21 \times 31$ (the first seven rows at each $Re_\tau$ in table~\ref{tab:production}). At this grid resolution, the difference in $DR$ between the LES and DNS was found to be within $2\%$, and similarly good agreement was found for the mean velocity profile. However, for the Reynolds stresses and spectra at $Re_\tau = 4000$ we used a finer grid resolution with $\Delta^+_x \times \Delta^+_z \simeq 14 \times 21$ (the last two rows in table~\ref{tab:production}).
Our nominal LES filter width $\Delta^+ = \left( \Delta^+_x \Delta^+_y \Delta^+_z \right)^{1/3}$ is $7 \lesssim \Delta^+ \lesssim 34$ for the coarser grid, and $5 \lesssim \Delta^+ \lesssim 22$ for the finer grid. However, given our anisotropic grid, we estimate our effective filter width from the two-dimensional energy spectrograms (figures~\ref{fig:appendix_fig2}\textit{e,f}). Our maximum filter width is in the spanwise direction and is about $50$ and $35$ viscous units for the coarser and finer grids, respectively, equivalent to the cut-off wavenumbers $k^+_{\Delta_z} \simeq 0.12$ and $0.18$. These wavenumbers are $6$ and $9$ times larger than our maximum actuation wavenumber $\kappa^+_x = 0.02$. We estimate our cut-off frequency from Taylor's frozen turbulence hypothesis~\citep{taylor1938spectrum}. The most challenging zone in terms of resolution is the buffer region ($y^+ \simeq 10$) with the smallest energetic eddies. If we take the convective speed of $10 u_{\tau_0}$ in this region, our cut-off frequencies are  $\omega^+_{\Delta_z}\simeq 1.2$ and $1.8$ for the coarser and finer grids, respectively, which are $6$ and $9$ times larger than our maximum actuation frequency $\omega^+ = \pm 0.2$.

%We conclude the suitability of this grid for Reynolds stresses and spectra in Appendix~\ref{sec:aliasing} through comparison with our self-generated DNS data at $Re_\tau = 590$ ($\Delta^+_x \times \Delta^+_z \simeq 7 \times 4$). 

% The domain size must be chosen such that the resolved height $y^+_\mathrm{res}$ is greater than the extent of the disturbed flow due to the surface modification.

In terms of the domain size, the cases at $Re_\tau = 951$ used a full-domain with $L_x \times L_z \simeq 6.6 h \times 3.2 h$ (figure~\ref{fig:geometry}\textit{c}), which is sufficiently large to resolve the first and second-order statistics across the entire channel~\citep{lozano2014}. However, at $Re_\tau = 4000$ each full-domain calculation is about $500$ times more expensive than that at $Re_\tau = 951$, and so the domain size was reduced to $L_x \times L_z \simeq 2.0h \times 0.6h$ (figure~\ref{fig:geometry}\textit{a}). As a consequence, the flow is only resolved up to a fraction of the channel height $y^+_\mathrm{res} \simeq 0.4 L^+_z$~\citep{chung2015}, shown by the grey shaded zones in figure~\ref{fig:geometry}. For a reduced domain calculation, the user decides the resolved height $y^+_\mathrm{res}$, with the constraint that it must fall somewhere in the logarithmic region. Then the domain size is obtained from the prescriptions of \cite{chung2015} and \cite{macdonald2017}. For the travelling wave actuation (\ref{eq:wallmotion}), the prescriptions are $L^+_z \simeq 2.5 y^+_\mathrm{res}, L^+_x \gtrsim \max(3L^+_z, 1000, \lambda^+)$, where $\lambda$ is the travelling wavelength. \cite{macdonald2017,macdonald2018direct} used the reduced-domain approach with $60 \lesssim y^+_\mathrm{res} \lesssim 250$ for turbulent flows over roughness. \cite{endrikat2020} used the same approach with $y^+_\mathrm{res} \simeq 100$ for turbulent flows over riblets. \cite{jimenez1991} who used this approach for the first time resolved the flow up to $y^+_\mathrm{res} \simeq 80$. They named this approach ``minimal flow unit''. Here, with $L_x \times L_z \simeq 2.0h \times 0.6h$ (figure~\ref{fig:geometry}\textit{a}) at $Re_\tau = 4000$, we resolve a substantial fraction of the inner layer up to $y^+_\mathrm{res} \simeq 1000$. Therefore, we name our reduced domain the ``medium domain'' to highlight its relatively larger size compared to the minimal flow unit. \cite{gatti2016} also used the medium domain size of $L_x \times L_z \simeq 1.4h \times 0.7h$ with $y^+_\mathrm{res} \simeq 250$ to study the travelling wave (\ref{eq:wallmotion}). In Appendix~\ref{sec:domain}, we assess the suitability of the medium domain size (figure~\ref{fig:geometry}\textit{a}) by comparing the results with those obtained using a larger domain size (figure~\ref{fig:geometry}\textit{b}) for selected cases from table~\ref{tab:production}.

% \cite{gatti2016} similarly reduced the domain size for $Re_\tau \approx 1000$. 

% In previous computations using reduced domains (roughness \citealt{macdonald2019}; riblets \citealt{endrikat2020}; streamwise travelling wave surface motion \citealt{gatti2016}), $100 \lesssim y^+_\mathrm{res} \lesssim 260$, and it was assumed that the surface modification did not disturb the flow beyond $y^+_\mathrm{res} \simeq 100 - 200$. Here, with $L_x \times L_z \simeq 2.0h \times 0.6h$ (figure~\ref{fig:geometry}\textit{a}) at $Re_\tau = 4000$, we resolve the flow up to $y^+_\mathrm{res} \simeq 1000$, several times larger than that given by these recommendations. \cite{macdonald2019} and \cite{endrikat2020} called the reduced domain configuration the ``minimal domain'' following the name first given by \cite{jimenez1991}, but we name our reduced domain the ``medium domain'' to highlight its relatively larger size compared to the conventional minimal domain size. 

\subsection{Calculation of the skin-friction coefficient}\label{sec:cf_dr}

\begin{figure}
  \centering
  \includegraphics[width=.9\textwidth,trim={{0.1\textwidth} {0.05\textwidth} {0.05\textwidth} {0.0\textwidth}},clip]{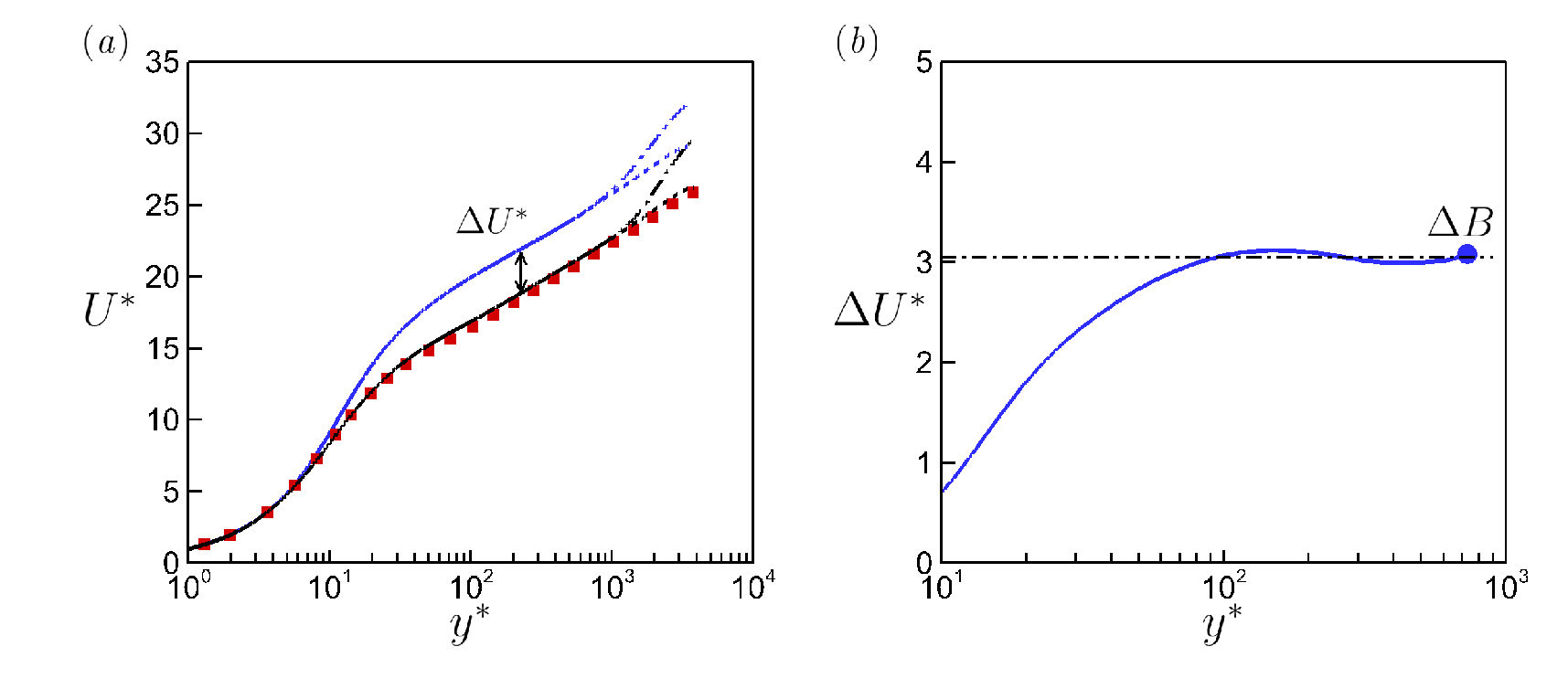}
 \caption{(\textit{a}) Profiles of the mean velocity $U^*$ for the LES of the actuated case at $Re_\tau = 4000, A^+ = 12, \kappa^+_x = 0.02$ and $\omega^+ = -0.05$ ({\color{blue}\solid, \dasheddotted}), and LES of the non-actuated case at $Re_\tau = 4000$ (\solid, \dasheddotted). \ar{The viscous-scaled quantities $U^*$ and $y^*$ are scaled by the actual values of $u_\tau$ for each case. The resolved portion of each LES profile $(y^* \lesssim 750)$ is shown with a solid line, and the unresolved portion $(y^* \gtrsim 750)$ is shown with a dashed-dotted line. The unresolved portion of each profile appears as a fictitious wake and is due to the medium domain size (figure~\ref{fig:geometry}\textit{a}). We reconstruct the unresolved portion using the composite profile for channel flow by \cite{nagib2008} (the dashed lines for $y^* \gtrsim 750$).} We compare the \ar{resolved ({\color{black}\solid}) and reconstructed ({\color{black}\dashed}) portions of the} non-actuated LES with the DNS of \cite{lozano2014} at $Re_\tau = 4200$ ({\color{red}\filsqr}). (\textit{b}) Difference between the actuated and non-actuated profiles $\Delta U^* = U^*_\mathrm{act} - U^*_\mathrm{non{\text -}act}$ (blue and black profiles in \textit{a}) up to the maximum resolved height $y^*_\mathrm{res} \simeq 750$. To reconstruct the actuated profile beyond $y^*_\mathrm{res} \simeq 750$ using the composite profile suggested by \cite{nagib2008}, we set the log-law shift $\Delta B$ as the value of $\Delta U^*$ at $y^*_\mathrm{res}$.}  
  \label{fig:Cf_reconstruct}
\end{figure}

To compute $DR$ (\ref{eq:dr}), we need the skin-friction coefficient \ar{$C_f \equiv 2\overline{\tau_w}/(\rho U^2_b) \equiv 2/{U^*_b}^2$ for both the actuated and the non-actuated cases.}
Here, $U^*_b = \int_0^{h^*} U^* dy^*/h^*$ is the viscous-scaled bulk velocity. For the cases at $Re_\tau = 951$ with the full domain size the $U^*$ profile is resolved across the whole channel and $U^*_b$ can be found directly. However, for the cases at $Re_\tau = 4000$ with the medium domain size, the $U^*$ profile is resolved only up to \ar{$y^*_\mathrm{res} \simeq 750 - 1000$}. Two of these high Reynolds number profiles are shown in figure~\ref{fig:Cf_reconstruct}(\textit{a}): the actuated case with $A^+=12$, $\kappa^+_x = 0.02$ and $\omega^+ = -0.05$ (\ar{blue lines}), and the non-actuated case (\ar{black lines}). \ar{The resolved portion of the LES profile below $y^*_\mathrm{res}$ is shown with a solid line, and the unresolved portion above $y^*_\mathrm{res}$ with a dashed-dotted line.} We also overlay the DNS of the non-actuated full-domain channel flow at $Re_\tau = 4200$ by \cite{lozano2014} (\ar{red squares}). \ar{For the non-actuated LES, the resolved portion up to $y^*_\mathrm{res} \simeq 1000$ (solid black line) accurately reproduces the non-actuated DNS. However, the unresolved portion beyond $y^*_\mathrm{res}$ (black dashed-dotted line) departs from the non-actuated DNS due to the reduced domain size.  } 
%This is due to the reduced medium domain size ($2.0h \times 0.6h$, figure~\ref{fig:geometry}\textit{a}) \ar{for the LES case} compared to the full-domain size ($6.6h \times 3.2h$, figure~\ref{fig:geometry}\textit{c}) \ar{for the reference DNS case. If we directly integrate the $U^*$ profile from the reduced domain LES case (i.e.\ integrate the black solid and dashed-dotted lines) the resulting $C_{f_0}$ is $0.00331$, while the DNS $C_{f_0}$ is $0.00361$, hence an $8\%$ underestimation of the DNS $C_{f_0}$.}

This issue has been addressed previously by
%\ar{The departure of the $U^*$ profiles beyond $y^*_\mathrm{res}$ is discussed in the previous investigations using reduced domains for channel flow simulations~}
\ar{\cite{chung2015} \cite{macdonald2017}, \cite{endrikat2021}, and \cite{endrikat2020}. For accurate prediction of $U^*_b$, hence $C_f$, it was found that the resolved height $y^*_\mathrm{res}$ must fall inside the logarithmic region, and it needs to be larger than the extent of the disturbed flow due to the surface modification. If $y^*_\mathrm{res}$ satisfies these criteria, the $U^*$ profile is resolved up to a portion of the log region, similar to the LES cases shown in figure~\ref{fig:Cf_reconstruct}(\textit{a}). Beyond $y^*_\mathrm{res}$, the unresolved portion of the log region and the outer region is assumed to be universal and so it can be reconstructed based on previous work. Here, we reconstruct the unresolved portions using the composite profile for full-domain channel flow~\citep{nagib2008}.} 
\begin{comment}
\begin{equation}
 U^* = \frac{1}{\kappa} \ln \left( y^* \right) + B + \Delta B + \frac{2 \Pi}{\kappa} \mathcal{W} \left( \frac{y}{h} \right), \quad y^* \ge y^*_\mathrm{res} \label{eq:composite}
\end{equation}
where we choose $\kappa = 0.4$ following \cite{pirozzoli2016} and $B= 5.2$ to make the profiles continuous at $y^*_\mathrm{res}$. For the outer wake function $\mathcal{W} (y/h)$, we use the formulation for channel flow by \cite{nagib2008} (see their equation 12). We set the velocity wake parameter $\Pi = 0.05$, as suggested by \cite{nagib2008} (see their figure 6).
\end{comment}
Figure~\ref{fig:Cf_reconstruct}(\textit{a}) demonstrates that for the non-actuated case at $Re_\tau=4000$, we obtain good agreement between the reconstructed profile for LES (\ar{dashed black line}) and DNS. Therefore, to obtain $U^*_b$ we integrate the resolved $U^*$ profile up to $y^*_\mathrm{res}$ and the reconstructed profile beyond $y^*_\mathrm{res}$. We find that $C_{f_0}$ using this corrected $U^*_b$ is only $1\%$ different than the value obtained from DNS.

We follow the same approach to reconstruct the actuated $U^*$ profile (\ar{dashed blue line in} figure~\ref{fig:Cf_reconstruct}\textit{a}). However, we need to add the log-law shift $\Delta B$ in the composite profile  to make the resolved and reconstructed profiles continuous at $y^*_\mathrm{res}$. We find $\Delta B$ by plotting the velocity difference between the actuated and non-actuated profiles $\Delta U^* = U^*_\mathrm{act} - U^*_\mathrm{non{\text -}actuated}$ (figure~\ref{fig:Cf_reconstruct}\textit{b}). As seen in figure~\ref{fig:Cf_reconstruct}(\textit{b}), $\Delta U^*$ reaches almost a plateau beyond $y^* \simeq 100$. We set $\Delta B$ as the value of $\Delta U^*$ at $y^*_\mathrm{res} \simeq 750$. Note that since the actuated $u_\tau$ is smaller than the non-actuated $u_{\tau_o}$, \ar{$y^*_\mathrm{res}$ for the actuated case is about $750$, but for the non-actuated case is about $1000$.} We calculate $U^*_b$ for the actuated case by integrating the resolved portion of the profile up to $y^*_\mathrm{res}$ (\ar{solid blue line}) and the reconstructed portion beyond $y^*_\mathrm{res}$ (\ar{dashed blue line}).

\ar{Another way of calculating $U^*_b$ (hence $C_f$) from the reduced domain is to integrate the composite profile from $y=0$ to $h$ (e.g.\ see 4.2 in \citealt{macdonald2019}), which assumes that the viscous sublayer and buffer layer make a negligible contribution to $U^*_b$. We believe that our present approach  is more accurate as it considers the complex variation of $U^*$ in the viscous sublayer and buffer layer. We only use the composite profile in the log-region and beyond.}

\section{Results}\label{sec:results}

\subsection{Drag reduction map as a function of frequency and wavelength}\label{sec:dr_map}

\ar{Figures~\ref{fig:dr_map_compare}(\textit{a,b}) display the maps of $DR(\omega^+,\kappa^+_x)$ at $Re_\tau = 951$ and $4000$ from the computations listed in table~\ref{tab:production}. At each $Re_\tau$, we have $56$ $DR$ data points. To generate the maps, we perform bilinear interpolation of our $DR$ data points onto a uniform $20 \times 20$ grid over the parameter space $0 \le \kappa^+_x \le 0.02$ and $-0.2 \le \omega^+ \le +0.2$. At $Re_\tau = 951$, the maximum $DR$ of $35.4\%$ at $(\omega^+,\kappa^+_x) = (0.05,0.021)$ is in close agreement with the DNS of \cite{gatti2016} at $Re_\tau \simeq 950$, where the maximum $DR$ was found to be $38.8\%$ at $(\omega^+,\kappa^+_x) = (0.05, 0.0195)$. At $Re_\tau = 4000$, the maximum $DR$ decreases to $27.5\%$ at the same actuation parameters $(\omega^+,\kappa^+_x) = (0.05,0.021)$. At each Reynolds number, $DR$ changes more drastically by changing $\omega^+$ than by changing $\kappa^+_x$.

\begin{figure}
  \centering
  \includegraphics[width=\textwidth,trim={{0.04\textwidth} {0.05\textwidth} {0.04\textwidth} {0.0\textwidth}},clip]{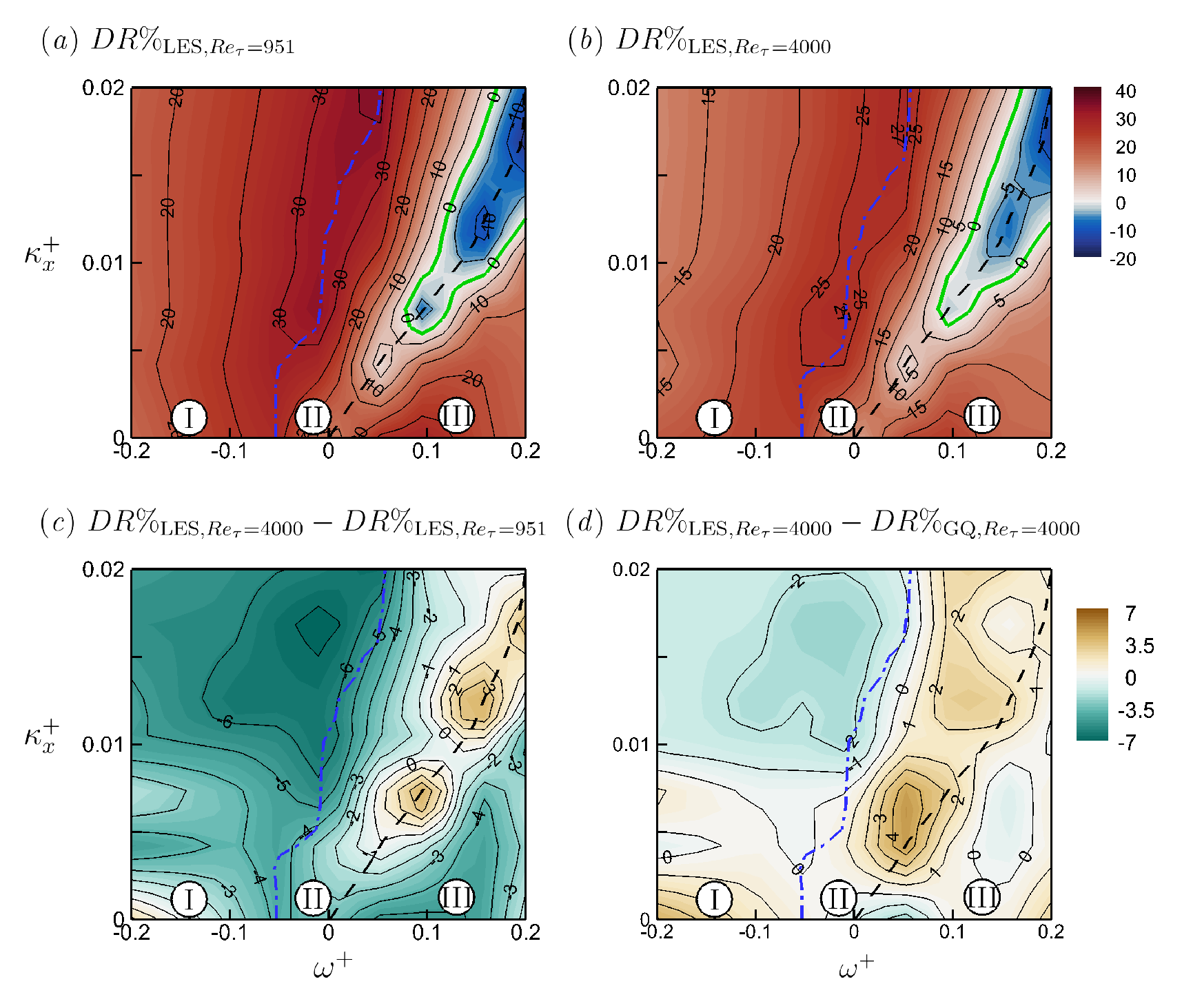}
\caption{(\textit{a,b}) Maps of $DR$ for $A^+ =12$ at (\textit{a}) $Re_\tau = 951$ and (\textit{b}) $Re_\tau = 4000$. The local maximum $DR$ ({\color{blue}\dasheddotted}) and the local minimum $DR$ (\dashed) for $\kappa^+_x > 0$ are indicated for clarity. We label the region on the left side of ({\color{blue}\dasheddotted}) with I, between ({\color{blue}\dasheddotted}) and (\dashed) with II and the right side of (\dashed) with III. (\textit{c}) Map of the difference in $DR$ between $Re_\tau = 4000$ and $Re_\tau = 951$. (\textit{d}) Map of the difference in $DR$ between $Re_\tau = 4000$ and GQ's prediction~\citep{gatti2016} at the same Reynolds number. In all (\textit{a,b,c,d}), the contour fields and the contour lines show the same quantity.  For (\textit{a,b}) the contour lines grow from $-20\%$ to $40\%$, and for (\textit{c,d}) the contour lines grow from $-7\%$ to $+7\%$.}  
  \label{fig:dr_map_compare}
\end{figure}

When $\kappa^+_x = 0$, there is no travelling wave (plane wall oscillation), and the variation of $DR$ is symmetric between $\omega^+ <0$ and $\omega^+ > 0$. In this case, two equal local maxima (at $\omega^+ \simeq \pm 0.05$) and a local minimum (at $\omega = 0$) emerge. When $\kappa^+_x >0$, a travelling wave is generated, and the variation of $DR$ is asymmetric between $\omega^+ < 0$ and $\omega^+ > 0$. In this case, at each $\kappa^+_x$ only one local maximum (blue dashed-dotted curve in figure~\ref{fig:dr_map_compare}) and one local minimum (black dashed curve in figure~\ref{fig:dr_map_compare}) appear in $DR$. These observations are in agreement with \cite{quadrio2009} and \cite{gatti2016}.
Overall, within our parameter space, the map of $DR$ consists of three distinct regions. Region I to the left of local maximum $DR$ (blue dashed-dotted curve) where $\omega^+ \lesssim 0$ (upstream travelling wave); in this region $DR> 0$, hence drag reduction. Region II represents the crossover from the local maximum to the local minimum $DR$ (between the blue dashed-dotted curve and the black dashed curve). For $\kappa^+_x \lesssim 0.007$, the local minimum $DR$ is positive, however, for $\kappa^+_x \gtrsim 0.007$ the local minimum $DR$ becomes negative (hence a drag increase). Increase in $\kappa^+_x$ beyond $0.007$ leads to a larger drag increase area, and the local minimum $DR$ becomes more negative; \cite{quadrio2009} and \cite{gatti2016} observe similar trends. \cite{quadrio2009} find that the local minimum $DR$ follows the line $\omega^+/\kappa^+_x \simeq 10$. In other words, maximum drag increase occurs when the travelling wave speed is about $10u_{\tau_0}$, which is nearly the same as the convective speed of the near-wall flow structures. Similarly, in figures~\ref{fig:dr_map_compare}(\textit{a,b}) the black dashed curve that marks the local minimum $DR$ follows $\omega^+/\kappa^+_x \simeq 10$. Region III covers the right of local minimum $DR$ (black dashed curve) where $\omega^+ > 0$ (downstream travelling wave); in this region, increase in $\omega^+$ increases $DR$. 
% In \S\, \ref{sec:stokes_layer} and \ref{sec:deltas_DR}, we will show that these three zones emerge out of changes in the Stokes layer dynamics, leading to changes in the near-wall turbulence. We also note that these dynamics are not expected to change sharply across the bounds of these regions (blue dashed-dotted curve and black dashed curve), and that the transitions will be smooth.

In figure~\ref{fig:dr_map_compare}(\textit{c}), we display the difference in $DR$ as the Reynolds number changes from 4000 to 951.  For most of the $(\omega^+, \kappa^+_x)$ space, $DR$ is lower at the higher Reynolds number. Only within the range $0.005 \lesssim \kappa^+_x \lesssim 0.020, +0.1 \lesssim \omega^+ \lesssim +0.2$ we observe the opposite trend. This region coincides with the drag increasing range ($DR < 0$) with $\omega^+ > 0$. This observation is consistent with GQ's model \eqref{eq:gq_model}, where $DR <0$ (hence $\Delta B < 0$) predicts an increase in $DR$ as Reynolds number increases. To make these comparisons more quantitative, in figure~\ref{fig:dr_map_compare}(\textit{d}) we show the difference in $DR$ between our results and GQ's model at $Re_\tau = 4000$. To predict $DR$, the model (\ref{eq:gq_model}) requires $C_{f_0}$ and the value of the log-law shift $\Delta B$ for each set of actuation parameters $(A^+, \kappa^+_x, \omega^+)$. 
% $\Delta B$ can be obtained at any fully turbulent Reynolds number, assuming that it is Reynolds number independent. By having $\Delta B$ in (\ref{eq:gq_model}), we can predict $DR$ at any Reynolds number for the same set of $(A^+, \kappa^+_x, \omega^+)$.  
For $C_{f_0}$, we use Dean's power-law correlation~\citep{dean1978} which agrees well with the DNS data given by \cite{macdonald2019}. We can obtain $\Delta B$ from a low Reynolds number simulation for the same set of $(A^+, \kappa^+_x, \omega^+)$ because $\Delta B$ is assumed to be Reynolds number independent. Therefore, we use our results at $Re_\tau = 951$, where for each $(\omega^+, \kappa^+_x)$ we find $\Delta B$ from the velocity difference $\Delta U^*$ at $y^* = 200$ (similar to figure~\ref{fig:Cf_reconstruct}\textit{b}). We choose $y^* = 200$ as it is far enough from the wall to fall into the log region, but not too far to fall into the wake region ($y/h \gtrsim 0.3$ according to \citealt{pope2000}). By having $\Delta B$ at each $(\omega^+, \kappa^+_x)$ and having $C_{f_0}$ at $Re_\tau = 4000$, we can reconstruct the $DR$ map based on GQ's model. 

Figure~\ref{fig:dr_map_compare}(\textit{d}) shows the overall good performance of GQ's model for this range of Reynolds numbers. In region I, the difference in $DR$ between LES and GQ's model is less than $2\%$, i.e.\ $|DR\%_\mathrm{LES,Re_\tau = 4000} - DR\%_\mathrm{GQ,Re_\tau = 4000}| \lesssim 2\%$. This is a very good agreement considering that $DR$ varies between $15\%$ and $28\%$ in region I. In regions II and III, we observe some slight differences in $DR$ between LES and GQ's model, especially in region II in the drag increasing range. In this range, the difference in $DR$ between LES and GQ's model reaches $4\%$, which is the same order as $DR$ (see figure~\ref{fig:dr_map_compare}\textit{b}). In region III, for $\omega^+ \gtrsim +0.1$ again we observe good agreement between LES and GQ's model (less than $2\%$ difference). In the following sections, we investigate the reasons behind the different performance of GQ's model in regions I, II and III related to the changes in the Stokes layer dynamics and the near-wall turbulence in each of these regions.}

\subsection{Mean velocity profiles}\label{sec:error_gq_model}

\begin{figure}
  \centering
  \includegraphics[width=.9\textwidth,trim={{0.05\textwidth} {0.02\textwidth} {0.05\textwidth} {0.0\textwidth}},clip]{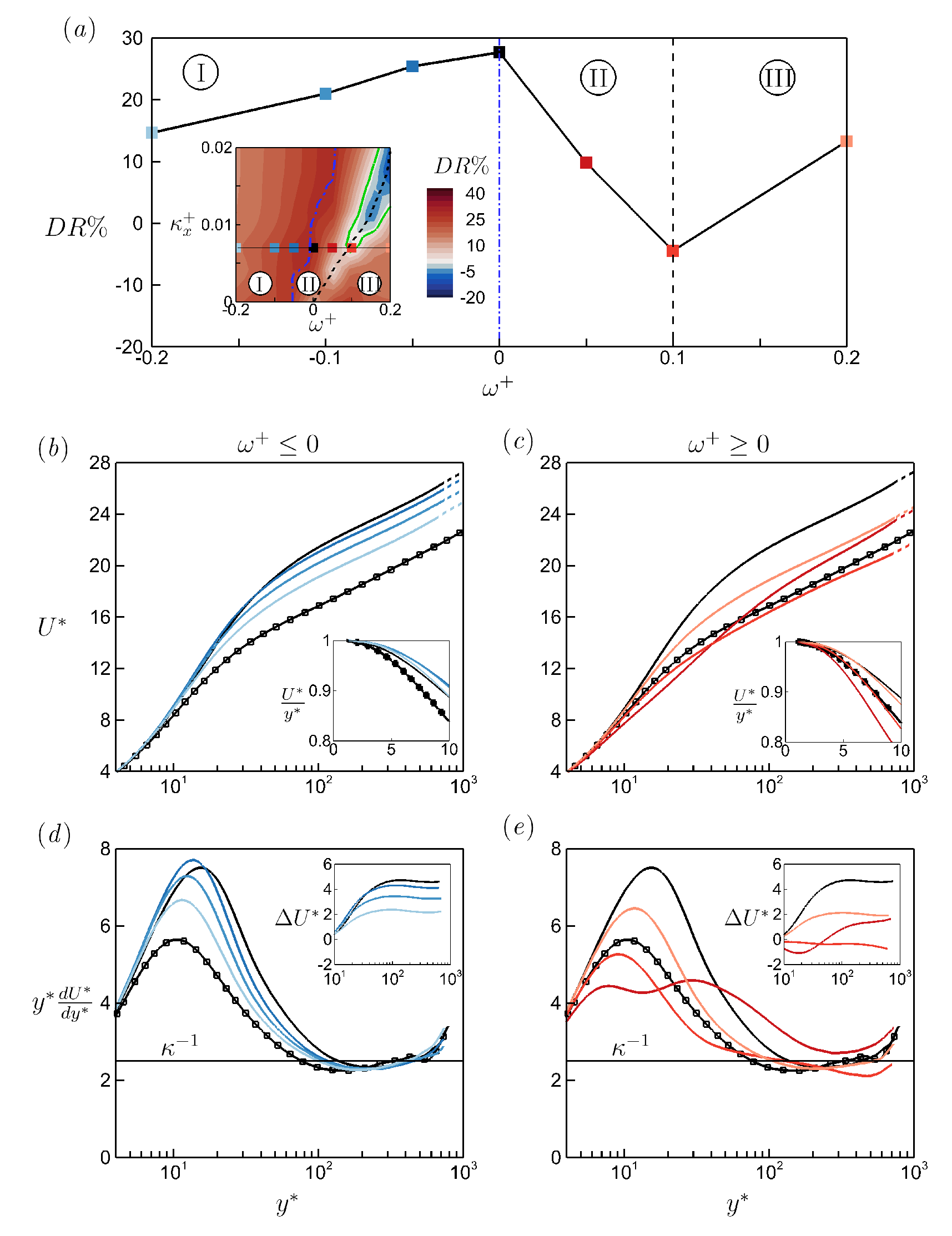}  
\caption{Variation of $DR$ and the mean velocity profiles $U^*$ at $Re_\tau = 4000$, $A^+ = 12$, $\kappa^+_x = 0.007$, and $-0.2 \le \omega^+ \le +0.2$. (\textit{a}) Variation of $DR$ with $\omega^+$; the inset shows the location of the data points on the $DR$ map. The lines ({\color{blue}\dasheddotted}) and (\dashed) are the local maximum and minimum $DR$. (\textit{b,c}) Variation of the $U^*$ profiles with $\omega^+$ for (\textit{b}) upstream travelling wave ($\omega^+ \le 0$) and (\textit{c}) downstream travelling wave ($\omega^+ \ge 0$); the profile (\lsqr) corresponds to the non-actuated case and the profiles with no symbol correspond to the actuated cases. For each profile, the solid line is the resolved portion and the dashed line is the reconstructed portion following \cite{nagib2008}. For each case, the colour of its $U^*$ profile in (\textit{b,c}) is consistent with the colour of its $DR$ datapoint in (\textit{a}). In (\textit{b,c}), the inset plots the same profiles in terms of $U^*/y^*$ versus $y^*$. (\textit{d,e}) Diagnostic function $y^* dU^*/dy^*$ for the profiles in (\textit{b,c}); the inset shows the velocity difference $\Delta U^* = U^*_\mathrm{act} - U^*_\mathrm{non{\text -}act}$ between each actuated profile $U^*_\mathrm{act}$ and the non-actuated profile $U^*_\mathrm{non{\text -}act}$.}  
  \label{fig:DR_stats_kxp0_00238}
\end{figure}

To obtain an overall picture of the mean velocity behaviour in regions I, II and III (see figure~\ref{fig:DR_stats_kxp0_00238}), we consider the seven runs conducted \ar{at} $Re_\tau = 4000, A^+ = 12$ and $ \kappa^+_x = 0.007$ for $\omega^+$ ranging from $-0.2$ to $+0.2$. \ar{In figure~\ref{fig:DR_stats_kxp0_00238}(\textit{a}), we identify the} selected values of $\omega^+$ (filled squares) on the $DR$ map along with the local maximum $DR$ ({\color{blue}\dasheddotted}) and the local minimum $DR$ (\dashed). 
%\ar{The selected cases fall into all regions I, II and III.  Our findings based on the investigation of these cases can explain the distinct physics that occur in different regions of our parameter space}. 
%For the selected $\omega^+$, we plot their values of $DR$ in figure~\ref{fig:DR_stats_kxp0_00238}(\textit{a}) and t
The corresponding velocity profiles are shown in \ar{figures~\ref{fig:DR_stats_kxp0_00238}(\textit{b},\textit{c})} for $\omega^+ \le 0$ (upstream travelling waves) up to the local maximum $DR$ \ar{(region I)} and $\omega^+ \ge 0$ (downstream travelling waves) beyond the local maximum $DR$ \ar{(regions II, III)}, respectively. 
% Some values of $\omega^+$ in figure~\ref{fig:DR_stats_kxp0_00238}(\textit{c}) fall between the local maximum and the local minimum $DR$ \ar{(region II)}.

When $\omega^+ \le 0$, the log region of the actuated profiles is shortened and shifted above the non-actuated counterpart (figure~\ref{fig:DR_stats_kxp0_00238}\textit{b}), corresponding to a positive $DR$. The shortening of the log region is due to the thickening of the viscous sublayer. We show the viscous sublayer thickening in the inset of figure~\ref{fig:DR_stats_kxp0_00238}(\textit{b}), in that the actuated profiles of $U^*/y^*$ are closer to unity for a greater wall distance compared to their non-actuated counterpart. We show the shortening of the log region in figure~\ref{fig:DR_stats_kxp0_00238}(\textit{d}) by plotting the diagnostic function $y^* dU^*/dy^*$. The log region appears as a plateau with the value of $\kappa^{-1} \simeq 2.5$. For the non-actuated case, the plateau appears for $100 \lesssim y^* \lesssim 600$. This range is consistent with the DNS of channel flow by \cite{lozano2014} and \cite{lee2015} at $Re_\tau = 4200$ and $5200$, respectively (see figure~3\textit{a} in \citealt{lee2015}). For the actuated cases, the plateau is narrowed further (i.e.\ log region is shortened) as $DR$ increases. We quantify the shift in the log region by plotting $\Delta U^*$ (inset of figure~\ref{fig:DR_stats_kxp0_00238}\textit{d}). The magnitude of the shift increases as $DR$ increases. These observations are also reported in the previous turbulent drag reduction studies, including turbulent flow with the spanwise wall oscillation~\citep{di2002,touber2012,hurst2014}, turbulent flow with the streamwise travelling wave~\citep{hurst2014,gatti2016}, turbulent flow of a polymer solution~\citep{ptasinski2003,white2008}, and turbulent flow over piezoelectrically excited travelling waves~\citep{musgrave2019}. \cite{gatti2016} derived their predictive model (\ref{eq:gq_model}) based on similar observations of the velocity profiles in region I, and as a result GQ's prediction works well in this region (figure~\ref{fig:dr_map_compare}\textit{d}). The behaviour of the profiles in region I is consistent with the ISA pathway, where only the inner-scale eddies up to the buffer region are actuated.

% When $\omega^+ \le 0$ \ar{(region I)}, the actuated profiles are shifted above their non-actuated counterpart (figure~\ref{fig:DR_stats_kxp0_00238}\textit{b}), corresponding to a positive $DR$. As $\omega^+$ increases from $-0.2$ to $0$, the magnitude of the shift (and the corresponding $DR$) increases until it reaches its maximum value at $\omega^+ = 0$. \ar{The profiles of $\Delta U^*$ (figure~\ref{fig:DR_stats_kxp0_00238}\textit{d}) better highlight this systematic increase in the logarithmic shift. This behaviour is similar to that seen in} 

%  These studies report the thickening of the viscous sublayer with a shortening and shifting of the logarithmic portion. The same thickening of the viscous sublayer in the present cases is indicated in the inset of figure~\ref{fig:DR_stats_kxp0_00238}(\textit{b})\ar{, in that the profiles of $U^*/y^*$ for the actuated cases are closer to unity for a greater wall distance compared to their non-actuated counterpart.}

% In figures~\ref{fig:DR_stats_kxp0_00238}(\textit{d,e}), $\Delta U^*$ of all cases reach a plateau by $y^* \simeq 200$, except the case with $\omega^+ = +0.05$ (figure~\ref{fig:DR_stats_kxp0_00238}\textit{e}). For this case, $\Delta U^*$ approaches a plateau of $1.7$ by $y^* \simeq 700$, but does not reach it.

In region II, we observe a sudden drop in $DR$ as $\omega^+$ changes from 0 to $+0.1$ (figure~\ref{fig:DR_stats_kxp0_00238}\textit{a}), with a corresponding decrease in the logarithmic shift (figure~\ref{fig:DR_stats_kxp0_00238}\textit{c}).
A distinct feature of region II is the high level of distortion in the $U^*$ profile, which is particularly severe at $\omega^+ = +0.05$. For this case, the diagnostic function tends towards the plateau $\kappa^{-1}$, but does not reach it. Similarly, $\Delta U^*$ for this case approaches a plateau of $1.7$ by the resolved height $y^*_\mathrm{res} \simeq 750$, but does not reach it (inset in figure~\ref{fig:DR_stats_kxp0_00238}\textit{e}). This is our most challenging case for computing $DR$ using our approach in \S\ \ref{sec:cf_dr} (figure~\ref{fig:Cf_reconstruct}). For accurate calculation of $DR$, $\Delta U^*$ needs to reach a plateau by the resolved height $y^*_\mathrm{res} \simeq 750$, i.e.\ the resolved height must fall into the logarithmic region. In Appendix \ref{sec:domain}, we deliberately consider this challenging case for domain size study. We double the domain length and width compared to the medium domain (figure~\ref{fig:geometry}\textit{b}), extending the resolved height to $y^*_\mathrm{res} \simeq 1500$. The difference in $DR$ is $1.4\%$ between the medium domain and the large domain (table~\ref{tab:domai_study}). Further, the large domain reinforces the approach of $\Delta U^*$ to a plateau of $1.7$ (the inset in figure~\ref{fig:domain_study}\textit{b}). To our knowledge, such significant levels of distortion in the $U^*$ profile have not been seen before in previous studies of flows over drag-reducing or drag-increasing surfaces. For example, in rough wall turbulent flows $\Delta U^*$ is almost constant for $y^* \gtrsim 30$ (e.g.\ figure~6 in \citealt{chan2015} or figure~3 in \citealt{macdonald2017}), while in turbulent flows over riblets $\Delta U^*$ is almost constant for $y^* \gtrsim 100$ (e.g.\ figure~2 in \citealt{endrikat2020}). In \S\ \ref{sec:stokes_layer} and \ref{sec:deltas_DR}, we discuss the physics behind the highly distorted mean velocity profiles (figures~\ref{fig:DR_stats_kxp0_00238}\textit{c,e}).

In region III, when $\omega^+$ increases to $+0.2$ (figure~\ref{fig:DR_stats_kxp0_00238}\textit{c}), $DR$ increases to $13\%$ and the $U^*$ profile behaves similarly to that seen in region I. A well-defined logarithmic shift appears beyond $y^* \simeq 100$ with viscous sublayer thickening.

%The mean velocity profiles demonstrate that the inner-scaled actuation can disturb the flow for distances of hundreds of viscous units away from the wall. This high level of actuation \ar{predominantly occurs in region II between the local maximum and the local minimum $DR$ (between {\color{blue}\dasheddotted} and \dashed in figure~\ref{fig:dr_map_compare}). In the following sections, we show that such a protrusive actuation is due to the Stokes layer that does not exist in turbulent flows over roughness or riblets.}

\subsection{Turbulence statistics}\label{sec:statistics}

\begin{figure}
  \centering
  \includegraphics[width=.9\textwidth,trim={{0.04\textwidth} {0.05\textwidth} {0.04\textwidth} {0.0\textwidth}},clip]{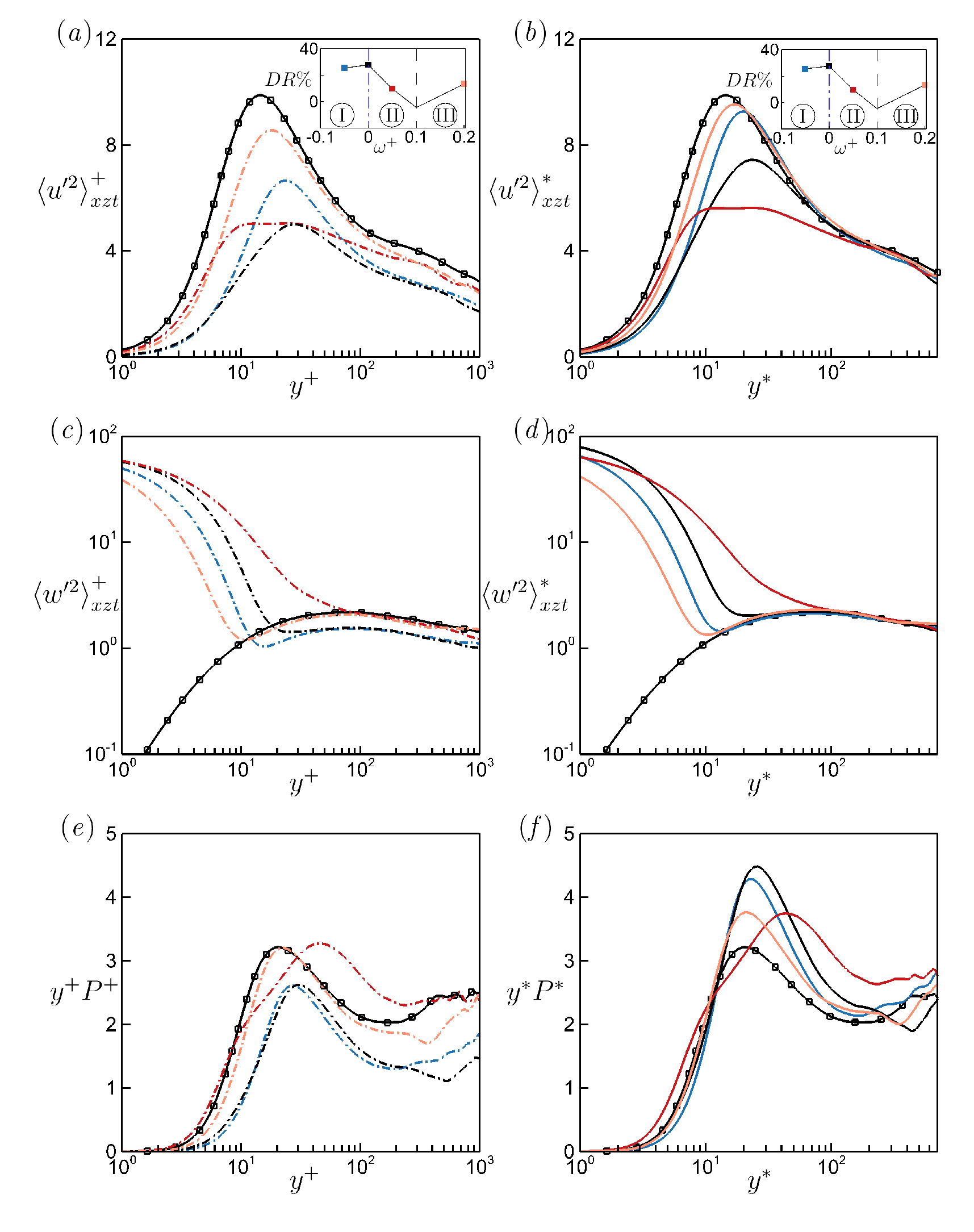}
\caption{Profiles of Reynolds stresses and turbulence production for four cases from figure~\ref{fig:DR_stats_kxp0_00238} at $Re_\tau = 4000, A^+ = 12, \kappa^+_x = 0.007$ and $\omega^+ = -0.05, 0, +0.05, +0.20$. The insets in (\textit{a,b}) indicate the considered values of $\omega^+$ and their drag reduction values. Line and symbol colours are consistent with figure~\ref{fig:DR_stats_kxp0_00238}. In each panel, only the resolved portion of the profiles are shown ($y^+ \lesssim 1000, y^* \lesssim 700$). The black lines with symbols correspond to the non-actuated case. (\textit{a,c,e}) plot the actuated profiles (dashed-dotted lines) scaled by the non-actuated $u_{\tau_0}$ (superscripted with $+$); (\textit{b,d,f}) plot the actuated profiles (solid lines) scaled by the actuated $u_\tau$ (superscripted with $*$). (\textit{a--d}) Reynolds stress profiles for (\textit{a,b}) the streamwise velocity $\left< u'^2 \right>_{xzt}$ and (\textit{c,d}) the spanwise velocity $\left< w'^2 \right>_{xzt}$. (\textit{e,f}) Premultiplied production of turbulent kinetic energy.}
  \label{fig:rms_kxp0_00700_prime}
\end{figure}

We now assess the behaviour of the Reynolds stress distributions at $Re_\tau=4000$ (figures~\ref{fig:rms_kxp0_00700_prime}\textit{a--d}) and the turbulent kinetic energy production $P = -\left< u' v' \right>_{xzt} dU/dy$ (figures~\ref{fig:rms_kxp0_00700_prime}\textit{e,f}), where $\left< ... \right>_{xzt}$ denotes averaging over $xz$-plane and time. We highlight four cases from figure~\ref{fig:DR_stats_kxp0_00238} ($A^+ = 12, \ \kappa^+_x = 0.007$), where we vary $\omega^+$ from  $-0.05$ to $+0.20$. As indicated earlier, we employ a finer grid resolution for these cases to properly resolve the Reynolds stresses (see Appendix~\ref{sec:aliasing}). 
%For the total Reynolds stress, we calculate the total fluctuating velocity (e.g.\ $u' = u - U$) by subtracting the $xz$-plane and time averaged velocity from the instantaneous one. Then, we average the square of the fluctuating velocity over the $xz$-plane and time (e.g.\ $\langle u'^2 \rangle_{xzt}$). 
We plot the profiles scaled by the non-actuated $u_{\tau_0}$ (dashed-dotted lines, figures~\ref{fig:rms_kxp0_00700_prime}\textit{a,c}) and by the actuated $u_\tau$ (solid lines, figures~\ref{fig:rms_kxp0_00700_prime}\textit{b,d}).
Scaling by $u_{\tau_0}$ is comparable to scaling by the bulk velocity $U_b$ \citep{gatti2016} because the bulk velocity $U_b$ is the same between the actuated and non-actuated cases. Any difference between the outer-scaled actuated and non-actuated profiles reflects the overall response of turbulence to the wall oscillation (\ref{eq:wallmotion}).

Scaling by the non-actuated $u_{\tau_0}$ (`$+$' superscript), as in figures~\ref{fig:rms_kxp0_00700_prime}(\textit{a,c,e}), indicates that the wall oscillation attenuates the $\langle u'^2 \rangle^+_{xzt}$ levels up to the resolved height $y^+_\mathrm{res} \simeq 1000$. The cases with the highest $DR$ ($\omega^+ = -0.05, 0$ in figure~\ref{fig:rms_kxp0_00700_prime}\textit{a}) show the highest level of attenuation in their $\langle u'^2 \rangle^+_{xzt}$. Additionally, for these cases the inner peak of $\langle u'^2 \rangle^+_{xzt}$ is farther from the wall. Consistently, the viscous sublayer is thickened and the buffer layer is shifted away from the wall (figure~\ref{fig:DR_stats_kxp0_00238}\textit{b}). 
% The stronger the level of attenuation, the higher the drag reduction, and the peak in $\langle u'^2 \rangle^+_{xzt}$ moves away from the wall while being further attenuated. This trend occurs in parallel with the thickening of the viscous sublayer and the shifting of the buffer layer (figure~\ref{fig:DR_stats_kxp0_00238}\textit{b}). At $\omega^+ = 0$, with the highest $DR$ (black square in the inset), the $\langle u'^2 \rangle^+_{xzt}$ profile is attenuated the most (black dashed-dotted line), and the viscous sublayer is the thickest (black line in figure~\ref{fig:DR_stats_kxp0_00238}\textit{b}). 
In contrast to the behaviour of $\langle u'^2 \rangle^+_{xzt}$, the $\langle w'^2 \rangle^+_{xzt}$ profiles are amplified near the wall.  According to \cite{quadrio2011} and \cite{touber2012}, this amplification is due to the Stokes layer that forms as a result of the spanwise wall motion. The pre-multiplied turbulent kinetic energy production $y^+ P^+$ (figure~\ref{fig:rms_kxp0_00700_prime}\textit{e}) also displays the attenuation of turbulence that accompanies increasing $DR$. All these trends are similar to previous studies on spanwise wall oscillation at lower Reynolds numbers \citep{quadrio2011,touber2012}. 

Scaling by the actuated $u_\tau$ (`$*$' superscript) is equivalent to inner scaling, which highlights the extent up to which the actuated profiles depart from the non-actuated profile. For $\langle u'^2 \rangle^*_{xzt}$ and $\langle w'^2 \rangle^*_{xzt}$ (figures~\ref{fig:rms_kxp0_00700_prime}\textit{b,d}), the actuated cases agree with the non-actuated case at distances far from the wall, but near the wall the actuated $\langle u'^2 \rangle^*_{xzt}$ levels are attenuated, while the $\langle w'^2 \rangle^*_{xzt}$ levels are amplified.  For $\omega^+ = +0.05$ (in region II), the point where the actuated profiles begin to depart from the non-actuated counterpart occurs at $y^* \simeq 100$, considerably farther than for the other cases $(y^* \lesssim 30)$.  The same case yields the strongest level of near-wall amplification for $\langle w'^2 \rangle^*_{xzt}$ (the red profile in figures~\ref{fig:rms_kxp0_00700_prime}\textit{c,d}) and the highest level of distortion in mean velocity (red profile in figures~\ref{fig:DR_stats_kxp0_00238}\textit{c,e}).

Regardless of the scaling used, as $\langle w'^2 \rangle_{xzt}$ is amplified near the wall, $\langle u'^2 \rangle_{xzt}$ is attenuated, the viscous sublayer is thickened and $DR$ is increased. This trend occurs in regions I ($\omega^+ = -0.05, 0$) and III ($\omega^+ = +0.2$). In region II ($\omega^+ = +0.05$), however, there is an excessive amplification of $\left< w'^2 \right>_{xzt}$ near the wall, a thinning of the viscous sublayer and a drop in $DR$.  
%As mentioned earlier, the near-wall amplification of $\left< w'^2 \right>_{xzt}$ indicates the amplification of the Stokes layer. In the following sections, we apply triple decomposition to study the interaction of the Stokes layer with the background turbulence. We aim to uncover how the strength of the Stokes layer modifies the near-wall turbulence, which in turn affects the wall drag. We mostly present the quantities based on $u_\tau$ scaling, as we are interested to study the level of departure in the turbulence statistics from the universal non-actuated behaviour, owing to the Stokes layer. In Part 2, we mostly present the quantities based on $u_{\tau_0}$ scaling, as we are interested to study the overall response of turbulence to the wall actuation. Nevertheless, our conclusions from Parts 1 and 2 are valid regardless of scaling by $u_\tau$ or $u_{\tau_0}$.

\subsection{Stokes layer: an important source of inner-scaled actuation}\label{sec:stokes_layer}

As indicated earlier, the near-wall amplification of $\left< w'^2 \right>_{xzt}$ is related to the growth of the Stokes layer. We now apply triple decomposition to more precisely uncover how the strength of the Stokes layer modifies the near-wall turbulence, which in turn affects the wall drag. We primarily consider $u_\tau$ scaling, as we are interested in the level of departure from the non-actuated behaviour. In Part 2, we mostly use $u_{\tau_0}$ scaling, as we are interested to study the overall response of turbulence to the wall actuation. Nevertheless, the conclusions from Parts 1 and 2 are valid regardless of the scaling.

Because the flow is subjected to a harmonic forcing (\ref{eq:wallmotion}), the instantaneous flow can be triply \ar{decomposed} similar to \cite{touber2012}, as in
\begin{align}
 f(x,y,z,t) &= \left< f \right>_{xzt}(y) + \underbrace{\tilde{f}(x,y,t) + f''(x,y,z,t)}_{f'(x,y,z,t)} \tag{3.1\textit{a}} \label{eq:w_decomp} \\[-1mm]
\tilde{f}(x,y,t) &= \frac{1}{N}\sum_{n=0}^{N-1}\left< f \right>_z (x,y,t+nT_{osc}) - \left< f \right>_{xzt}(y) \tag{3.1\textit{b}} \label{eq:w_phase} \\[-1mm]
\left< f'^2 \right>_{xzt} &= \left< \tilde{f}^2 \right>_{xt} + \left< f''^2 \right>_{xzt} \tag{3.1\textit{c}} \label{eq:harm_stoch}
\end{align}
where $f$ indicates the quantity of interest, i.e.\ $u,v$ or $w$. In (\ref{eq:w_decomp}), the total fluctuations $f'$ is decomposed into the harmonic contribution $\tilde{f}$ and the stochastic (turbulent) contribution $f''$. The harmonic contribution $\tilde{f}$ is obtained by phase averaging the spanwise averaged field $\left< f \right>_z$ in time over the number of periods $N$, and then subtracting the mean vertical profile $\left< f \right>_{xzt}$. Accordingly, the total Reynolds stress $\left< f'^2 \right>_{xzt}$ is decomposed into its harmonic component $\left< \tilde{f}^2\right>_{xt}$ associated with the Stokes layer dynamics and its turbulent (stochastic) component $\left< f''^2\right>_{xzt}$ (\ref{eq:harm_stoch}). 

\begin{figure}
  \centering
  \includegraphics[width=.9\textwidth,trim={{0.05\textwidth} {0.05\textwidth} {0.05\textwidth} {0.0\textwidth}},clip]{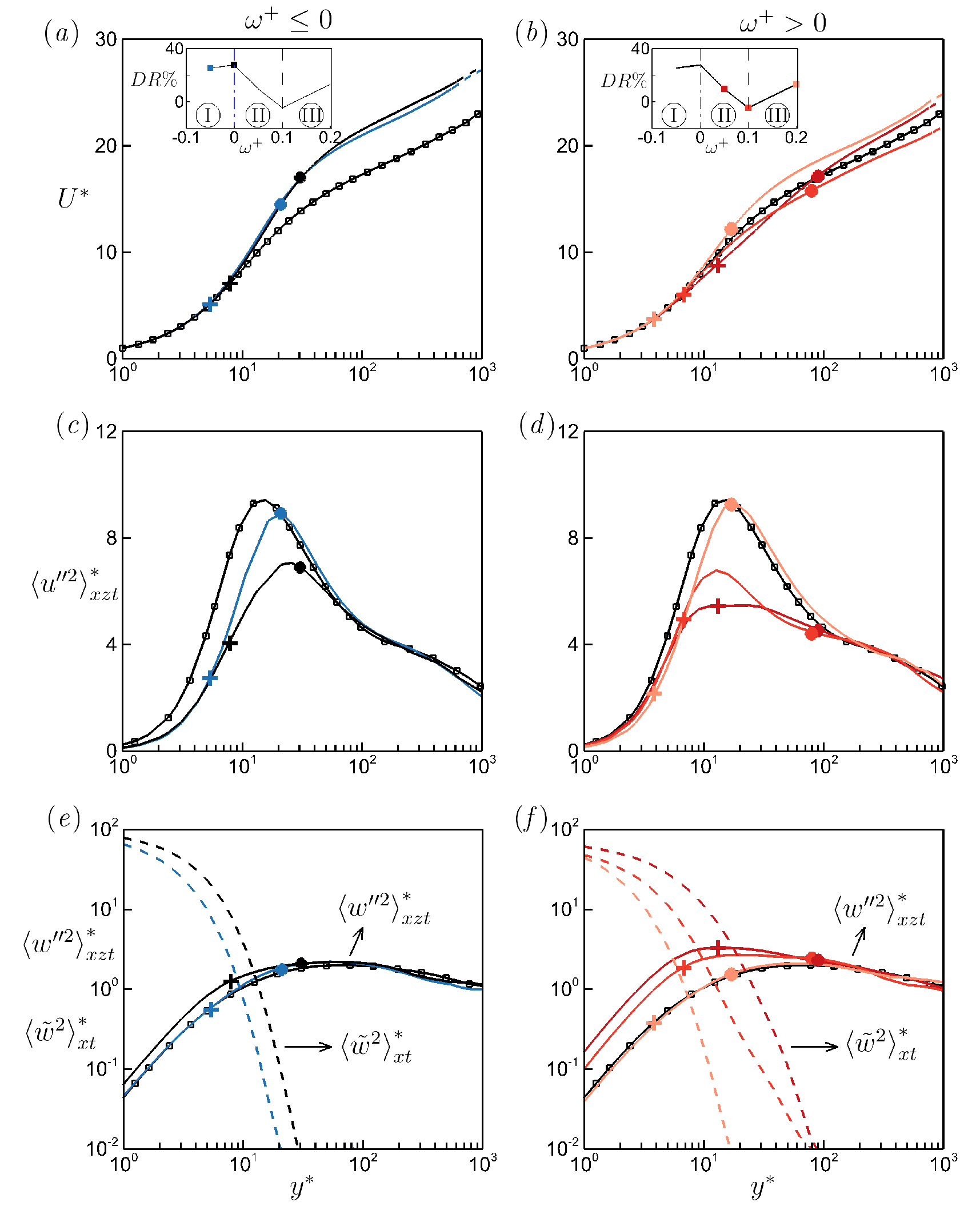}
\caption{Profiles of velocity statistics at $Re_\tau=4000$ for given cases as in figures~\ref{fig:DR_stats_kxp0_00238} and \ref{fig:rms_kxp0_00700_prime} ($A^+ = 12, \kappa^+_x = 0.007$). Line legends are consistent with figures~\ref{fig:DR_stats_kxp0_00238} and \ref{fig:rms_kxp0_00700_prime}. In each panel, only the resolved portion of the profiles are shown corresponding to $y^* \le 1000$. (\textit{a,c,e}) correspond to $\omega^+ \le 0$, and (\textit{b,d,f}) correspond to $\omega^+ > 0$. (\textit{a,b}) $U^*$ profiles; the insets indicate the value of $\omega^+$ and its $DR$ for each profile. In (\textit{c--f}) the Reynolds stress profiles are presented in terms of the turbulent component (\ar{solid lines}) and the harmonic component (\ar{dashed lines}) following (\ref{eq:w_decomp}, \ref{eq:w_phase}). (\textit{c,d}) turbulent component of the streamwise velocity $\left< u''^2 \right>^*_{xzt}$. (\textit{e,f}) turbulent component $\left< w''^2 \right>^*_{xzt}$ and harmonic component $\left< \tilde{w}^2 \right>^*_{xt}$ for the spanwise velocity. On each actuated profile, the cross symbol (+) marks the Stokes layer thickness $\delta^*_S$, and the bullet symbol ($\bullet$) marks the \ar{protrusion} height \ar{$\ell^*_{0.01}$} due to the Stokes layer.}
  \label{fig:rms_kxp0_00700}
\end{figure}

In figure~\ref{fig:rms_kxp0_00700}, we plot these two components for the cases given in figures~\ref{fig:DR_stats_kxp0_00238} and \ref{fig:rms_kxp0_00700_prime} ($A^+ = 12, \kappa^+_x = 0.007$, $Re_\tau=4000$). For reference, figures~\ref{fig:rms_kxp0_00700}(\textit{a,b}) show the considered $U^*$ profiles (as in figures~\ref{fig:DR_stats_kxp0_00238}\textit{b,c}). Figures~\ref{fig:rms_kxp0_00700}(\textit{c,d}) display $\left<u''^2\right>^*_{xzt}$, the stochastic component of the streamwise Reynolds stress. By comparing figure~\ref{fig:rms_kxp0_00700_prime}\textit{b} with figures~\ref{fig:rms_kxp0_00700}\textit{c,d}, we see that  $\left<u'^2\right>^*_{xzt} \simeq \left<u''^2\right>^*_{xzt}$, indicating that the harmonic (Stokes layer) component makes a negligible contribution.
For the spanwise velocity, however, the harmonic component $\left< \tilde{w}^2\right>^*_{xt}$ contributes significantly to the total spanwise Reynolds stress $\left<w'^2\right>^*_{xzt}$ close to the wall (see figures~\ref{fig:rms_kxp0_00700}\textit{e,f}). 
%{\arr The harmonic component} $\left< \tilde{w}^2 \right>^*_{xt}$ represents the Reynolds stress due to the Stokes layer, and {\arr the stochastic component} $ \left< w''^2 \right>^*_{xzt}$ represents the Reynolds stress due to the background turbulence. 
At $y^* \sim \mathcal{O}(1)$, the harmonic component  is about three orders of magnitude larger than the turbulent component, while at $y^* \sim \mathcal{O}(10)$ they have comparable magnitudes. Figures~\ref{fig:rms_kxp0_00700}(\textit{e,f}) indicate that the rate of decay in $\left< \tilde{w}^2 \right>^*_{xt}$, hence the protrusion of the Stokes layer, strongly depends on $\omega^+$. Further, the level of distortion in the $U^*$ profiles (figures~\ref{fig:rms_kxp0_00700}\textit{a,b}) strongly depends on the rate of decay in $\left< \tilde{w}^2 \right>^*_{xt}$. Interestingly, in region II (figure~\ref{fig:rms_kxp0_00700}\textit{f}) the decay rate in $\left< \tilde{w}^2 \right>^*_{xt}$ is noticeably slower compared to regions I and III, implying the presence of a more protrusive Stokes layer. Accordingly, the $U^*$ profile in region II is distorted to the highest level. The turbulent stress profiles are also shown in figure~\ref{fig:spectra_kxp0_00700}, where they are accompanied by the turbulent kinetic energy profiles $\left< \mathcal{K} \right>_{xzt}$, which follow the same trends.
  
To quantify the protrusion of the Stokes layer (figures~\ref{fig:rms_kxp0_00700}\textit{e,f}), we calculate two length scales from the spanwise Reynolds stress profiles. The first is the laminar Stokes layer thickness $\delta^*_S$ that is featured in Stokes' second problem~\citep{batchelor2000}. Following \cite{quadrio2011}, we define $\delta^*_S$  as the height $y^*$ where the amplitude of $\tilde{w}$ decays to $A e^{-1}$ (i.e.\ where $\left< \tilde{w}^2 \right>^*_{xt} = \textstyle{\frac{1}{2}}{A^*}^2 e^{-2}$). In figure~\ref{fig:rms_kxp0_00700}, we mark $\delta^*_S$ on each profile with a cross symbol. The second length scale $\ell^*_{0.01}$ is new, and it is defined as the height where $\left< \tilde{w}^2 \right>^*_{xt} = 0.01$. 
%We arrive at this definition based on the strength of the Reynolds stress due to the Stokes layer ($\left< \tilde{w}^2 \right>^*_{xt}$) relative to the background turbulence ($\left< w''^2 \right>^*_{xzt}$). Considering figures~\ref{fig:rms_kxp0_00700}(\textit{e,f}), for all cases $\left< w''^2 \right>^*_{xzt} \sim \mathcal{O}(1)$ for most of the domain height $y^* \gtrsim 10$. Therefore, 
Our choice for the threshold of $\left< \tilde{w}^2 \right>^*_{xt} = 0.01$ is based on the observation that $\left< w''^2 \right>^*_{xzt} \sim \mathcal{O}(1)$ in the buffer and log regions (also reported by \citealt{lee2015} and \citealt{baidya2021}). In other words, we define $\ell^*_{0.01}$ as the height where the Stokes layer stress $\left< \tilde{w}^2 \right>^*_{xt}$ drops to about $1\%$ of the spanwise turbulent stress $\left< w''^2 \right>^*_{xzt}$. In figure~\ref{fig:rms_kxp0_00700}, we mark $\ell^*_{0.01}$ on each profile with a bullet symbol.

The key difference between $\delta^*_S$ and $\ell^*_{0.01}$ is that we mark $\delta^*_S$ where the Stokes layer stress $\left< \tilde{w}^2 \right>^*_{xt}$ is a small fraction of its maximum value at the wall ${A^*}^2/2$. Thus, we ignore the background turbulence in this definition. However, we mark $\ell^*_{0.01}$ where the Stokes layer stress $\left< \tilde{w}^2 \right>^*_{xt}$ is a small fraction of the turbulent stress $\left< w''^2 \right>^*_{xzt}$, hence considering the background turbulence in this definition. In figures~\ref{fig:rms_kxp0_00700}(\textit{c--f}), $\ell^*_{0.01}$ coincides well with the distance where the actuated $\left< u''^2 \right>^*_{xzt}$ and $\left< w''^2 \right>^*_{xzt}$ profiles depart from the non-actuated counterpart. However, $\delta^*_S$ underestimates the actual protrusion by the Stokes layer due to its ignorance of the background turbulence. For instance, for the case with $\omega^+ = 0$ at $y^* = \delta^*_S$ (black cross symbol in figure~\ref{fig:rms_kxp0_00700}\textit{e}), $\left< \tilde{w}^2 \right>^*_{xt} \simeq 8 \left< w''^2 \right>^*_{xzt}$, i.e.\ the Stokes layer is $8$ times stronger than the background turbulence. However, at $y^* = \ell^*_{0.01}$ (black bullet symbol) $\left< \tilde{w}^2 \right>^*_{xt} \simeq 0.01 \left< w''^2 \right>^*_{xzt}$, i.e.\ the Stokes layer is $100$ times weaker than the background turbulence. We propose, therefore, that $\ell^*_{0.01}$ is a more suitable measure for reflecting the entire penetration of the Stokes layer into the turbulent field.

% For the cases considered here, $\ell^*_{0.01}$ is the height $y^*$ where $\left< \tilde{w}^2 \right>^*_{xt}$ (Stokes layer stress) is nominally $1\%$ of $\left< w''^2 \right>^*_{xzt}$ (turbulent stress), and so it provides a measure of the protrusion height of the Stokes layer into the turbulent field. 

%  Considering figures~\ref{fig:rms_kxp0_00700}(\textit{c--f}), $\ell^*_{0.01}$  coincides well with the distance where the actuated $\left< u''^2 \right>^*_{xzt}$ and $\left< w''^2 \right>^*_{xzt}$ profiles depart from the non-actuated counterpart. 
 
 In regions I and III, the level of protrusion by the Stokes layer $\ell^*_{0.01}$, as well as the departure height in the $\left< u''^2 \right>^*_{xzt}$ and $\left< w''^2 \right>^*_{xzt}$ profiles, stay below $20 - 30$ viscous units. As a result, the mean velocity profiles in regions I and III (figures~\ref{fig:rms_kxp0_00700}\textit{a,b}) yield a well-defined logarithmic shift beyond $y^* \simeq 100$ with viscous sublayer thickening. However, in region II there is a large increase in $\ell^*_{0.01}$ and the departure in the $\left< u''^2 \right>^*_{xzt}$ and $\left< w''^2 \right>^*_{xzt}$ profiles also starts at larger distance from the wall. For example, for $\omega^+ = +0.05$ in region II ( figures~\ref{fig:rms_kxp0_00700}\textit{d,f}), $\ell^*_{0.01} \simeq 80$, which also closely marks the point where the actuated $\left< u''^2 \right>^*_{xzt}$ and $\left< w''^2 \right>^*_{xzt}$ profiles depart from their non-actuated counterpart. As a result, the mean velocity profile for $\omega^+ = +0.05$ in region II (figure~\ref{fig:rms_kxp0_00700}\textit{b}) is highly distorted up to $y^* \simeq 200 - 300$.

% In general, we see that the Stokes layer thickness $\delta^*_S$ does not reflect the actual protrusion of the Stokes layer into the flow. 
% The Stokes layer thickness marks the height where the Stokes layer stress $\left< \tilde{w}^2\right>^*_{xt}$ decays to a fraction of its strength at the wall. However, in the presence of turbulence, the Stokes layer stress at $\delta_S^*$ is several times larger than the turbulent stress $ \left< w''^2 \right>^*_{xzt}$. For instance, for the case where $\omega^+ = 0$ at $y^* = \delta^*_S$ (figure~\ref{fig:rms_kxp0_00700}\textit{e}), it is larger by a factor of 8.  We propose, therefore, that $\ell^*_{0.01}$ is a more suitable measure for the full Stokes layer penetration into the turbulent field.

\begin{figure}
  \centering
  \includegraphics[width=\textwidth,trim={{0.0\textwidth} {0.03\textwidth} {0.0\textwidth} {0.0\textwidth}},clip]{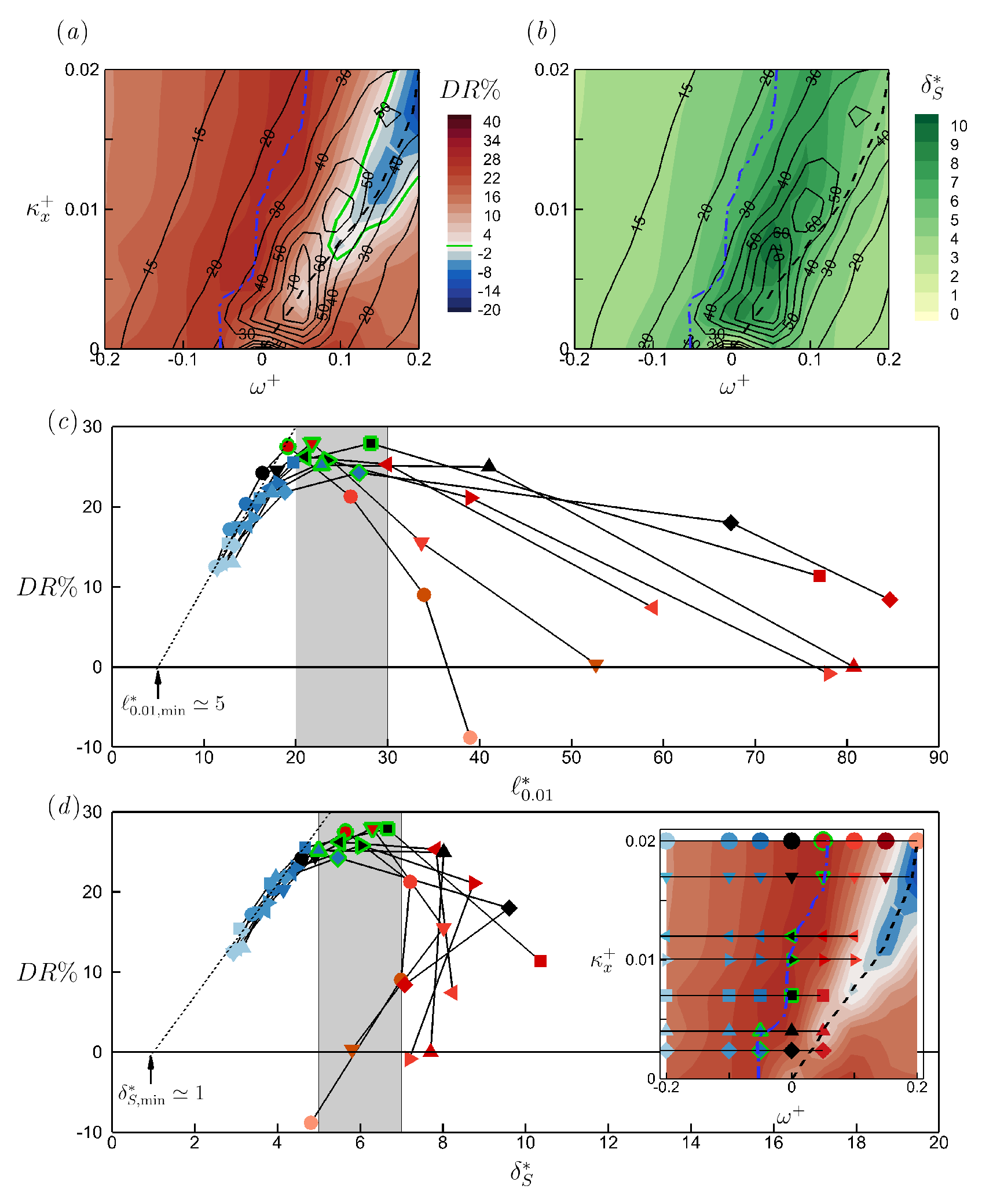}
\caption{(\textit{a}) Comparison between the map of drag reduction $DR$ (contour field) and the protrusion height by the Stokes layer $\ell^*_{0.01}$ (contour lines) for our considered parameter space at $Re_\tau = 4000$. (\textit{b}) Comparison between the map of Stokes layer thickness $\delta^*_S$ (contour field) and $\ell^*_{0.01}$ (contour lines) for the same cases as in (\textit{a}). The lines ({\color{blue}\dasheddotted}) and (\dashed) are the local maximum and minimum $DR$ (same as in figure~\ref{fig:dr_map_compare}\textit{b}). (\textit{c,d}) plot $DR$ versus $\ell^*_{0.01}$ and $DR$ versus $\delta^*_S$, respectively, for the same data as in (\textit{a,b}); $\kappa^+_x = 0.00238$ ($\blackdiam$), $0.004$ ($\blacktriangle$), $0.007$ (\filsqr), $0.010$ ($\blacktriangleright$), $0.012$ ($\blacktriangleleft$), $0.017$ ($\blacktriangledown$), $0.021$ ($\bullet$). At each $\kappa^+_x$, we plot the cases only in regions I and II (see the map in \textit{d}), with the maximum $DR$ case highlighted with a green outline. The grey regions in (\textit{c,d}) $(20 \le \ell^*_{0.01} \le 30, 5 \le \delta^*_S \le 7)$ shade the range of maximum $DR$ at each $\kappa^+_x$. The linear dotted lines in (\textit{c,d}) fit the data for $\ell^*_{0.01} \lesssim 20$ (\textit{c}) and $\delta^*_S \lesssim 5$ (\textit{d}). The fitting lines also locate the minimum values for $\ell^*_{0.01,\mathrm{min}} \simeq 5$ (\textit{c}) and $\delta^*_{S,\mathrm{min}} \simeq 1$ (\textit{d}) to achieve drag reduction.}
  \label{fig:DR_deltas}
\end{figure}

Furthermore, we can draw a connection between the protrusion height $\ell^*_{0.01}$ and the level of drag reduction.
%{\arr Our findings so far, highlight the drastic changes in the near-wall turbulence as a result of the drastic changes in the protrusion of the Stokes layer $(\ell^*_{0.01})$.} \ar{In fact, the Stokes layer alters the near-wall turbulence which in turn affects the drag. Therefore, we expect $DR$ to be related to $\ell^*_{0.01}$. Figures~\ref{fig:rms_kxp0_00700}(\textit{a,b}) show some relation between $DR$ (the insets) and $\ell^*_{0.01}$ (bullet symbols). In region I (figure~\ref{fig:rms_kxp0_00700}\textit{a}), from $\omega^+ = -0.05$ to zero, $\ell^*_{0.01}$ increases from $20$ to $30$ and $DR$ increases from $25\%$ to its maximum $28\%$. Therefore, $\ell^*_{0.01}$ has a favourable effect on $DR$ in this region. On the other hand, in region II (figure~\ref{fig:rms_kxp0_00700}\textit{b}), from $\omega^+ = 0$ to $+0.10$, $\ell^*_{0.01}$ increases from $30$ to $90$, while $DR$ drops from $28\%$ to $-5\%$. Thus, $\ell^*_{0.01}$ has an unfavourable effect on $DR$ in this region. {\arr In region III, $\ell^*_{0.01}$ drops to below $20$ and $DR\%$ increases to $13\%$.}
In figures~\ref{fig:DR_deltas}(\textit{a,b}), we overlay the map of $\ell^*_{0.01}$ onto the maps of $DR$ and $\delta^*_S$. 
%These maps highlight that our observations of figures~\ref{fig:rms_kxp0_00700}(\textit{a,b}) regarding the relation between $DR, \ell^*_{0.01}$ and $\delta^*_S$ are applicable to our entire parameter space of $(\omega^+, \kappa^+_x)$. 
In region I (left side of the blue dashed-dotted line), $\ell^*_{0.01} \lesssim 30$ and $\delta^*_S \lesssim 7$. In this region, an increase in $\ell^*_{0.01}$ and $\delta^*_S$ leads to an increase in $DR$.  For upstream travelling waves ($\omega^+ < 0$), therefore, the growing protrusion of the Stokes layer has a favourable effect on $DR$. In contrast, in region II (between the blue dashed-dotted line and the black dashed line), $DR$ drops by increasing $\ell^*_{0.01}$ and $\delta^*_S$. Another difference between regions I and II, is in the relation between $\ell^*_{0.01}$ and $\delta^*_S$. In region I, $\ell^*_{0.01}$ and $\delta^*_S$ are proportional to each other with $\ell^*_{0.01}\approx 4\delta^*_S$. However, in region II this proportional relation is broken and $\ell^*_{0.01}$ can reach as high as $8\delta^*_S$.
At each $\kappa^+_x$, the maximum $DR$ (the blue dashed-dotted line) coincides with the optimal range $20 \lesssim \ell^*_{0.01} \lesssim 30$ $(5 \lesssim \delta^*_S \lesssim 7)$. In figure~\ref{fig:DR_deltas}(\textit{c}), we plot $DR$ versus $\ell^*_{0.01}$ for our simulation cases in regions I and II. Also, following \cite{quadrio2011} (their figure~9), we plot $DR$ versus $\delta^*_S$ for the same cases (figure~\ref{fig:DR_deltas}\textit{d}). These plots confirm that the maximum $DR$ coincides with $20 \lesssim \ell^*_{0.01} \lesssim 30$ and $5 \lesssim \delta^*_S \lesssim 7$ (shaded in grey). Further, for $\ell^*_{0.01} \lesssim 20$ ($\delta^*_S \lesssim 5$), $\ell^*_{0.01} \simeq 4 \delta^*_S$ and $DR$ increases linearly with $\ell^*_{0.01}$ and $\delta^*_S$ (see the fitting dotted lines in figures~\ref{fig:DR_deltas}\textit{c,d}). Following \cite{quadrio2011}, if we extrapolate the linear fits to $DR = 0$, we obtain $\ell^*_{0.01,\mathrm{min}}\simeq 5$ and $\delta^*_{S,\mathrm{min}}\simeq 1$; these values indicate the minimum limits for drag reduction to occur. 

The linear relation between $DR, \ell^*_{0.01}$ and $\delta^*_S$ is limited to region I. In region II when $DR$ drops, this linear relation is broken. The observed trends for $DR$ versus $\delta^*_S$ and the minimum limit of $\delta^*_{S,\mathrm{min}}\simeq 1$ (figure~\ref{fig:DR_deltas}\textit{d}) are also reported by \cite{quadrio2011} (their figure~9). \cite{quadrio2011} calculated $\delta^*_S$ from the laminar $\left< \tilde{w}^2 \right>^*_{xt}$ profile based on Stokes layer solution. Here, however, we calculate $\delta^*_S$ from the actual $\left< \tilde{w}^2 \right>^*_{xt}$ profile by phase averaging the simulation data. The variation in $DR$ versus $\ell^*_{0.01}$ is similar to $DR$ versus $\delta^*_S$ up to region I, in terms of the linear trends, an optimal thickness for the maximum $DR$ and a minimum thickness for the occurrence of $DR$ ($\delta^*_{S,\mathrm{min}}, \ell^*_{0.01, \mathrm{min}}$). However, in region II when the linear trends are broken, we observe noticeable differences between $DR$ versus $\delta^*_S$ and $DR$ versus $\ell^*_{0.01}$. In region II, there does not appear to be a consistent relation between $DR$ and $\delta^*_S$ (the red symbols in figure~\ref{fig:DR_deltas}\textit{d}). In other words, we cannot find a threshold for $\delta^*_S$ beyond which $DR$ drops. For instance, for the case with $\kappa^+_x = 0.02, \omega^+ = +0.2$, $DR$ drops to $-10\%$ but $\delta^*_S \simeq 5$ which is within the optimal range $(5 \lesssim \delta^*_S \lesssim 7)$. In contrast, there is a much stronger connection between $DR$ and $\ell^*_{0.01}$, even in region II (figure~\ref{fig:DR_deltas}\textit{c}). For all cases, increasing $\ell^*_{0.01}$ beyond $30$ decreases $DR$. As a result, the value of $\ell^*_{0.01}$ can be used to determine whether we are in region I ($\ell^*_{0.01} \lesssim 30$) or region II ($\ell^*_{0.01} \gtrsim 30$).

% We further confirm this in figure~\ref{fig:DR_deltas}(\textit{c}) by plotting $DR$ versus $\ell^*_{0.01}$ for the cases at $Re_\tau = 4000$ that fall into regions I and II. At each $\kappa^+_x$, the case with the maximum $DR$ (highlighted with a green outline) has $20 \lesssim \ell^*_{0.01} \lesssim 30$ (shaded in grey).

% What is surprising is when $\ell^*_{0.01} > 30$ (e.g.\ $\omega^+ = +0.05, +0.10$ in figure~\ref{fig:rms_kxp0_00700}\textit{d}), the turbulent Reynolds stress $\left< u''^2 \right>^*_{xzt}$ has a higher attenuation level compared to when $\ell^*_{0.01} \le 30$ (e.g.\ $\omega^+ = -0.05, 0$ in figure~\ref{fig:rms_kxp0_00700}\textit{c}), however, $DR$ is lower for the more attenuated cases with $\ell^*_{0.01} > 30$. Naturally, we might expect $DR$ to be higher when $\left< u''^2 \right>^*_{xzt}$ has a higher attenuation level. We investigate this question in the next section.

\subsection{Interaction between the Stokes layer and the near-wall turbulence}\label{sec:deltas_DR}

\begin{landscape}
\begin{figure}
  \centering
  \includegraphics[width=1.4\textwidth,trim={{0.0\textwidth} {0.0\textwidth} {0.0\textwidth} {0.0\textwidth}},clip]{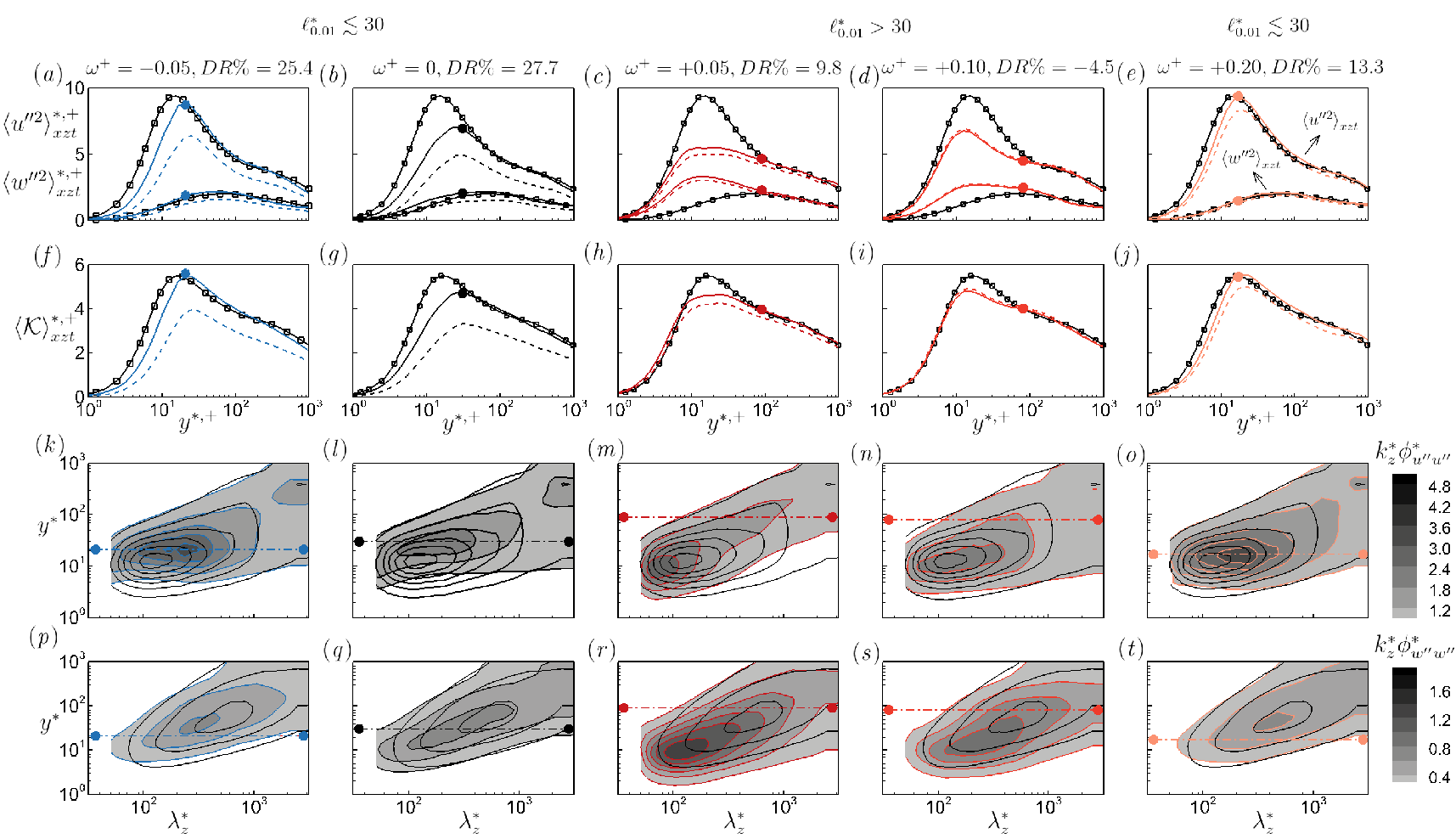}
  \caption{Profiles of the turbulence statistics (\textit{a--j}) and pre-multiplied spectrograms (\textit{k--t}) at $Re_\tau = 4000$ for the non-actuated case and the actuated cases with $A^+ = 12, \kappa^+_x = 0.007$ and different values of $\omega^+$ (same cases as in figure~\ref{fig:rms_kxp0_00700}); $\omega^+ = -0.05$ (\textit{a,f,k,p}), $\omega^+ = 0$ (\textit{b,g,l,q}), $\omega^+ = +0.05$ (\textit{c,h,m,r}), $\omega^+ = +0.10$ (\textit{d,i,n,s}) and $\omega^+ = +0.20$ (\textit{e,j,o,t}). (\textit{a--e}) profiles of turbulent stresses for the streamwise and spanwise velocity $\left< u''^2 \right>_{xzt}, \left< w''^2 \right>_{xzt}$. (\textit{f--j}) profiles of turbulent kinetic energy $\left< \mathcal{K} \right>_{xzt} = \left( \left< u''^2 \right>_{xzt} + \left< v''^2 \right>_{xzt} + \left< w''^2 \right>_{xzt} \right)/2$. Throughout (\textit{a--j}), black lines with symbols correspond to the non-actuated case; lines with no symbol correspond to the actuated case scaled by the actuated $u_\tau$ (solid line) and the non-actuated $u_{\tau_o}$ (dashed line). Pre-multiplied spectrograms for the turbulent part of the streamwise velocity $k^*_z \phi^*_{u'' u''}$ (\textit{k--o}) and spanwise velocity $k^*_z \phi^*_{w'' w''}$ (\textit{p--t}). The filled contours correspond to the actuated cases and the line contours correspond to the non-actuated case. \ar{The contour lines for $k^*_z \phi^*_{u'' u''}$ (\textit{k--o}) change from $0.6$ to $4.8$ with an increment of $0.6$, and for $k^*_z \phi^*_{w'' w''}$ (\textit{p--t}) change from $0.2$ to $1.8$ with an increment of $0.2$.} We locate \ar{$\ell^*_{0.01}$} with a bullet point.} 
  \label{fig:spectra_kxp0_00700}
\end{figure}
\end{landscape}

\begin{landscape}
\begin{figure}
  \centering
  \includegraphics[width=1.5\textwidth,trim={{-0.07\textwidth} {0.0\textwidth} {0.0\textwidth} {0.0\textwidth}},clip]{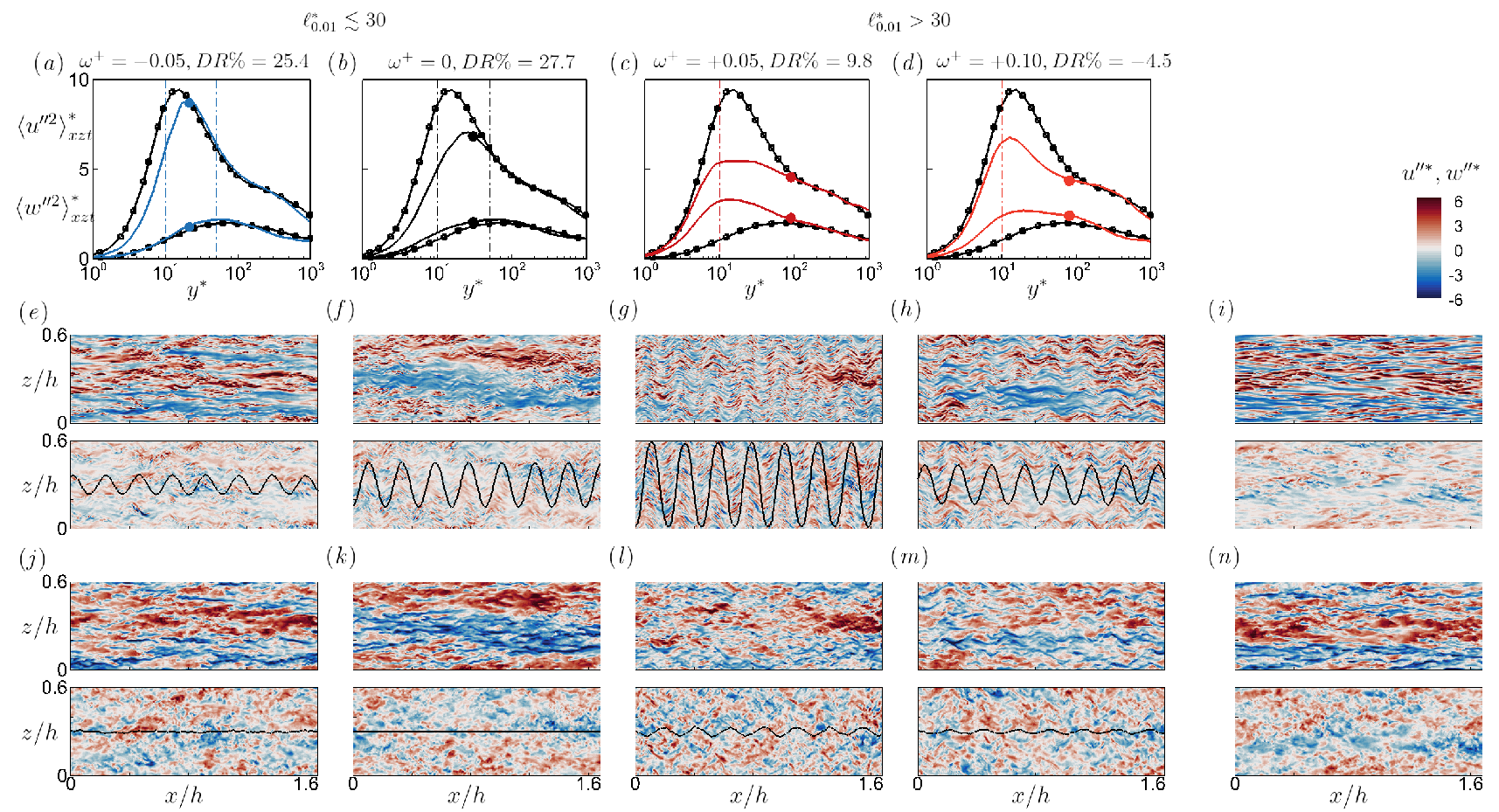}
  \put(-17.2,1.4){$y^* = 50$}
 \put(-17.2,4.2){$y^* = 10$} 
\caption{Profiles of the turbulence statistics (\textit{a--d}) and visualisation of the near-wall turbulence (\textit{e--n}) at $Re_\tau = 4000$ for the non-actuated case (\textit{i,n}) and the actuated cases (\textit{e--h, j--m}) with $A^+ = 12, \kappa^+_x = 0.007$ and different vales of $\omega^+$ (same cases as in figures~\ref{fig:rms_kxp0_00700} and \ref{fig:spectra_kxp0_00700}); $\omega^+ = -0.05$ (\textit{a,e,j}), $\omega^+ = 0$ (\textit{b,f,k}), $\omega^+ = +0.05$ (\textit{c,g,l}) and $\omega^+ = +0.10$ (\textit{d,h,m}). (\textit{a--d}) profiles of turbulent stresses for the streamwise and spanwise velocity components $\left< u''^2 \right>^*_{xzt}, \left< w''^2 \right>^*_{xzt}$ (same profiles as in figures~\ref{fig:spectra_kxp0_00700}\textit{a--d}); lines correspond to the actuated cases and lines with symbols correspond to the non-actuated case. Visualisation of the near-wall turbulence at (\textit{e--i}) $y^* = 10$ and (\textit{j--n}) $y^* = 50$ (located with vertical dashed-dotted lines in \textit{a--d}). In each of (\textit{e--n}), the upper field shows the turbulent streamwise velocity $u''^*$ and the lower field shows the turbulent spanwise velocity $w''^*$. We overlay the spanwise and phase-averaged spanwise velocity $\tilde{w}^*$ (as solid curves) onto the $w''^*$ field.
} 
  \label{fig:flowviz_DR}
\end{figure}
\end{landscape}

%Thus far, we have established a close relationship between the levels of Stokes layer protrusion, turbulence modification, and drag reduction. There appeared to be an optimal range $20 \lesssim \ell^*_{0.01} \lesssim 30$ where $DR$ reaches a maximum. Below this range an increase in $\ell^*_{0.01}$ leads to an increase in $DR$, while beyond this range an increase in $\ell^*_{0.01}$ leads to a decrease in $DR$.  
In a turbulent flow with spanwise wall oscillation, \cite{touber2012} similarly report that an overly protrusive Stokes layer leads to the degradation of $DR$. 
They proposed that the attenuation of $\left< u''^2 \right>_{xzt}$ and amplification of $\left< w''^2 \right>_{xzt}$ are based on the periodic realignment of the near-wall streaks. To examine this proposal further, we consider energy spectrograms and  near-wall flow visualisations (figures~\ref{fig:spectra_kxp0_00700} and \ref{fig:flowviz_DR}).  We focus on the same cases as in figure~\ref{fig:rms_kxp0_00700}, where $Re_\tau = 4000$, $A^+ = 12$, and $\kappa^+_x = 0.007$.

For the cases with \ar{$\ell^*_{0.01} \lesssim 30$}, the streamwise pre-multiplied spectrograms $k^*_z \phi^*_{u'' u''}$ (figures~\ref{fig:spectra_kxp0_00700}\textit{k,l,o}) show the attenuation of $u''^2$ below \ar{$y^* \simeq \ell^*_{0.01}$} (i.e.\ within the Stokes layer). For these cases, increasing \ar{$\ell^*_{0.01}$} (hence strengthening the Stokes layer) attenuates $u''$ over a wider range of wavelength $\lambda^*_z$ and height $y^*$ (e.g.\ compare figure~\ref{fig:spectra_kxp0_00700}\textit{k} with \ref{fig:spectra_kxp0_00700}\textit{l}). At the same time, the energetic peak in $k^*_z \phi^*_{u'' u''}$ is shifted to a higher $y^*$ and a higher $\lambda^*_z$. This attenuation is apparent in the visualisations of the instantaneous velocity fields of $u''$ at $y^* = 10$ for the cases with $\ell^*_{0.01} \lesssim 30$ (figures~\ref{fig:flowviz_DR}\textit{e,f}). On the $w''$ fields, we overlay the spanwise and phase-averaged $\tilde{w}^*$ (solid black curves) as a measure of the Stokes motion. As \ar{$\ell^*_{0.01}$} increases from $20$ at $\omega^+ = -0.05$ (figure~\ref{fig:flowviz_DR}\textit{e}) to $30$ at $\omega^+ = 0$ (figure~\ref{fig:flowviz_DR}\textit{f}), \ar{the Stokes motion becomes stronger and} the energy level in the $u''$ field is decreased compared to the non-actuated counterpart (figure~\ref{fig:flowviz_DR}\textit{i}). At the same time, the spanwise spacing between the high-speed streaks increases by increasing \ar{$\ell^*_{0.01}$}. Overall, for $\ar{\ell^*_{0.01}} \lesssim 30$, the Stokes layer dampens the level of turbulence within $y^* \lesssim \ar{\ell^*_{0.01}}$, hence acting favourably towards increasing $DR$.

For the cases with \ar{$\ell^*_{0.01} > 30$}, the Stokes layer is excessively strong and protrusive. As a result, the near-wall flow structures meander, following the Stokes motion. This meandering is observed in the $u''$ and $w''$ fields for the cases with \ar{$\ell^*_{0.01} > 30$} at $y^* = 10$ (figures~\ref{fig:flowviz_DR}\textit{g,h}). Even at $y^* = 50$ for the same cases (figures~\ref{fig:flowviz_DR}\textit{l,m}), we can see the protrusion of the Stokes motion (solid black curves) and evidence of meandering in the $u''$ fields. This meandering is also evident in the spectrograms. For instance, \ar{for} $\omega^+ = +0.05$ \ar{with $\ell^*_{0.01} \simeq 90$} (figures~\ref{fig:spectra_kxp0_00700}\textit{m,r}), the meandering flow structures at $y^* \simeq 10$ \ar{(visualised in figure~\ref{fig:flowviz_DR}\textit{g})} manifest as an energetic peak in the $k^*_z \phi^*_{w'' w''}$ spectrogram (figure~\ref{fig:spectra_kxp0_00700}\textit{r}) at $(\lambda^*_z, y^*) \simeq (100, 10)$\ar{; this peak} coincides with the peak in the $k^*_z \phi^*_{u'' u''}$ spectrogram (figure~\ref{fig:spectra_kxp0_00700}\textit{m}). \cite{touber2012} relate the attenuation of $\left< u''^2 \right>_{xzt}$ and amplification $\left< w''^2 \right>_{xzt}$ (e.g.\ figures~\ref{fig:flowviz_DR}\textit{c,d}) to this meandering behaviour and argue that the strong Stokes shear strain periodically re-orients the streaks. As a result, energy is transferred from $u''$ to $w''$, and the anisotropy between $u''$ and $w''$ is reduced. Considering the flow visualisation at $y^* = 10$ for $\omega^+ = +0.05$ with $\ell^*_S \simeq 90$ (figure~\ref{fig:flowviz_DR}\textit{g}), we see a strong resemblance between the $u''$ and $w''$ fields in terms of the energy level and structure, which support the \ar{reduction} in anisotropy.

A noticeable difference between the cases with \ar{$\ell^*_{0.01} \lesssim 30$} and those with $\ell^*_{0.01} > 30$ is the wall distance of the maximum turbulence activity given by \ar{the} location of the energetic peaks in $k^*_z \phi^*_{u'' u''}$ and $k^*_z \phi^*_{w'' w''}$. For the cases with \ar{$\ell^*_{0.01} \lesssim 30$} (figures~\ref{fig:spectra_kxp0_00700}\textit{k,l,o}), the energetic peak in $k^*_z \phi^*_{u'' u''}$ is lifted away from the wall to a $y^*$ distance that coincides with \ar{$\ell^*_{0.01}$}. However, for the cases with \ar{$\ell^*_{0.01} > 30$}, the energetic peaks in $k^*_z \phi^*_{u'' u''}$ (figures~\ref{fig:spectra_kxp0_00700}\textit{m,n}) and $k^*_z \phi^*_{w'' w''}$ (figures~\ref{fig:spectra_kxp0_00700}\textit{r,s}) instead reside near the wall at $y^* \simeq 10$, well below \ar{$\ell^*_{0.01} \simeq 90$}.  It appears that when \ar{$\ell^*_{0.01} > 30$}, a near-wall cycle of streaks with high turbulence activity is generated within the Stokes layer. Contrast this \ar{behaviour} to the case when \ar{$\ell^*_{0.01} \lesssim 30$} where the turbulence is damped within the Stokes layer and the cycle of turbulence generation is lifted away from the wall. 

Overall, through flow visualisations and spectrograms we could explain the physics behind the trends in $DR$ versus $\ell^*_{0.01}$ (figure~\ref{fig:DR_deltas}). When $\ell^*_{0.01} \lesssim 30$, turbulence is damped within $y^* \lesssim \ell^*_{0.01}$. The level of damping increases by increasing $\ell^*_{0.01}$. As a result, $DR$ increases by increasing $\ell^*_{0.01}$, with the maximum $DR$ attained when $\ell^*_{0.01} \simeq 30$. However, when $\ell^*_{0.01} > 30$, the Stokes layer becomes excessively strong. In this situation, a near-wall cycle of turbulence is generated at $y^* \simeq 10$ that meanders following the Stokes motion. As a result, $DR$ drops by increasing $\ell^*_{0.01}$.

%\noindent 

\subsection{Power performance analysis}\label{sec:power}

While drag reduction is an important performance parameter for many applications, the efficiency of the  flow control effort is often even more important. Here we use the concept of net power saving $NPS$:
\begin{equation}
 NPS = \frac{P^+_0 - \left( P^+ + P^+_\mathrm{in} \right)}{P^+_0} = DR - \frac{P^+_\mathrm{in}}{P^+_0} \tag{3.4} \label{eq:nps}
\end{equation}
where $P^+ = \left( 1 - DR \right) U^+_b$ is the pumping power required to drive the flow through the actuated channel, $P^+_0$ is the non-actuated analogue of $P^+$, and $P^+_\mathrm{in}$ is the input power required to oscillate the wall actuation mechanism (\ref{eq:wallmotion}) while neglecting any mechanical losses. A positive $NPS$ indicates that the total power cost of the actuated case is less than the total cost of its non-actuated counterpart. 
%In other words, the power saving due to drag reduction, $(P^+_0 - P^+)/P^+_0 = DR$, outweighs the required input power, $P^+_\mathrm{in}/P^+_0$.  
We are also interested in 
\ar{assessing the accuracy of generalised Stokes layer (GSL) theory~\citep{quadrio2011} for estimating $P^+_\mathrm{in}$.  In Part 2 \citep{Chandran2022part2}, we use this theory to estimate $NPS$ for our experimental data.}  Here, in Part I, our actuation frequencies fall into the ISA regime. In Part 2, the data fall into both the ISA and OSA regimes. 

The input power is given as follows, as first proposed by \cite{baron1995} for an oscillating plane, and then used by \cite{quadrio2011}, \cite{gatti2013} and \cite{marusic2021} for a travelling wave.
\begin{equation}
 P^+_\mathrm{in} = \frac{1}{T^+_\mathrm{avg} L^+_x L^+_z}\int_{t^+}^{t^+ + T^+_\mathrm{avg}} \int_0^{L^+_x} \int_0^{L^+_z} w^+_s \left( \frac{\partial w^+}{\partial y^+}\Big|_{y^+ = 0} \right) dx^+ dz^+ dt^+  \tag{3.2} \label{eq:pin}
\end{equation}
where all the quantities are normalised by $\nu$ and the non-actuated $u_{\tau_o}$ (hence superscripted with \ar{a cross symbol}). In (\ref{eq:pin}), $T_\mathrm{avg}$ is the averaging time, $w_s$ is the instantaneous wall velocity (\ref{eq:wallmotion}) and $\partial w^+/\partial y^+ |_{y^+=0}$ is the instantaneous wall-normal gradient of the spanwise velocity at the wall.

\begin{figure}
  \centering
  \includegraphics[width=\textwidth,trim={{0.04\textwidth} {0.03\textwidth} {0.0\textwidth} {0.0\textwidth}},clip]{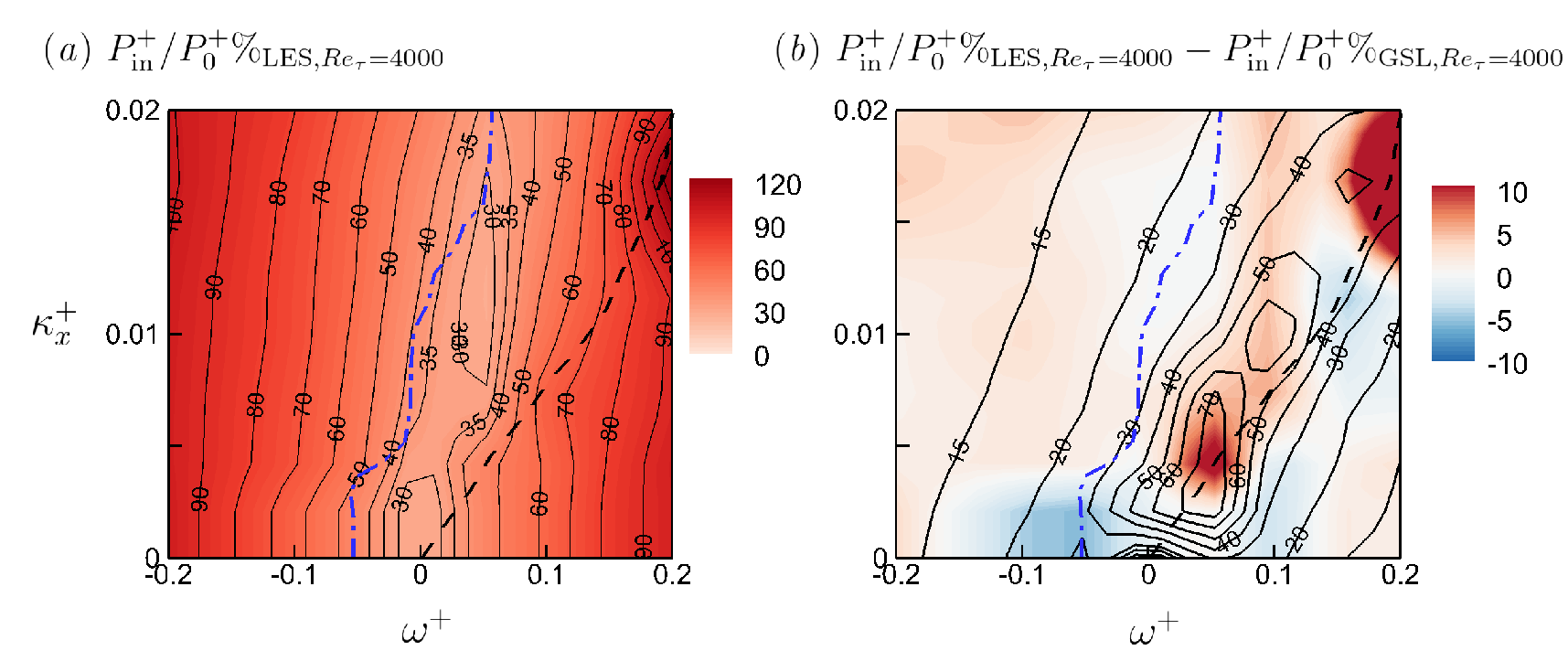}
\caption{(\textit{a}) Map of $P^+_\mathrm{in}$ at $Re_\tau = 4000$. The filled contour and the line contours show the same quantity. (\textit{b}) Filled contour is the difference in calculation of $P^+_\mathrm{in}$ at $Re_\tau = 4000$ between LES and its theoretical estimation from the generalised Stokes layer (GSL) theory~\citep{quadrio2011}; line contours give the Stokes layer protrusion height $\ell^*_{0.01}$ (same as in figure~\ref{fig:DR_deltas}\textit{a}). 
%(\textit{c}) Net power saving $NPS$ for LES at $Re_\tau = 4000$. (\textit{d}) Difference in $NPS$ between LES at $Re_\tau = 4000$ and LES at $Re_\tau = 951$. In each of (\textit{c,d}), the filled contour and the line contours show the same quantity. All the quantities with + superscript are scaled by $\nu$ and the non-actuated $u_{\tau_o}$. ({\color{blue}\dasheddotted}) and (\dashed) locate the local maximum and the local minimum $DR$, respectively.
} 
  \label{fig:power}
\end{figure}

In \ar{figure~\ref{fig:power}(\textit{a})}, we present the map of  $P^+_\mathrm{in}/P^+_0$ as computed over our parameter space of $(\omega^+, \kappa^+_x)$ at \ar{$Re_\tau = 4000$}.  The map is much more symmetric about $\omega^+ = 0$ compared to $DR$ (figure~\ref{fig:dr_map_compare}). \ar{We also see that substantially more power is required at higher actuation frequencies. For example, $P^+_\mathrm{in}/P^+_0 \%$ can reach up to $100\%$ when $\omega^+ \simeq \pm 0.2$. \ar{In region II,} between the local maximum and the local minimum $DR$ \ar{(between the blue dashed-dotted line and the black dashed line)}, $P^+_\mathrm{in}/P^+_0$ decreases to about $30\% - 35\%$. }

We can use \eqref{eq:pin} only if we have an estimate for $\partial w^+/\partial y^+ |_{y^+=0}$.  In most experimental studies, including Part 2 of the present study, this quantity is unavailable and some estimate needs to be made instead.  In Part 2 we use GSL theory, \ar{which gives the} instantaneous spanwise velocity for a laminar flow with  wall actuation.  That is,
\begin{align}
  &{\color{white}=}w^+(x^+,y^+,t^+) = \tag{3.3} \label{eq:gs_theory}  \\
\nonumber &{\color{white}=} A^+ \mathcal{R}\left\{ C e^{i (\kappa^+_x x^+ - \omega^+ t^+)}\mbox{Ai}\left[ e^{\pi i/6}\left( \kappa^+_x [1-DR ] \right)^{1/3} \left( y^+ - \frac{\omega^+}{\kappa^+_x [1-DR]} - \frac{i \kappa^+_x}{1-DR} \right) \right]  \right\}
\end{align}
where $C = \left\{ \mbox{Ai} \left[ i e^{i \pi/3} \left( \kappa^+_x \left[ 1-DR  \right]  \right)^{1/3} (\omega^+/\kappa^+_x + i \kappa^+_x)/[1-DR ] \right] \right\}^{-1}$, Ai is the Airy function of the first kind, and $\mathcal{R}\{ ... \}$ is the real part of the argument.  To use (\ref{eq:gs_theory}) for a turbulent flow, one needs to assume that \ar{1)} the Stokes layer preserves its laminar structure near the wall, i.e.,  $\tilde{w}$ is the same in the laminar and turbulent flow, and 2) the turbulent spanwise velocity is negligible near the wall, i.e., $w'' \simeq 0$.

We now compare the results for $P^+_\mathrm{in}$ using GSL to the results obtained using LES, so as to verify the validity of using GSL estimates in experiments. Figure~\ref{fig:power}(\textit{b}) shows the difference between the pumping power obtained from the LES \ar{at $Re_\tau = 4000$ ($P^+_\mathrm{in}/P^+_0 \%_{\mathrm{LES}, Re_\tau = 4000}$)} and that estimated using GSL theory \ar{at the same $Re_\tau$ ($P^+_\mathrm{in}/P^+_0 \%_{\mathrm{GSL}, Re_\tau = 4000}$)}. Overall, the differences are small, especially in the range $\omega^+ < 0$ where \ar{they differ less than $3\%$}. Only in region II with $\omega^+ > 0$ (between the blue dashed-dotted line and the black dashed line) do the differences approach $10\%$, especially along the minimum drag reduction line (black dashed line). The overlay of \ar{the Stokes layer protrusion height $\ell^*_{0.01}$ (contour lines)} indicates that the region where the power differences are significant coincides with the region where $\ell^*_{0.01}$ is large. In other words, the error in using GSL theory (laminar Stokes layer assumption) is largest when the Stokes layer is most protrusive. 

\begin{figure}
  \centering
  \includegraphics[width=.92\textwidth,trim={{0.05\textwidth} {0.05\textwidth} {0.05\textwidth} {0.0\textwidth}},clip]{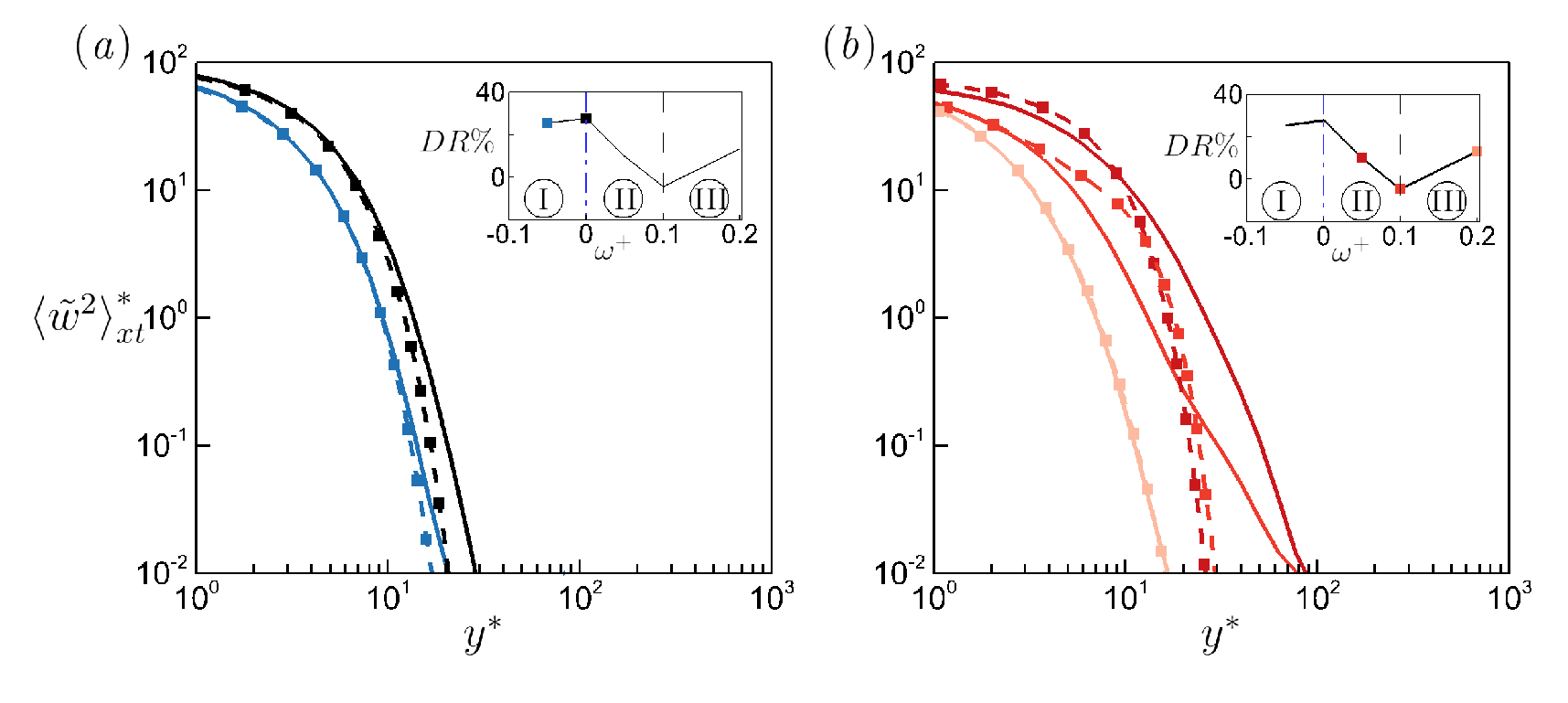}
\caption{Comparison of the phase-averaged (harmonic) Reynolds stress profiles $\left< \tilde{w}^2 \right>^*_{xt}$ between LES (solid lines) and the laminar solution from the generalised Stokes layer (GSL) theory (dashed lines with symbols) ($A^+ = 12, \ \kappa^+_x = 0.007$ at $Re_\tau = 4000$). (\textit{a}) $\omega^+ = -0.05, 0$. (\textit{b}) $\omega^+ = +0.05, +0.10, +0.20$.  The insets plot $DR$ for the selected cases.}  
  \label{fig:GSL_LES}
\end{figure}

This behaviour is further substantiated by figure~\ref{fig:GSL_LES}, which compares the phase-averaged (harmonic) Reynolds stress profiles $\left< \tilde{w}^2 \right>^*_{xt}$ between LES (solid lines) and its laminar solution (\ref{eq:gs_theory}) from  GSL theory (dashed lines with symbols). We see that in region I with $\omega^+ \le 0$ (figure~\ref{fig:GSL_LES}\textit{a}), the agreement is reasonably good; we obtain better agreement with $\omega^+ = -0.05$ ($\ell^*_{0.01} \simeq 20$) than with $\omega^+ = 0$ ($\ell^*_{0.01} \simeq 30$). However, in region II (figure~\ref{fig:GSL_LES}\textit{b}), when $\omega^+ = +0.05$ ($\ell^*_{0.01} \simeq 90$) and $\omega^+ = +0.10$ ($\ell^*_{0.01} \simeq 80$), we observe significant departures between LES and the GSL theory. For instance, for $\omega^+ = +0.05$ at $y^* \simeq 20$, $\left< \tilde{w}^2 \right>^*_{xz}$ from LES is $2.2$ but from GSL is $0.2$. This is a significant difference considering that the background turbulent stress $\left< w''^2 \right>^*_{xzt} \sim \mathcal{O}(1)$. In region III with $\omega^+ = +0.20$ where $\ell^*_{0.01} \simeq 17$, we see a return of the good agreement between LES and GSL theory.

Our observations regarding the differences between $P_\mathrm{in}$ from the simulation and that from the GSL theory (figure~\ref{fig:power}) are similar to those reported by \cite{quadrio2011} (their figure~7); they report close agreement between the GSL theory and the turbulence simulation in the drag-decreasing range, but report noticeable differences in the drag-increasing range. They explain this behaviour through the timescale $\mathcal{T}^+ \equiv 2\pi/(\omega^+ - \kappa^+_x \mathcal{U}^+_w)$, which represents the period of oscillation as observed by the near-wall eddies with the convective speed $\mathcal{U}^+_w \simeq 10$. As discussed in \S\ \ref{sec:dr_map}, in the drag-increasing range $\omega^+/\kappa^+_x \rightarrow \mathcal{U}^+_w$ leading to $\mathcal{T}^+ \rightarrow \infty$. In other words, the spanwise oscillation becomes too slow that close to the wall, the $u$- and $w$-momentum equations are coupled together. However, GSL theory assumes that these equations are decoupled. Here, we add a new explanation based on the protrusion of the Stokes layer. As discussed in \S\ \ref{sec:deltas_DR} (figure~\ref{fig:flowviz_DR}), in the drag-increasing range, the Stokes layer is too protrusive and a near-wall cycle of turbulence is embedded within the Stokes layer. As a result, near the wall, all the terms of the momentum equation (2.1\textit{b}) are active. However, the GSL theory neglects the advection (non-linear) terms from the $w$-momentum equation. The departure of the $\left< \tilde{w}^2 \right>^*_{xt}$ profiles from the GSL solution (figure~\ref{fig:GSL_LES}\textit{b}) supports the activation of these terms.

We can now explore the net power saving ($NPS$).  \ar{Figure~\ref{fig:nps}(\textit{a})} demonstrates that for our considered parameter space $NPS$ is mostly negative. The highest (best) $NPS$ is $0.5\%$ at $(\omega^+, \kappa^+_x) \simeq (0.05, 0.012)$. \ar{In figure~\ref{fig:nps}(\textit{b}), we plot the map of the difference between $NPS$ from LES at $Re_\tau = 4000$ and its counterpart at $Re_\tau = 951$. If this difference is positive, $NPS$ increases with $Re_\tau$. Over a large portion of our parameter space, the difference is negative, i.e., $NPS$ becomes more negative with $Re_\tau$. However, for a small portion of region I with $\omega^+ <0$ and $\kappa^+_x \lesssim 0.0025$, the difference is positive. One experimental case reported by  \cite{marusic2021} falls into this region, with $\omega^+ = -0.044$ and $\kappa^+_x = 0.0014$ (see their figures~3\textit{a,b}).  The $NPS$ of this case was negative, but it increased with $Re_\tau$ in accordance with our analysis.

\cite{quadrio2009}, similar to figure~\ref{fig:nps}(\textit{a}), generate a map of $NPS$ for their travelling wave study at $Re_\tau = 200$ (their figure~5). They report $NPS>0$ within the range $0 \lesssim \omega^+ \lesssim +0.05, 0.002 \lesssim \kappa^+_x \lesssim 0.025$. This range coincides with the range where $DR \gtrsim 40\%$. \cite{gatti2013} generate a similar map at $Re_\tau = 1000$ (their figure~9). They also observe $NPS>0$ within the same range of $(\omega^+,\kappa^+_x)$. However, the level of $NPS > 0$ is lower at $Re_\tau = 1000$ compared to $Re_\tau = 200$. Considering (\ref{eq:nps}), we speculate that the decrease in $NPS>0$ from $Re_\tau = 200$ to $1000$ is due to the decrease in $DR$. We observe a similar trend in figure~\ref{fig:nps}(\textit{b}). Within the range of $(\omega^+,\kappa^+_x)$ where $DR$ is maximum (blue dashed-dotted line), $NPS$ decreases by increasing $Re_\tau$ from $1000$ to $4000$.

\begin{figure}
  \centering
  \includegraphics[width=\textwidth,trim={{0.04\textwidth} {0.03\textwidth} {0.0\textwidth} {0.0\textwidth}},clip]{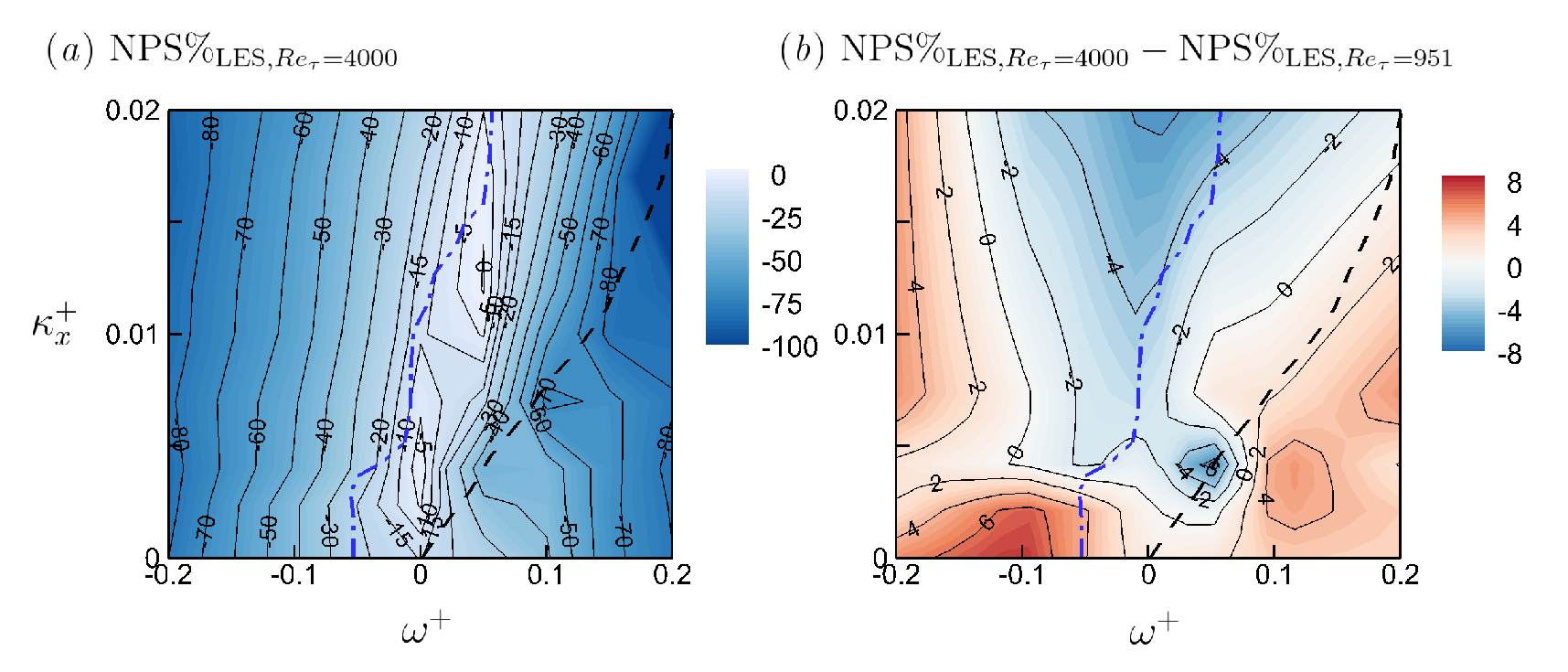}  
\caption{(\textit{a}) Net power saving $NPS$ for LES at $Re_\tau = 4000$. (\textit{b}) Difference in $NPS$ between LES at $Re_\tau = 4000$ and LES at $Re_\tau = 951$. In each of (\textit{a,b}), the filled contour and the line contours show the same quantity. All the quantities with + superscript are scaled by $\nu$ and the non-actuated $u_{\tau_o}$. ({\color{blue}\dasheddotted}) and (\dashed) locate the local maximum and the local minimum $DR$, respectively.}  
  \label{fig:nps}
\end{figure}

%, in agreement with figures~\ref{fig:GSL_LES}(\textit{c,d}). 
%An extrapolation of $NPS$ for this case to high Reynolds numbers highlights that $NPS$ becomes positive when $Re_\tau \gtrsim \mathcal{O}(10^7)$. Therefore, 
Overall, for our considered parameter space, $NPS$ is negative and predominantly decreases with Reynolds number. As discussed in \S\,\ref{sec:intro}, our parameter space  falls into the inner-scaled actuation (ISA) regime. In Part 2, we conduct experiments with some actuation parameters in the outer-scaled actuation (OSA) regime, which yield positive values for the $NPS$ that actually increase with Reynolds number. }

\section{Conclusions}\label{sec:conclusions}

Turbulent drag reduction was considered using spanwise wall oscillation based on  streamwise travelling waves at friction Reynolds numbers $Re_\tau = 951$ and $4000$ using wall-resolved large-eddy simulation in a channel flow. We conducted parametric studies at both Reynolds numbers with a fixed actuation amplitude $A^+ = 12$, for wavenumbers and frequencies within the range $0.002 \le \kappa^+_x \le 0.02$ and $-0.2 \le \omega^+ \le +0.2$, covering upstream ($\omega^+ < 0$) and downstream ($\omega^+ > 0$) travelling waves. Our actuation parameters fall into the inner-scaled actuation (ISA) regime, where only the near-wall scales are actuated. 

% For the cases with a downstream travelling wave, the logarithmic portion of the actuated profile may appear beyond $200$ viscous units above the surface, particularly for $0 \lesssim \omega^+ \lesssim +0.10$. The distant formation of the logarithmic profile is related to the protrusive nature of the Stokes layer. 

We find that GQ's model for the variation of drag reduction with Reynolds number performs well if the logarithmic shift in the velocity profile is accurately calculated. The present travelling wave actuation can highly distort the mean velocity profile and extend the beginning of the logarithmic region beyond $200$ viscous units above the surface. We find that such a high level of distortion is related to the protrusive Stokes layer. 
Accordingly, we propose a length scale $\ell_{0.01}$ for the protrusion height, where the Reynolds stress due to the Stokes layer drops to 1\% of the Reynolds stress due to the background turbulence. We find that depending on $\ell_{0.01}$, hence the Stokes layer protrusion, the $DR$ map over the parameter space of $(\omega^+,\kappa^+_x)$ can be categorised into two regions. When $\ell_{0.01}$ is less than $30$ viscous units, increasing \ar{$\ell_{0.01}$} leads to an increase in $DR$. %As a result, at each wavenumber, the maximum $DR$ is achieved when \ar{$\ell_{0.01}$ is $20 - 30$} viscous units. 
In this regime, the viscous sublayer is thickened and the logarithmic region appears at a point about $100$ viscous units above the wall. The Stokes layer acts to attenuate the turbulence below $\ell_{0.01}$ and lifts  the cycle of turbulence generation away from the wall. Increasing \ar{$\ell_{0.01}$} in this regime, further attenuates the turbulence and leads to higher $DR$. 
% All the upstream travelling wave frequencies ($\omega^+ < 0$), the standing wave ($\omega^+ = 0$) and high downstream travelling wave frequencies ($\omega^+ \gtrsim +0.2$) fall into this range.
When \ar{$\ell_{0.01}$} exceeds $30$ viscous units, however, increasing the Stokes layer thickness leads to a drop in $DR$. \ar{In this regime, the logarithmic region appears beyond $200$ viscous units above the wall. The decrease of $DR$ in this regime is due to the} Stokes layer becoming strong enough to cause a meandering of the near-wall turbulence, rather than attenuating it. That is, a cycle of  near-wall streaks appear within 10 viscous units that follow the Stokes oscillatory motion. 

% We find that the change in drag reduction $DR$ depends strongly on the change in $\ell_{0.01}$, hence the Stokes layer protrusion level.

% These cases occur in a portion of the downstream travelling wave frequencies ($0 \lesssim \omega^+ \lesssim +0.1$). 

Our power cost analysis showed that generalised Stokes layer theory agrees reasonably well with the LES data, so that it can be used with some confidence in cases where the gradient of the velocity at the wall is not accessible, as in most experiments. In addition, for our considered range of $\omega^+$ and $\kappa^+_x$ at $Re_\tau = 4000$ the net-power savings ($NPS$) was always negative. In other words, the power cost necessary to oscillate the near-wall fluid exceeds the power savings by the drag reduction. We speculate that negative $NPS$ is inevitable in the ISA pathway at least at high Reynolds numbers. We confirm this speculation in Part 2, where we investigate the ISA and OSA pathways experimentally at $Re_\tau$ up to $\mathcal{O}(10^4)$.
 
To afford the parametric study conducted here, we employed a reduced simulation domain size. This setup was found to be suitable for the ISA pathway, especially for studying $DR$, Stokes layer dynamics and the near-wall turbulence. However, the present configuration cannot resolve the outer scale eddies, which become important in the OSA pathway. This aspect will also be investigated in Part 2, where the inner and outer scale eddies are captured through experimental techniques and for higher Reynolds numbers.
 
 %We demonstrate the emergence of the OSA pathway and the effects of inertially dominated eddies on $DR$. Performing wall-resolved LES at $Re_\tau \sim \mathcal{O}\left( 10^4 \right)$ is computationally expensive, with the computational cost increasing $\propto Re^2_\tau$~\citep{piomelli2002,larsson2016}. Therefore, only wind tunnel experiments at high Reynolds numbers are capable of studying OSA. 
%In Part 2, we extend our earlier experimental investigation~\citep{marusic2021} by employing particle-image velocimetry (PIV) in addition to hot-wire anemometry (HW) and drag-balance (DB). We study the following aspects in Part 2.
%\begin{enumerate}[label=\arabic*.]
% \item $\:$ We study many sets of actuation parameters $(A^+, \kappa^+_x, \omega^+)$ in the OSA pathway $(T^+_{osc}>350)$, especially at $Re_\tau \simeq$ 9,700 and 12,800. In \cite{marusic2021}, we only studied two sets of $(A^+,\kappa^+_x,\omega^+)$ in the OSA pathway. We also study several sets of actuation parameters in the ISA pathway at $Re_\tau \simeq$ 4,500 and 6000. We compare the two pathways together through the analysis of the mean velocity, Reynolds stresses, spectrograms and flow visualisation. In the ISA pathway, our experimental observations agree with our LES observations.
 %\item $\:$ We assess $NPS$ of all the considered actuation parameters. We achieve $\mbox{$NPS$} >0$ for all the OSA parameters. In agreement with our LES observation, we find that the ISA pathway leads to $\mbox{$NPS$}<0$.
%\end{enumerate}}

\vspace{0.5cm}

\noindent \textbf{Acknowledgements}\\
The research was funded through the Deep Science Fund of Intellectual Ventures. We acknowledge Dr.\ Daniel Chung for providing insightful comments, and sharing his DNS solver and computing resources during the early stages of this work. Computing resources were provided through the Spartan High-Performance Computing service at The University of Melbourne, ARCHER2 UK National Supercomputing Service (https://www.archer2.ac.uk), the Pawsey Supercomputing Centre with funding from the Australian Government and the Government of Western Australia, and the National Computing Infrastructure (NCI), which is supported by the Australian Government.\\

\noindent \textbf{Declaration of interests.} The authors report no conflict of interest.

\appendix
\black{\section{Validation of LES and grid resolution study for drag reduction}\label{sec:grid_drag}
We perform several validation studies for LES. In this appendix, we focus on the accuracy of the dynamic Smagorinsky subgrid-scale model~\citep{germano1991} in predicting $DR$. We also assess the proper grid resolution for predicting $DR$. In Appendix~\ref{sec:aliasing}, we perform a grid resolution study for the Reynolds stresses and their spectra.

Our first validation study is summarised in figure~\ref{fig:les_validation}. We compare $DR$ between LES, the experimental data of \cite{marusic2021} and the DNS data of \cite{gatti2016}. All sets of data have matched actuation parameters $A^+ \simeq 12, \kappa^+_x \simeq 0.0014, \omega^+ \simeq -0.044$. For the LES cases (table~\ref{tab:dr_grid} above the separating line), we change $Re_\tau$ from $951$ to $6000$. The LES cases at $Re_\tau = 951$ and $6000$ are comparable with the DNS of \cite{gatti2016} and experiments of \cite{marusic2021}, respectively. All the LES cases have the viscous-scaled grid size $\Delta^+_x \times \Delta^+_z \simeq 60 \times 30$. We use the full domain size (figure~\ref{fig:geometry}\textit{c}) at $Re_\tau = 951, 2000$ and $4000$ (red bullet), and the medium domain size (figure~\ref{fig:geometry}\textit{a}) at $Re_\tau = 4000$ and $6000$ (black circle).

 \begin{table}
\centering
 \begin{tabular}[t]{ccccccc}
  \black{$Re_\tau$} & $\kappa^+_x$  & $\omega^+$ & $L_x,L_z$ & $N_x , N_y, N_z$ & $\Delta^+_x , \Delta^+_z$ & \black{Sym.}   \\  \\
  \black{$951$} & \black{$0.0014$} & \black{$-0.044$} & \black{$9.44 h, 3.14 h$} & \black{$144,48 , 96$}  & \black{$62 , 33$} & {\color{red}$\bullet$}   \\
  \black{$2000$} & \black{$0.0014$} & \black{$-0.044$} & \black{$6.73 h, 3.14 h$} & \black{$224,96 , 192$}  & \black{$60 , 33$} & {\color{red}$\bullet$}   \\ 
  \black{$4000$} & \black{$0.0014$} & \black{$-0.044$} & \black{$6.73 h, 3.14 h$} & \black{$448,192 , 384$}  & \black{$60 , 33$} & {\color{red}$\bullet$}   \\ 
  \black{$4000$} & \black{$0.0014$} & \black{$-0.044$} & \black{$2.24 h, 0.63 h$} & \black{$144,192 , 72$}  & \black{$62 , 35$} & $\circ$ \\
  \black{$6000$} & \black{$0.0014$} & \black{$-0.044$} & \black{$2.24 h, 0.63 h$} & \black{$216,288 , 108$}  & \black{$62 , 34$} & $\circ$ \\ \hline  
  \black{$951$} & $0.0347$ & $-0.20, -0.05, -0.01, +0.06, +0.12, +0.28$ & $6.86 h, 3.14 h$ & $108,48 , 96$  & $60 , 31$ & {\color{red}$\blacklozenge$}   \\
  \black{$951$} & $0.0347$ & $-0.20, -0.05, -0.01, +0.06, +0.12, +0.28$ & $6.86 h, 3.14 h$ & $288,48 , 96$  & $23 , 31$ & {\color{blue}$\blacklozenge$}  \\
  \black{$951$} & $0.0347$ & $-0.20, -0.05, -0.01, +0.06, +0.12, +0.28$ & $6.86 h, 3.14 h$ & $384,64 ,128$  & $17 , 23$ & {\color{ForestGreen}$\blacklozenge$} \\ \\  
  \black{$951$} & $0.0208$ & $-0.10, 0, +0.05, +0.07, +0.15, +0.20$ & $6.35 h, 3.14 h$ & $256,48 , 96$  & $24 , 31$ & {\color{blue}$\blacklozenge$}  \\
  \black{$951$} & $0.0208$ & $-0.10, 0, +0.05, +0.07, +0.15, +0.20$ & $6.35 h, 3.14 h$ & $384,64 ,128$  & $16 , 23$ & {\color{ForestGreen}$\blacklozenge$} \\   
\end{tabular}
\caption{\black{Summary of the LES cases for validation. The cases above the separating line have fixed actuation parameters $A^+, \kappa^+_x, \omega^+$ and grid resolution $\Delta^+_x \times \Delta^+_z$, but $Re_\tau$ changes from $951$ (first row) to $6000$ (fifth row). These cases are compared with the DNS of \cite{gatti2016} and experiments of \cite{marusic2021} at matched actuation parameters and Reynolds number (figure~\ref{fig:les_validation}). The cases below the separating line have fixed $Re_\tau = 951$ and $A^+ = 12$, but $\kappa^+_x, \omega^+$ and grid resolution change. Each row consists of six cases with fixed $Re_\tau, A^+, \kappa^+_x$ and grid resolution, but $\omega^+$ is different for each case. These cases are for validation against the DNS of \cite{gatti2016} at selected actuation parameters (figure~\ref{fig:appendixa_fig1}).}}  
\label{tab:dr_grid}
\end{table}

\begin{figure}
  \centering
  \includegraphics[width=\textwidth,trim={{0.05\textwidth} {0.09\textwidth} {0.0\textwidth} {0.0\textwidth}},clip]{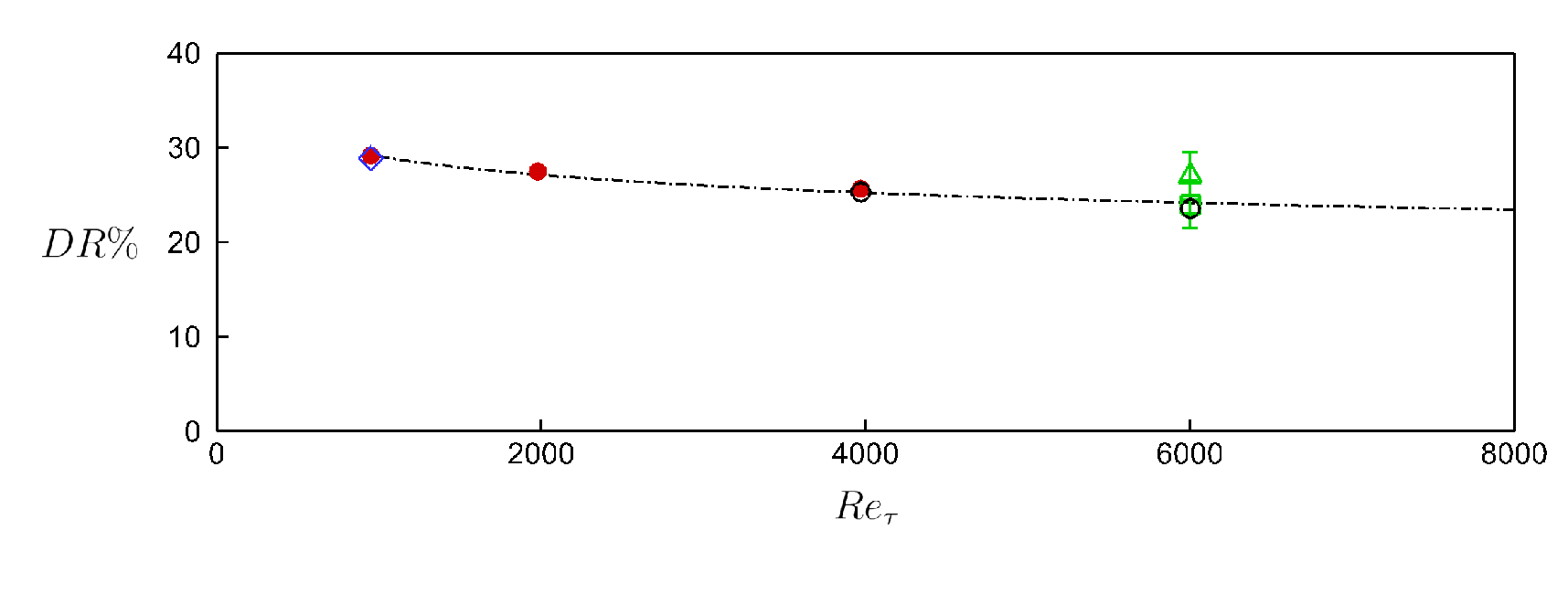}
\caption{\black{Comparison of $DR$ from the LES runs in table~\ref{tab:dr_grid} ({\color{red}$\bullet$}, $\circ$) with the DNS of \cite{gatti2016} at $Re_\tau = 951$ ({\color{blue}$\diamond$}), and experiment of \cite{marusic2021} at $Re_\tau = 6000$ using hot-wire anemometry (${\color{green}\square}$) and drag balance ({\color{green}\trian}). All the data points from different techniques have matched actuation parameters $A^+ = 12, \kappa^+_x = 0.0014, \omega^+ = -0.044$. For the LES, we use full domain (figure~\ref{fig:geometry}\textit{c}) at $Re_\tau = 951, 2000, 4000$ ({\color{red}$\bullet$}), and medium domain (figure~\ref{fig:geometry}\textit{a}) at $Re_\tau = 4000, 6000$ ($\circ$). We overlay GQ's predictive model for $DR$ (\dasheddotted).}}  
  \label{fig:les_validation}
\end{figure}

Considering figure~\ref{fig:les_validation}, at $Re_\tau = 951$ we obtain good agreement between LES (red bullet) and DNS of \cite{gatti2016} (blue diamond), and at $Re_\tau = 6000$ we obtain good agreement between LES (black circle) and the experimental data of \cite{marusic2021} from the hot-wire anemometry (green square) and drag balance (green triangle). These agreements support the accuracy of the dynamic Smagorinsky model~\citep{germano1991} for LES. At $Re_\tau = 4000$, we obtain less than $1\%$ difference between the LES case with the medium domain (black circle) and the case with the full domain (red bullet). This agreement supports the suitability of the medium domain size for the actuation parameters considered here. We further demonstrate the accuracy of the medium domain size in Appendix~\ref{sec:domain}. All the data points from DNS, LES and experiments agree well with GQ's predictive model for $DR$ (dashed-dotted line). This agreement is because the actuation frequency $\omega^+ = -0.044$ ($T^+_{osc}\simeq 142$) falls into the inner-scaled actuation pathway ($T^+_{osc} < 350$). As discussed in \cite{marusic2021} and \S\ \ref{sec:intro}, GQ's model performs accurately in this pathway. 

Our second validation study is shown in figure~\ref{fig:appendixa_fig1}. We compare the present LES with DNS dataset of \cite{gatti2016} at matched $Re_\tau = 951$ over a range of actuation parameters within our parameter space of interest. We compare at $A^+ = 12, \kappa^+_x = 0.0347$ (figure~\ref{fig:appendixa_fig1}\textit{a}) and $A^+ = 12, \kappa^+_x = 0.0208$ (figure~\ref{fig:appendixa_fig1}\textit{b}) over the range $-0.20 \lesssim \omega^+ \lesssim +0.28$. Table~\ref{tab:dr_grid} lists the LES cases for this validation study. For $A^+ = 12, \kappa^+_x = 0.0347, -0.20 \le \omega^+ \le +0.28$, we perform LES with three grids $(\Delta^+_x \times \Delta^+_z)=(60\times 31)$, $(23\times 31)$, $(17\times 23)$. For $A^+ = 12, \kappa^+_x = 0.0208, -0.10 \le \omega^+ \le +0.20$, we perform LES with two grids $(\Delta^+_x \times \Delta^+_z)=(24\times 31)$, $(16\times 23)$. Figure~\ref{fig:appendixa_fig1} shows that the LES grid $\Delta^+_x \times \Delta^+_z \simeq 23 \times 31$ (blue diamond) yields good agreement with DNS for all the compared cases. Also, this LES grid yields grid convergence. Further grid refinement to $\Delta^+_x \times \Delta^+_z \simeq 16 \times 23$ (green diamond) does not significantly change $DR$. In our first validation study with the experiments (figure~\ref{fig:les_validation}), we employed the LES grid $\Delta^+_x \times \Delta^+_z \simeq 60 \times 31$. We also employed this grid for our second validation study with $A^+ = 12, \kappa^+_x = 0.0347, -0.20 \le \omega^+ \le +0.28$ (red diamond in table~\ref{tab:dr_grid} and figure~\ref{fig:appendix_fig1}\textit{a}). We observe that this grid performs accurately for the upstream travelling wave ($\omega^+ < 0$ in figure~\ref{fig:appendix_fig1}\textit{a}). This observation is consistent with our first validation study with $\omega^+ = -0.044$ (figure~\ref{fig:les_validation}). However, for the downstream travelling wave ($\omega^+>0$), the LES grid $\Delta^+_x \times \Delta^+_z \simeq 60 \times 31$ (red diamond) under-predicts $DR$. Further refinement to $\Delta^+_x \times \Delta^+_z \simeq 23 \times 31$ (blue diamond) improves the prediction of $DR$ for all the values of $\omega^+$.

We conclude that with the viscous-scaled grid resolution of $\Delta^+_x \times \Delta^+_z \simeq 23 \times 31$ (blue diamond) we can study $DR$ with high confidence. Therefore, we adopt this grid resolution to study $DR$ (table~\ref{tab:production}).}

\begin{figure}
  \centering
  \includegraphics[width=\textwidth,trim={{0.0\textwidth} {0.05\textwidth} {0.0\textwidth} {0.0\textwidth}},clip]{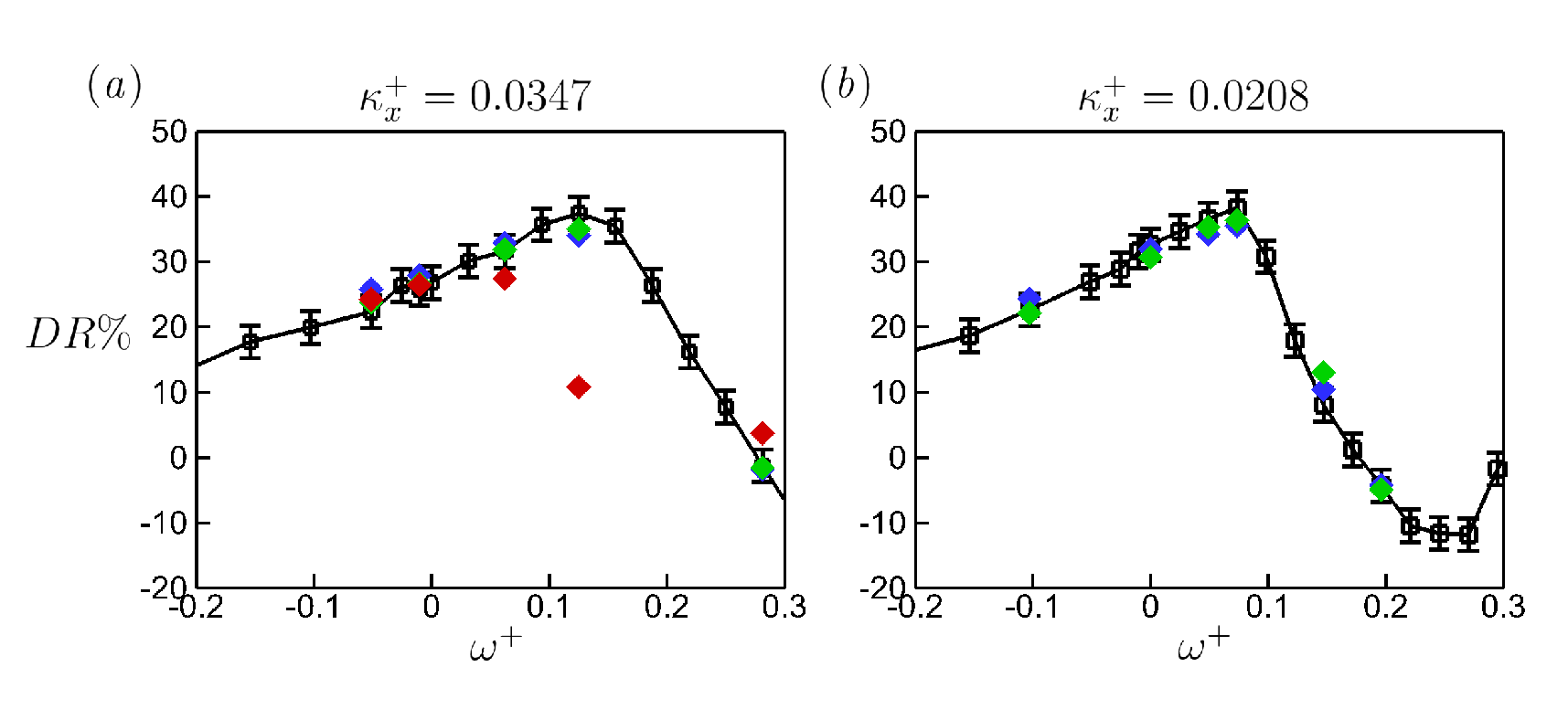}
\caption{Comparison between the DNS of \cite{gatti2016} ($\square$) and the LES of the present study at different grid resolutions: $\Delta^+_x \times \Delta^+_z \simeq 60 \times 31$ ({\color{red}$\blacklozenge$}), $ 23 \times 31$ ({\color{blue}$\blacklozenge$}), $16 \times 23$ ({\color{ForestGreen}$\blacklozenge$}). Table~\ref{tab:dr_grid} lists the simulation details for LES. Both DNS and LES cases are compared at matched $Re_\tau = 951$, $A^+ = 12$, and $\kappa^+_x, \omega^+$. (\textit{a,b}) show the comparison at $\kappa^+_x = 0.0347$ and $0.0208$, respectively. At each value of $\kappa^+_x$, comparison is made at six values of $\omega^+$ (listed in table~\ref{tab:dr_grid}).} 
  \label{fig:appendixa_fig1}
\end{figure}

\section{Grid resolution study for Reynolds stresses and spectra}\label{sec:aliasing}
Where the previous section determined the adequate grid resolution for calculating the drag reduction, we conduct a similar analysis to assess the proper grid spacing for resolving the Reynolds stresses and velocity spectra. These are the quantities that we investigate to explain the flow physics (\S\ \ref{sec:results}).
\begin{table}
\centering
 \begin{tabular}[t]{ccccccccc}
  case & $Re_\tau$ & $A^+$ & $\kappa^+_x$  & $\omega^+$ & $L_x,L_z$ & $N_x , N_y, N_z$ & $\Delta^+_x , \Delta^+_z$ & line   \\  \\
  coarse LES & $590$ & $12$ & $0.0014$ & $-0.044$ & $7.6 h, 3.14 h$ & $204,30 , 60$  & $22 , 31$ & {\color{red}\solid}   \\
  fine LES & $590$ & $12$ & $0.0014$ & $-0.044$ & $7.6 h, 3.14 h$ & $306, 45 , 90$  & $15 , 21$ & {\color{blue}\solid}  \\
  DNS & $590$ & $12$ & $0.0014$ & $-0.044$ & $7.6 h, 3.14 h$ & $608, 240 , 480$  & $7 , 4$ & {\color{green}\solid}  \\
\end{tabular}
\caption{Simulation cases for assessing the LES grid \black{for studying the} Reynolds stresses and their spectra (Appendix~\ref{sec:aliasing}). All the cases have the same $Re_\tau$, actuation parameters $(A^+, \kappa^+_x, \omega^+)$ and domain size $L_x, L_z$, where $h$ is the open-channel height. The top two cases are LES with coarse and fine grid resolutions, respectively. The third case is DNS.}  
\label{tab:aliasing}
\end{table}
\black{To evaluate the accuracy of LES for the Reynolds stresses and spectra, we generate a DNS dataset} $(\Delta^+_x \times \Delta^+_z \simeq 7 \times 4)$ in a full-domain open-channel flow with wall actuation (table~\ref{tab:aliasing}). To afford the DNS, we consider $Re_\tau = 590$ with the actuation parameters $(A^+, \kappa^+_x, \omega^+) = (12, 0.0014, -0.044)$. We perform two LES calculations that match the DNS case in terms of the domain size, $Re_\tau$ and actuation parameters, but have different grid resolutions (table~\ref{tab:aliasing}). We name the \black{LES case} with a coarser grid ($\Delta^+_x \times \Delta^+_z \simeq 22 \times 31$) ``coarse LES'', and the case with a finer grid ($\Delta^+_x \times \Delta^+_z \simeq 15 \times 21$) ``fine LES''. \black{Note that} the coarse LES case still has a fine grid for wall-resolved LES. Previous LES studies have employed a similar grid size to study a turbulent wall jet~\citep{banyassady2014} or separating turbulent boundary layer~\citep{wu2018}. Further, the coarse LES grid predicts $DR$ quite well (Appendix~\ref{sec:grid_drag}).

\begin{figure}
  \centering
  \includegraphics[width=\textwidth,trim={{0.0\textwidth} {0.05\textwidth} {0.0\textwidth} {0.0\textwidth}},clip]{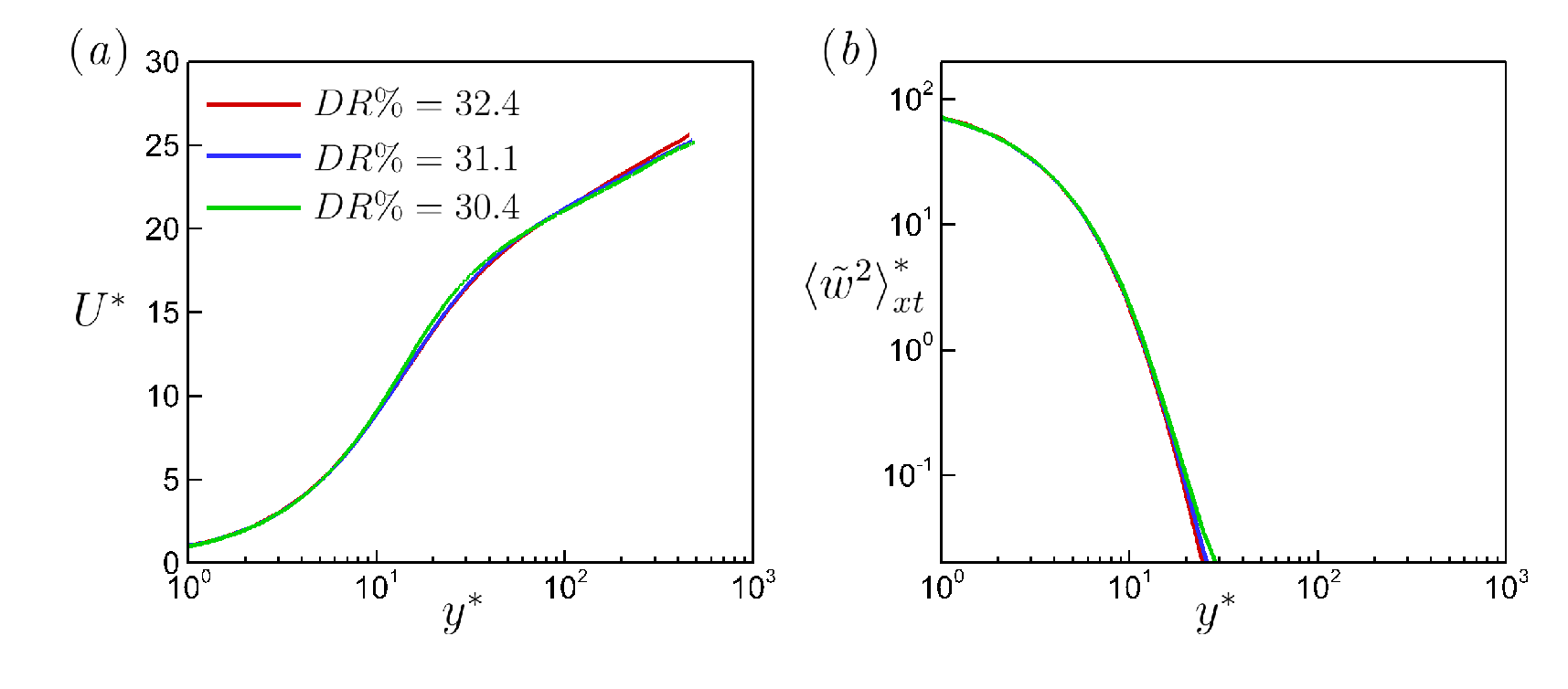}
\caption{Comparison between the coarse LES ($\Delta^+_x \times \Delta^+_z \simeq 22 \times 31$, {\color{red}\solid}), fine LES ($\Delta^+_x \times \Delta^+_z \simeq 14 \times 21$, {\color{blue}\solid}) and DNS ($\Delta^+_x \times \Delta^+_z \simeq 7 \times 4$, {\color{green}\solid}). All cases have the same $Re_\tau = 590$ and actuation parameters $(A^+, \kappa^+_x, \omega^+) = (12, 0.0014, -0.044)$, see table~\ref{tab:aliasing}. The comparison is based on (\textit{a}) mean velocity profiles $U^*$ and $DR$, and (\textit{b}) profiles of the Reynolds stress by the phase-averaged spanwise velocity \black{$\left< \tilde{w}^2 \right>^*_{xt}$}.} 
  \label{fig:appendix_fig1}
\end{figure}

\begin{figure}
  \centering
  \includegraphics[width=\textwidth,trim={{0.0\textwidth} {0.1\textwidth} {0.0\textwidth} {0.0\textwidth}},clip]{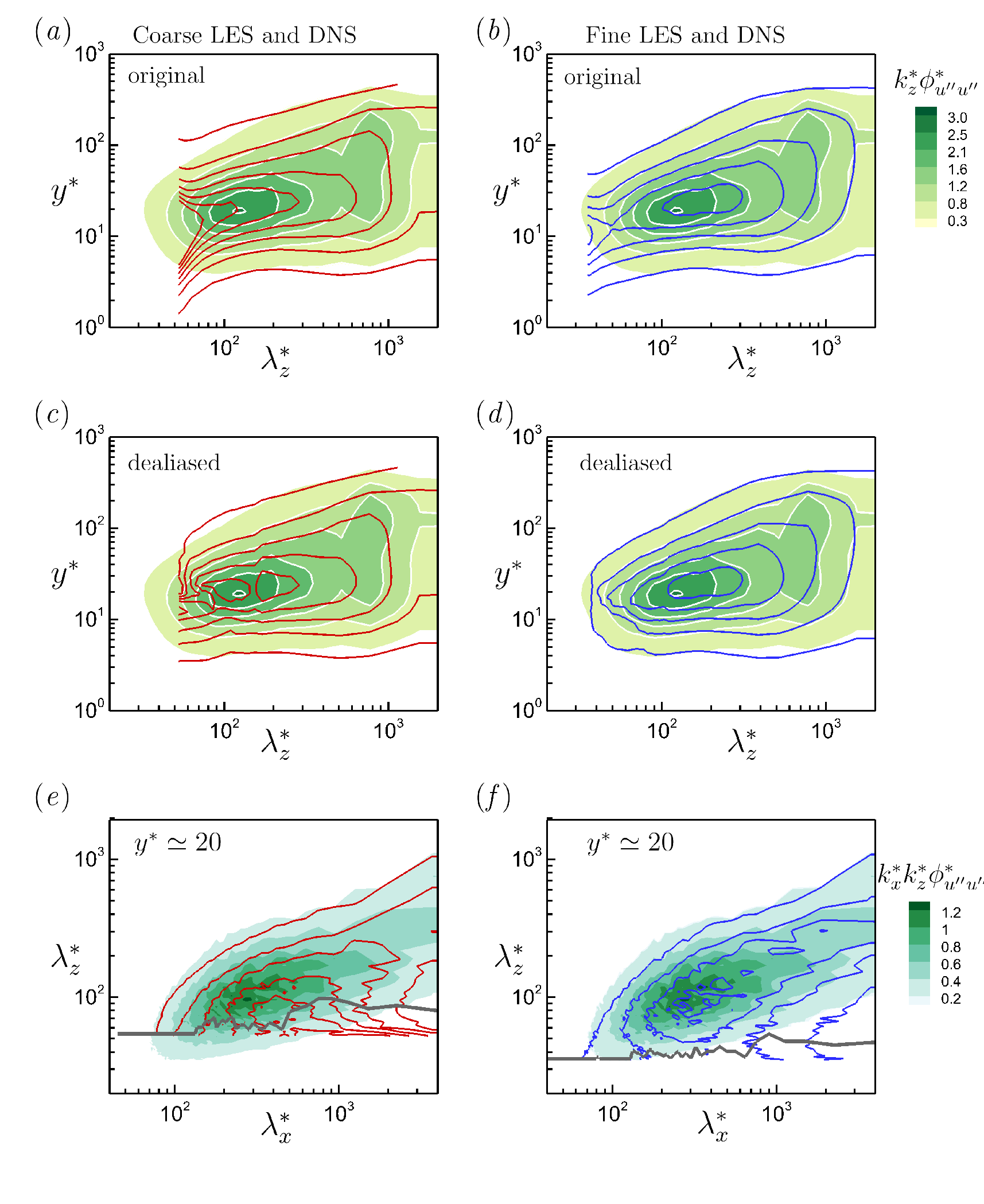}
\caption{Comparison between the coarse LES ($\Delta^+_x \times \Delta^+_z \simeq 22 \times 31$, {\color{red}\solid}), fine LES ($\Delta^+_x \times \Delta^+_z \simeq 14 \times 21$, {\color{blue}\solid}) and DNS ($\Delta^+_x \times \Delta^+_z \simeq 7 \times 4$, filled contour). All cases have matched $Re_\tau = 590$ and actuation parameters $(A^+, \kappa^+_x, \omega^+) = (12, 0.0014, -0.044)$, see table~\ref{tab:aliasing}. The comparison is made in terms of (\textit{a,b,c,d}) one-dimensional pre-multiplied spectrograms of the \black{turbulent part of the streamwise velocity $k^*_z \phi^*_{u''u''} (\lambda^*_z, y^*)$}, and (\textit{e,f}) two-dimensional pre-multiplied spectrograms of the \black{turbulent part of the streamwise velocity $k^*_x k^*_z \phi^*_{u''u''}(\lambda^*_x, \lambda^*_z)$} at $y^* \simeq 20$. (\textit{a,c,e}) \black{are} the comparison between the coarse LES and DNS, and (\textit{b,d,f}) \black{are} the comparison between the fine LES and DNS. (\textit{a,b}) compare the original spectrograms from the raw LES data (contour lines) with the DNS spectrogram (contour field). (\textit{c,d}) compare the dealiased spectrograms from LES (contour lines) with the DNS spectrogram (contour field). Dealiasing is performed through the two-dimensional spectrograms, e.g.\ by removing the scales below ({\color{gray}\thick}) in (\textit{e,f}). See the text for the details. The colourbar for (\textit{a-d}) is next to (\textit{b}), and the colourbar for (\textit{e,f}) is next to (\textit{f}).} 
  \label{fig:appendix_fig2}
\end{figure}

\begin{figure}
  \centering
  \includegraphics[width=\textwidth,trim={{0.0\textwidth} {0.05\textwidth} {0.0\textwidth} {0.0\textwidth}},clip]{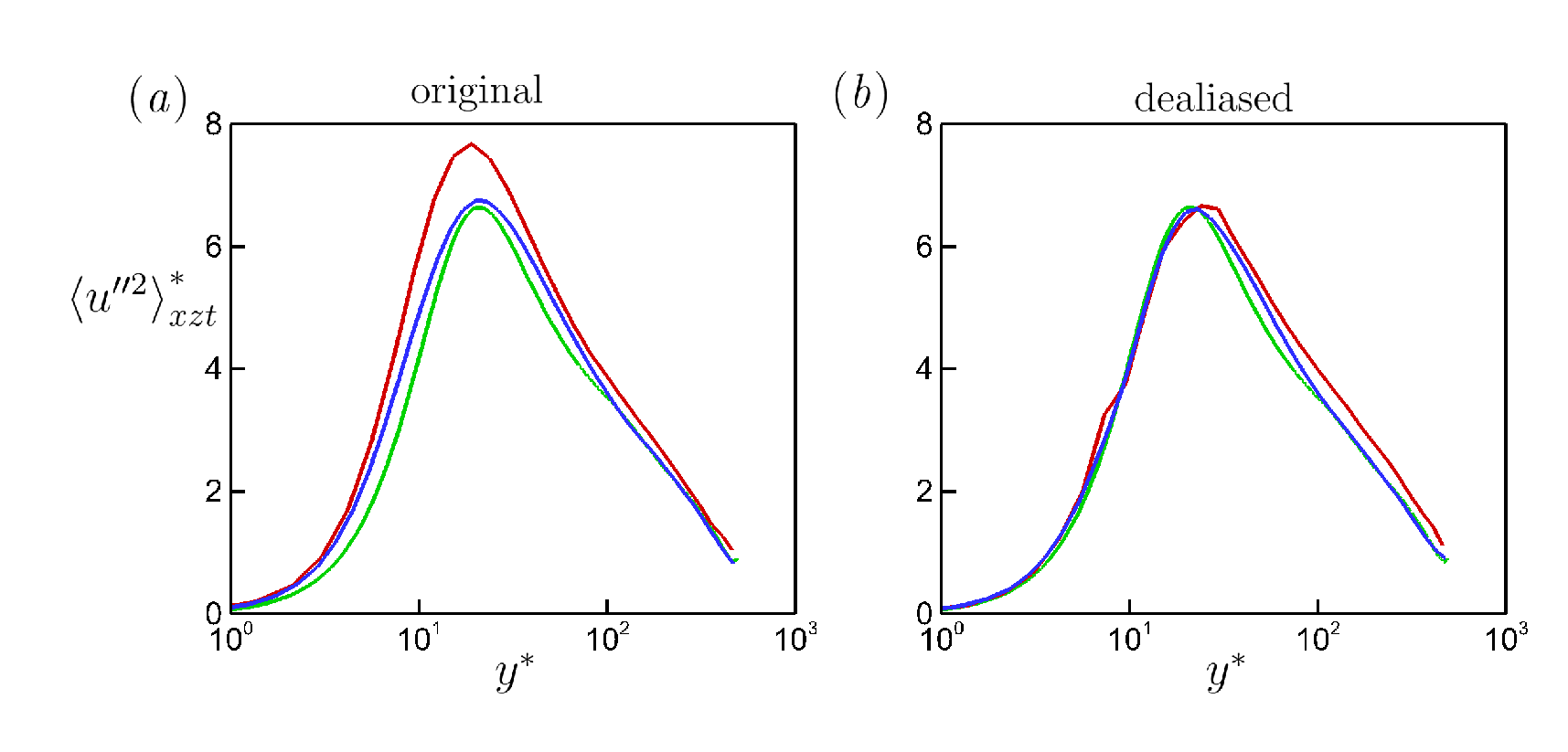}
\caption{Comparison between the coarse LES ($\Delta^+_x \times \Delta^+_z \simeq 22 \times 31$, {\color{red}\solid}), fine LES ($\Delta^+_x \times \Delta^+_z \simeq 14 \times 21$, {\color{blue}\solid}) and DNS ($\Delta^+_x \times \Delta^+_z \simeq 7 \times 4$, {\color{green}\solid}) in terms of the Reynolds stress profiles due to the \black{turbulent part of the streamwise velocity $\left< u''^2 \right>^*_{xzt}$}. All cases have the same $Re_\tau = 590$ and actuation parameters $(A^+, \kappa^+_x, \omega^+) = (12, 0.0014, -0.044)$, see table~\ref{tab:aliasing}. The LES profiles ({\color{red}\solid}, {\color{blue}\solid}) in (\textit{a}) are obtained from the raw LES data, and in (\textit{b}) are obtained by integrating the dealiased spectrograms (figures~\ref{fig:appendix_fig2}\textit{c,d}).} 
  \label{fig:appendix_fig3}
\end{figure}

In figures~\ref{fig:appendix_fig1} to \ref{fig:appendix_fig3}, we compare coarse and fine LES cases with DNS in terms of various parameters of interest. In figure~\ref{fig:appendix_fig1}, our comparison is based on the mean velocity \black{profiles} $U^*$ and $DR$ (figure~\ref{fig:appendix_fig1}\textit{a}), as well as the Reynolds stress \black{profiles} due to the phase-averaged spanwise velocity \black{$\left< \tilde{w}^2 \right>^*_{xt}$} (figure~\ref{fig:appendix_fig1}\textit{b}). \black{We use $\left< \tilde{w}^2 \right>^*_{xt}$ to calculate the protrusion height by the Stokes layer (\S\ \ref{sec:stokes_layer})}. Figure~\ref{fig:appendix_fig1} shows that $U^*, DR$ and \black{$\left< \tilde{w}^2 \right>^*_{xt}$} are predicted reasonably well with the coarse LES grid ($\Delta^+_x \times \Delta^+_z \simeq 22 \times 31$). \black{We also concluded in Appendix~\ref{sec:grid_drag} that the coarse LES grid predicts $DR$ quite well.} Therefore, we employ the coarse LES grid ($\Delta^+_x \times \Delta^+_z \simeq 22 \times 31$) to produce the maps of $DR$ (figure~\ref{fig:dr_map_compare}), \black{and} study the mean velocity profiles (figure~\ref{fig:DR_stats_kxp0_00238}), and \black{the protrusion height by the Stokes layer} (figure~\ref{fig:DR_deltas}).

However, studying the turbulent stresses and their spectra, requires the fine LES grid ($\Delta^+_x \times \Delta^+_z \simeq 14 \times 21$) as evidenced by figures~\ref{fig:appendix_fig2} and \ref{fig:appendix_fig3}. In figures~\ref{fig:appendix_fig2}(\textit{a,b}), we compare coarse LES with DNS (figure~\ref{fig:appendix_fig2}\textit{a}), and fine LES with DNS (figure~\ref{fig:appendix_fig2}\textit{b}). Our comparison is based on the one-dimensional pre-multiplied spectrogram for the \black{fluctuating streamwise velocity $k^*_z \phi^*_{u''u''} (\lambda^*_z, y^*)$}. The coarse LES spectrogram \black{(red contour lines in figure~\ref{fig:appendix_fig2}\textit{a})} is highly distorted for $\lambda^*_z \lesssim 100$. This is due to the aliasing error that energises the scales near the cut-off wavelength~\citep{kravchenko1997,park2004}. The aliasing error is clearer from the two-dimensional premultiplied spectrogram \black{$k^*_x k^*_z \phi^*_{u''u''}\left( \lambda^*_x, \lambda^*_z \right)$} at $y^* \simeq 20$ \black{(figure~\ref{fig:appendix_fig2}\textit{e})}; the coarse LES spectrogram \black{(red contour lines)} agrees well with the DNS spectrogram (filled contour) above \black{the breaking grey line}. However, below \black{the grey line,} the energy in the LES spectrogram starts to rise, while it must fall following the DNS spectrogram.

Refining the LES grid improves the spectrograms (figures~\ref{fig:appendix_fig2}\textit{b,d,f}). In figure~\ref{fig:appendix_fig2}(\textit{b}), we compare the one-dimensional spectrogram of the fine LES \black{(blue contour lines)} with DNS (filled contour). The range of scales affected by the aliasing error \black{is narrowed to $\lambda^*_z \lesssim 50$}. Attenuation of the aliasing error by the grid refinement is also evident in the two-dimensional spectrograms (\black{compare figure~\ref{fig:appendix_fig2}\textit{e} with \ref{fig:appendix_fig2}\textit{f}}). Further improvement is achieved by removing the aliased scales (dealiasing). We perform dealiasing through the \black{$k^*_x k^*_z \phi^*_{u''u''}\left( \lambda^*_x, \lambda^*_z \right)$} spectrogram at each $y^*$. \black{We can explain the dealiasing process through figures~\ref{fig:appendix_fig2}(\textit{e,f}).} At each $\lambda^*_x$, if aliasing error occurs, a local minimum appears in \black{$k^*_x k^*_z \phi^*_{u''u''}\left( \lambda^*_x, \lambda^*_z \right)$}. In figures~\ref{fig:appendix_fig2}(\textit{e,f}), we mark the local minima at all values of $\lambda^*_x$ and connect them together with \black{a grey line. Thus, the grey line} separates the healthy scales from the aliased scales. For dealiasing, we remove the aliased scales \black{below the grey line}. After dealiasing \black{$k^*_x k^*_z \phi^*_{u''u''}\left( \lambda^*_x, \lambda^*_z \right)$} at each $y^*$, we integrate it to reconstruct the dealiased one-dimensional spectrograms (figures~\ref{fig:appendix_fig2}\textit{c,d}). Accordingly, we integrate the dealiased one-dimensional spectrograms to reconstruct the dealiased Reynolds stress profiles \black{$\left< u''^2 \right>^*_{xzt}$} (figure~\ref{fig:appendix_fig3}\textit{b}). Comparing the original spectrograms from the raw LES data (figures~\ref{fig:appendix_fig2}\textit{a,b}) with the dealiased spectrograms (figures~\ref{fig:appendix_fig2}\textit{c,d}), highlights the improvement due to dealiasing. Similarly, comparing the original \black{$\left< u''^2 \right>^*_{xzt}$} profiles from the raw LES data (figure~\ref{fig:appendix_fig3}\textit{a}) with the dealiased \black{$\left< u''^2 \right>^*_{xzt}$} profiles (figure~\ref{fig:appendix_fig3}\textit{b}), highlights the improvement due to dealiasing\black{, especially for the fine LES case (blue line in figure~\ref{fig:appendix_fig3}\textit{b}).}

Overall, we conclude that the coarse LES grid ($\Delta^+_x \times \Delta^+_z \simeq 22 \times 31$) is suitable for studying $DR$, mean velocity profiles $U^*$ and \black{$\left< \tilde{w}^2 \right>^*_{xt}$} (for the Stokes layer dynamics). The fine LES resolution ($\Delta^+_x \times \Delta^+_z \simeq 14 \times 21$) with dealiasing is more suitable for studying the Reynolds stress profiles and their spectrograms.

\black{\section{Domain size study}\label{sec:domain}
In figure~\ref{fig:les_validation} we obtained very good agreement in $DR$ between the medium-domain simulation and the full-domain simulation for the case at $Re_\tau = 4000$ with $A^+ = 12, \kappa^+_x = 0.0014, \omega^+ = -0.044$. Here, we further study the domain size effect for some of our production cases at $Re_\tau = 4000$ (table~\ref{tab:domai_study}). We aim to show that the medium domain size is suitable for our parameter space of interest. We select three cases with $(\kappa^+_x, \omega^+) = (0.021, -0.1), (0.021, +0.1), (0.007, +0.05)$. The cases with $\kappa^+_x = 0.021$ fall at the upper bound of our range of interest for $\kappa^+_x$, and the case with $\kappa^+_x = 0.007$ falls within this range. Also, we consider cases with upstream travelling waves ($\omega^+ < 0$) and downstream travelling waves ($\omega^+ > 0$). We deliberately choose the case with $(\kappa^+_x, \omega^+) = (0.007, +0.05)$, because the wall actuation disturbs the flow to the highest extent (\S\ \ref{sec:results}, figures~\ref{fig:DR_stats_kxp0_00238}, \ref{fig:rms_kxp0_00700}, \ref{fig:DR_deltas}). In fact, this is the most challenging case for the application of the medium domain size among our production cases (table~\ref{tab:production}). For each case, we perform LES with the medium domain size (figure~\ref{fig:geometry}\textit{a}, $2.0h \times 0.6h, y^+_\mathrm{res} \simeq 1000$) and the large domain size (figure~\ref{fig:geometry}\textit{b}, $4.0h \times 1.2h, y^+_\mathrm{res} \simeq 2000$).

 \begin{table}
\centering
 \begin{tabular}[t]{cccccccccc}
  domain & $Re_\tau$ & $y^+_\mathrm{res}$ & $\kappa^+_x$  & $\omega^+$ & $L_x/h,,L_z/h$ & $N_x , N_y, N_z$ & $\Delta^+_x , \Delta^+_z$ & $DR$   \\  \\
  medium & $4000$ & $1000$ & $0.021$ & $- 0.1$ & $2.04,0.63$ & $384 , 192, 80$ & $21 , 31$ & $17.1$  \\
   large & $4000$ & $2000$ & $0.021$ & $- 0.1$ & $4.08,1.25$ & $768 , 192, 160$ & $21 , 31$ & $18.7$  \\ \\
   
  medium & $4000$ & $1000$ & $0.021$ & $+ 0.1$ & $2.04,0.63$ & $384 , 192, 80$ & $21 , 31$ & $21.1$  \\
   large & $4000$ & $2000$ & $0.021$ & $+ 0.1$ & $4.08,1.25$ & $768 , 192, 160$ & $21 , 31$ & $21.6$  \\ \\
   
  medium & $4000$ & $1000$ & $0.007$ & $+ 0.05$ & $2.04,0.63$ & $384 , 192, 80$ & $21 , 31$ & $11.3$  \\
   large & $4000$ & $2000$ & $0.007$ & $+ 0.05$ & $4.08,1.25$ & $768 , 192, 160$ & $21 , 31$ & $12.7$  \\    
\end{tabular}
\caption{Summary of the LES cases for the domain size study. For all cases, $A^+ = 12$. We consider three cases with $(\kappa^+_x, \omega^+) = (0.021, -0.1), (0.021, +0.1), (0.007, +0.05)$. For each case, we perform a medium domain simulation (figure~\ref{fig:geometry}\textit{a}, $L_x \times L_z \simeq 2.0h \times 0.6 h, y^+_\mathrm{res} \simeq 1000$), and a large domain simulation (figure~\ref{fig:geometry}\textit{b}, $L_x \times L_z \simeq 4.0h \times 1.2 h, y^+_\mathrm{res}\simeq 2000$).}  
\label{tab:domai_study}
\end{table}

\begin{figure}
  \centering
  \includegraphics[width=\textwidth,trim={{0.1\textwidth} {0.05\textwidth} {0.0\textwidth} {0.0\textwidth}},clip]{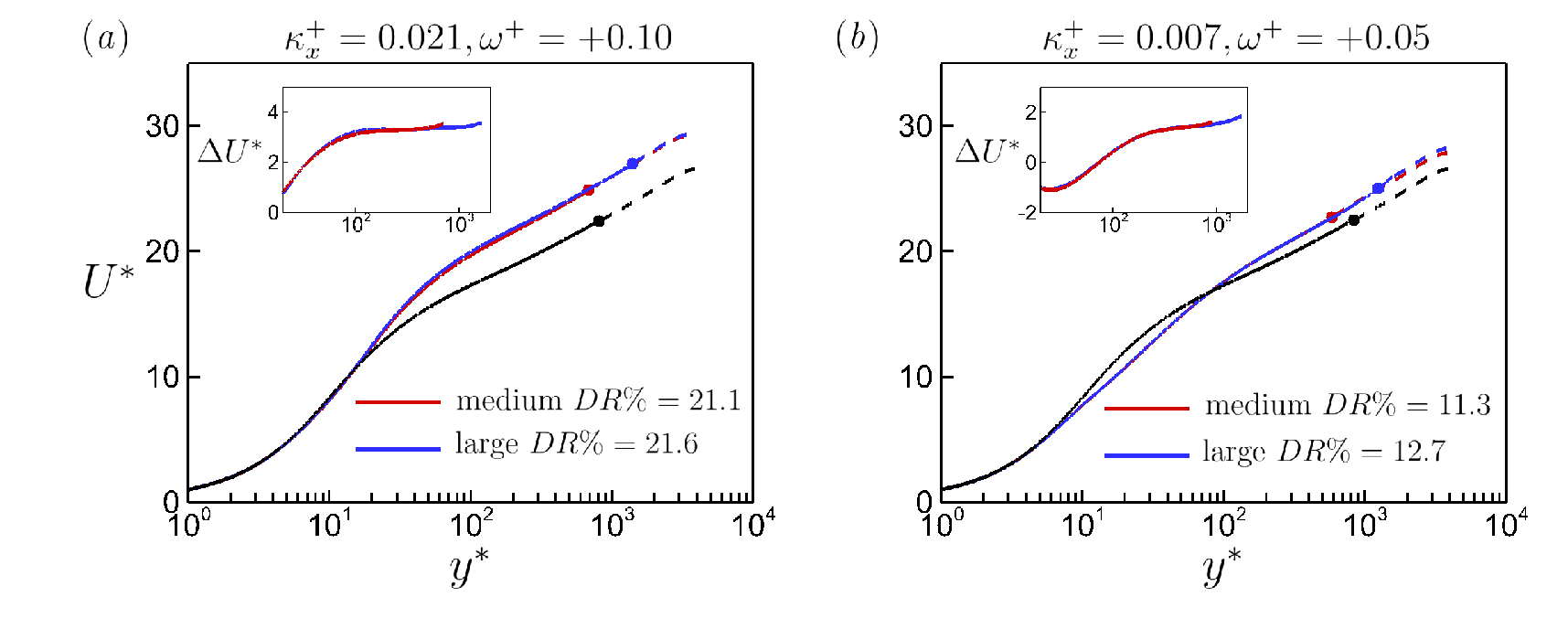}
\caption{Comparison of the mean velocity profiles $U^*$ between the medium domain simulation ({\color{red}\solid} $L_x \times L_z \simeq 2.0h \times 0.6 h, y^+_\mathrm{res} \simeq 1000$) and the large domain simulation ({\color{blue}\solid} $L_x \times L_z \simeq 4.0h \times 1.2 h, y^+_\mathrm{res}\simeq 2000$) for two actuated cases from table~\ref{tab:domai_study}: (\textit{a}) $Re_\tau = 4000, A^+ = 12, \kappa^+_x = 0.021, \omega^+ = +0.1$, and (\textit{b}) $Re_\tau = 4000, A^+ = 12, \kappa^+_x = 0.007, \omega^+ = +0.05$. In both (\textit{a,b}), we also plot the non-actuated case at $Re_\tau = 4000$ with the medium domain size (\solid). The profiles are presented in viscous units (scaled by their actual $u_\tau$ and kinematic viscosity $\nu$). The bullet points ({\color{red}$\bullet$}, {\color{blue}$\bullet$}, $\bullet$) locate the resolved height $y^*_\mathrm{res}$. The profiles ({\color{red}\dashed}, {\color{blue}\dashed}, \dashed) beyond $y^*_\mathrm{res}$ are reconstructed using the composite profile of \cite{nagib2008} (see \S\ \ref{sec:cf_dr}). The insets plot the velocity difference $\Delta U^* = U^*_\mathrm{act} - U^*_\mathrm{non{\text -}act}$.}  
  \label{fig:domain_study}
\end{figure}

We report the obtained $DR$ for each case in table~\ref{tab:domai_study}. The agreement in $DR$ between the medium domain and the large domain is quite good for all cases (within $1.6\%$ difference). We compute $DR$ (hence $C_f$), following \S\ \ref{sec:cf_dr}. First, we reconstruct the $U^*$ profile beyond $y^*_\mathrm{res}$ using \cite{nagib2008}'s composite profile, indicated with dashed line in figure~\ref{fig:domain_study}. Then, we obtain $C_f \equiv 2/{U^*_b}^2$ by integrating the resolved portion of the $U^*$ profile up to $y^*_\mathrm{res}$ (solid line in figure~\ref{fig:domain_study}) and its reconstructed portion beyond $y^*_\mathrm{res}$ (dashed line in figure~\ref{fig:domain_study}). Therefore, for the medium domain, $C_f$ is obtained by integrating the resolved $U^*$ profile up to $y^*_\mathrm{res}\simeq 1000$ and the reconstructed part beyond that. However, for the large domain size, the integrated $U^*$ profile consists of the resolved portion up to $y^*_\mathrm{res}\simeq 2000$ and the reconstructed portion beyond that. The close agreement in $DR$ between the medium domain and the large domain (table~\ref{tab:domai_study}), indicates the suitability of the medium domain size (hence sufficiency of resolving up to $y^*_\mathrm{res} \simeq 1000$). Beyond $y^*_\mathrm{res} \simeq 1000$ can be accurately reconstructed with the composite profile.

Further support for the suitability of the medium domain size is provided in figure~\ref{fig:domain_study}. We compare the profiles of the mean velocity $U^*$ and the velocity difference $\Delta U^*$ between the medium domain size (red solid line) and the large domain size (blue solid line) for two cases from table~\ref{tab:domai_study}; $\kappa^+_x = 0.021, \omega^+ = +0.1$ (figure~\ref{fig:domain_study}\textit{a}) and $\kappa^+_x = 0.007, \omega^+ = +0.05$ (figure~\ref{fig:domain_study}\textit{b}). For both actuated cases, the resolved portion of the profiles agree well between the medium domain and the large domain. We observe this agreement in the $U^*$ and $\Delta U^*$ profiles (the insets). For both cases, the logarithmic $U^*$ profile appears by $y^* \simeq 200$. This allows to use the composite profile beyond $y^* \simeq 200$. Overall, we conclude that the medium domain size ($L_x \times L_z \simeq 2.0h \times 0.6h$, figure~\ref{fig:geometry}\textit{a}) is suitable for our production simulations at $Re_\tau = 4000$ (table~\ref{tab:production}).}

\bibliography{Part1_FINAL}
\bibliographystyle{jfm}

\end{document}